\documentclass[fleqn,10pt]{article}
\pdfoutput=1
\usepackage{amsmath}
\usepackage{amssymb}
\usepackage{array}
\usepackage{calc}
\usepackage{longtable}
\usepackage{multirow}
\usepackage{amsfonts}
\usepackage{todonotes}
\usepackage{pstricks}
\usepackage{graphicx}
\usepackage{xspace}
\usepackage{units}

\numberwithin{equation}{section}
\usepackage{mciteplus}
\usepackage[pdfborder={0 0 0}]{hyperref}
\usepackage[format=hang,labelfont=bf,hypcap=true]{caption}
\usepackage{subcaption}
\usepackage{sectsty}
\allsectionsfont{\sffamily}
\subsubsectionfont{\mdseries\itshape\large}
\makeatletter
\DeclareRobustCommand*{\bfseries}{%
  \not@math@alphabet\bfseries\mathbf
  \fontseries\bfdefault\selectfont
  \boldmath
}
\makeatother
\let\spreprint\empty
\newcommand{\preprint}[1]{\def\spreprint{\protect#1}}
\let\sinstitute\empty
\newcommand{\institute}[1]{\def\sinstitute{\protect#1}}
\makeatletter
\renewcommand{\maketitle}{\begingroup
  \null\thispagestyle{empty}%
    \ifx\spreprint\empty
      \vskip 5ex
    \else
      \flushright\large\spreprint\vskip 2ex
    \fi
    \vskip 5ex
    \flushleft
      {\sffamily\bfseries\huge\@title}\vskip 6ex
      \@author\vskip 2ex
      \ifx\sinstitute\empty
      \else
        {\small\sinstitute}
      \fi
    \vskip 5ex
  \endgroup
}
\makeatother
\renewenvironment{abstract}{\begin{center}
  {\large\sffamily\bfseries Abstract: }
  \begin{minipage}[t]{0.75\textwidth}
}{\end{minipage}\end{center}\vskip 10ex}

\numberwithin{equation}{section}
\long\def\symbolfootnote[#1]#2{\begingroup%
\def\thefootnote{\fnsymbol{footnote}}\footnote[#1]{#2}\endgroup}
\newcommand{\abs}[1]{\left| #1\right|}
\newcommand{\rbr}[1]{\left( #1\right)}
\newcommand{\abr}[1]{\left< #1\right>}
\newcommand{\cbr}[1]{\left\{ #1\right\}}
\newcommand{\sbr}[1]{\left[ #1\right]}

\newcommand{\im}{\imath}
\newcommand{\jm}{\jmath}
\newcommand{\wt}[1]{\widetilde{#1}}

\newcommand{\args}[1]{\{\vec{#1}\}}

\newcommand{\dt}{{\rm d}}
\newcommand{\mc}[1]{\mathcal{#1}}
\newcommand{\mr}[1]{\mathrm{#1}}
\newcommand{\mb}[1]{\mathbb{#1}}
\newcommand{\bmap}[3]{b_{#1,#2}(#3)}
\newcommand{\rmap}[3]{r_{\widetilde{#1},\tilde{#2}}(#3)}

\newcommand{\dst}{\displaystyle}
\newcommand{\sst}{\scriptstyle}

\newcommand{\smallfeynmf}{
  \fmfset{thin}{0.7pt}
  \fmfset{thick}{1.25thin}
  \fmfset{arrow_len}{2mm}\fmfset{curly_len}{1.5mm}
  \fmfset{wiggly_len}{1.5mm}\fmfset{decor_size}{2mm}
  \fmfset{dot_len}{1mm}\fmfset{dot_size}{2thick}
  \fmfset{dash_len}{1.5mm}
  \unitlength=.5mm}

\newcommand{\dglapqa}[7]{
  \parbox{#3mm}{\begin{center}#4
      \begin{fmfgraph*}(#1,#2)
        \fmfleft{l1}
        \fmfright{dr1,r1,dr3,r2,dr2}
        \fmf{plain,label.side=right,label.dist=.07w,
          tension=2}{l1,v1}
        \fmf{fermion,label.side=left,label.dist=.05w,
          label=$\scriptstyle q$}{v1,r2}
        \fmf{plain}{v1,r1}
        \fmfv{decor.shape=circle,decor.size=.2w,decor.filled=0,
          label.dist=.15w,label.angle=120,
          label=$\scriptstyle f_q(#5,,#7)$}{v1}
        \fmffreeze
        \fmfi{plain}{vpath(__v1,__r1)shifted(2,0)rotated(5)}
        \fmfi{plain}{vpath(__v1,__r1)shifted(-16,3)rotated(-5)}
      \end{fmfgraph*}
    \end{center}} 
}
\newcommand{\dglapqba}[7]{
  \parbox{#3mm}{\begin{center}#4
      \begin{fmfgraph*}(#1,#2)
        \fmfleft{l1}
        \fmfright{dr1,r1,r3,r2,dr2}
        \fmf{plain,label.side=left,label.dist=.07w,
          tension=2}{l1,v1}
        \fmf{phantom}{v1,r2}
        \fmf{plain}{v1,r1}
        \fmfv{decor.shape=circle,decor.size=.2w,decor.filled=0,
          label.dist=.15w,label.angle=-115,
          label=$\scriptstyle f_q(#5/#6,,#7)$}{v1}
        \fmfv{decor.shape=circle,decor.size=.05w,
          label.dist=.1w,label.angle=150,
          label=$\scriptstyle P_{qq}(#6)$}{v2}
        \fmffreeze
        \fmf{fermion}{v1,v2}
        \fmf{fermion,label.side=left,label.dist=.05w,
          label=$\scriptstyle q$}{v2,r2}
        \fmf{gluon,tension=0}{v2,r3}
        \fmfi{plain}{vpath(__v1,__r1)shifted(2,0)rotated(5)}
        \fmfi{plain}{vpath(__v1,__r1)shifted(-16,3)rotated(-5)}
      \end{fmfgraph*}
    \end{center}} 
}
\newcommand{\dglapqbb}[7]{
  \parbox{#3mm}{\begin{center}#4
      \begin{fmfgraph*}(#1,#2)
        \fmfleft{l1}
        \fmfright{dr1,r1,r3,r2,dr2}
        \fmf{plain,label.side=left,label.dist=.07w,
          tension=2}{l1,v1}
        \fmf{phantom}{v1,r2}
        \fmf{plain}{v1,r1}
        \fmfv{decor.shape=circle,decor.size=.2w,decor.filled=0,
          label.dist=.15w,label.angle=-115,
          label=$\scriptstyle f_g(#5/#6,,#7)$}{v1}
        \fmfv{decor.shape=circle,decor.size=.05w,
          label.dist=.1w,label.angle=150,
          label=$\scriptstyle P_{gq}(#6)$}{v2}
        \fmffreeze
        \fmf{gluon}{v2,v1}
        \fmf{fermion,label.side=left,label.dist=.05w,
          label=$\scriptstyle q$}{v2,r2}
        \fmf{fermion,tension=0}{r3,v2}
        \fmfi{plain}{vpath(__v1,__r1)shifted(8,0)rotated(5)}
        \fmfi{plain}{vpath(__v1,__r1)shifted(-16,3)rotated(-5)}
      \end{fmfgraph*}
    \end{center}} 
}
\newcommand{\dglapga}[7]{
  \parbox{#3mm}{\begin{center}#4
      \begin{fmfgraph*}(#1,#2)
        \fmfleft{l1}
        \fmfright{dr1,r1,dr3,r2,dr2}
        \fmf{plain,label.side=right,label.dist=.07w,
          tension=2}{l1,v1}
        \fmf{gluon,label.side=right,label.dist=.09w,
          label=$\scriptstyle g$}{r2,v1}
        \fmf{plain}{v1,r1}
        \fmfv{decor.shape=circle,decor.size=.2w,decor.filled=0,
          label.dist=.15w,label.angle=120,
          label=$\scriptstyle f_g(#5,,#7)$}{v1}
        \fmffreeze
        \fmfi{plain}{vpath(__v1,__r1)shifted(8,0)rotated(5)}
        \fmfi{plain}{vpath(__v1,__r1)shifted(-16,3)rotated(-5)}
      \end{fmfgraph*}
    \end{center}} 
}
\newcommand{\dglapgba}[7]{
  \parbox{#3mm}{\begin{center}#4
      \begin{fmfgraph*}(#1,#2)
        \fmfleft{l1}
        \fmfright{dr1,r1,r3,r2,dr2}
        \fmf{plain,label.side=left,label.dist=.07w,
          tension=2}{l1,v1}
        \fmf{phantom}{v1,r2}
        \fmf{plain}{v1,r1}
        \fmfv{decor.shape=circle,decor.size=.2w,decor.filled=0,
          label.dist=.15w,label.angle=-115,
          label=$\scriptstyle f_q(#5/#6,,#7)$}{v1}
        \fmfv{decor.shape=circle,decor.size=.05w,
          label.dist=.1w,label.angle=150,
          label=$\scriptstyle P_{qg}(#6)$}{v2}
        \fmffreeze
        \fmf{fermion}{v1,v2}
        \fmf{gluon,label.side=right,label.dist=.09w,
          label=$\scriptstyle g$}{r2,v2}
        \fmf{fermion,tension=0}{v2,r3}
        \fmfi{plain}{vpath(__v1,__r1)shifted(8,0)rotated(5)}
        \fmfi{plain}{vpath(__v1,__r1)shifted(-16,3)rotated(-5)}
      \end{fmfgraph*}
    \end{center}} 
}
\newcommand{\dglapgbb}[7]{
  \parbox{#3mm}{\begin{center}#4
      \begin{fmfgraph*}(#1,#2)
        \fmfleft{l1}
        \fmfright{dr1,r1,r3,r2,dr2}
        \fmf{plain,label.side=left,label.dist=.07w,
          tension=2}{l1,v1}
        \fmf{phantom}{v1,r2}
        \fmf{plain}{v1,r1}
        \fmfv{decor.shape=circle,decor.size=.2w,decor.filled=0,
          label.dist=.15w,label.angle=-115,
          label=$\scriptstyle f_g(#5/#6,,#7)$}{v1}
        \fmfv{decor.shape=circle,decor.size=.05w,
          label.dist=.1w,label.angle=150,
          label=$\scriptstyle P_{gg}(#6)$}{v2}
        \fmffreeze
        \fmf{gluon}{v2,v1}
        \fmf{gluon,label.side=right,label.dist=.09w,
          label=$\scriptstyle g$}{r2,v2}
        \fmf{gluon,tension=0}{v2,r3}
        \fmfi{plain}{vpath(__v1,__r1)shifted(8,0)rotated(5)}
        \fmfi{plain}{vpath(__v1,__r1)shifted(-16,3)rotated(-5)}
      \end{fmfgraph*}
    \end{center}} 
}
\newcommand{\coldecga}[4]{
  \parbox{#3mm}{\begin{center}#4
      \begin{fmfgraph*}(#1,#2)
        \fmfstraight
        \fmfleft{d1,l1,d3,l2,d2}
        \fmfright{e1,r1,e3,r2,e2}
        \fmf{fermion}{l1,r1}
        \fmf{fermion}{r2,l2}
        \fmfv{decor.size=0,label.dist=.1w,label.angle=180,label=$i$}{d1}
        \fmfv{decor.size=0,label.dist=.1w,label.angle=180,label=$\bar\jmath$}{d2}
        \fmfv{decor.size=0,label.dist=.1w,label.angle=0,label=$\bar l$}{e1}
        \fmfv{decor.size=0,label.dist=.1w,label.angle=0,label=$k$}{e2}
      \end{fmfgraph*}
    \end{center}} 
}
\newcommand{\coldecgb}[4]{
  \parbox{#3mm}{\begin{center}#4
      \begin{fmfgraph*}(#1,#2)
        \fmfstraight
        \fmfleft{d1,l1,d3,l2,d2}
        \fmfright{e1,r1,e3,r2,e2}
        \fmf{plain,right=0.5}{l1,vl,l2}
        \fmf{plain,right=0.5}{r2,vr,r1}
        \fmf{dots,tension=0.5}{vl,vr}
        \fmfv{decor.size=0,label.dist=.1w,label.angle=180,label=$i$}{d1}
        \fmfv{decor.size=0,label.dist=.1w,label.angle=180,label=$\bar\jmath$}{d2}
        \fmfv{decor.size=0,label.dist=.1w,label.angle=0,label=$\bar l$}{e1}
        \fmfv{decor.size=0,label.dist=.1w,label.angle=0,label=$k$}{e2}
      \end{fmfgraph*}
    \end{center}} 
}
\newcommand{\bgrrr}[4]{
  \parbox{#3mm}{\begin{center}#4
      \begin{fmfgraph*}(#1,#2)
        \fmfcurved
        \fmfleft{l1}
        \fmfright{e1,e2,d1,r1,d2,e3,e4}
        \fmf{gluon}{v,l1}
        \fmf{phantom}{v,r1}
	\fmffreeze
        \fmf{gluon}{v,e1}
        \fmf{gluon}{v,e2}
        \fmf{gluon}{e3,v}
        \fmf{gluon}{e4,v}
        \fmffreeze
        \fmf{phantom}{v,f1,f2,m1,e2}
        \fmf{phantom}{v,g1,g2,m2,e3}
        \fmf{dots,right=0.3,tension=0.5}{m1,m2}
        \fmfv{decor.size=.25w,decor.shape=circle,decor.filled=30,label.dist=0w,label=$J^\mu$}{v}
        \fmfv{decor.size=0,label.dist=.05w,label.angle=60,label=$1$}{e4}
        \fmfv{decor.size=0,label.dist=.05w,label.angle=0,label=$2$}{e3}
        \fmfv{decor.size=0,label.dist=.05w,label.angle=0,label=$n{\sst -}2$}{e2}
        \fmfv{decor.size=0,label.dist=.05w,label.angle=-60,label=$n{\sst -}1$}{e1}
      \end{fmfgraph*}
    \end{center}} 
}
\newcommand{\bgrra}[4]{
  \parbox{#3mm	}{\begin{center}#4
      \begin{fmfgraph*}(#1,#2)
        \fmfcurved
        \fmfleft{l1}
        \fmfright{d7,d3,d4,e1,e2,d1,r1,d2,e3,e4,d5,d6,d8}
        \fmftop{h3,h4,f1,f2,h1,t1,h2,f3,f4,h5,h6}
        \fmfbottom{i3,i4,g1,g2,i1,b1,i2,g3,g4,i5,i6}
        \fmf{gluon,tension=1.5}{vx,l1}
        \fmf{phantom}{vx,r1}
	\fmffreeze
        \fmf{gluon,tension=3}{v2,vx}
        \fmf{gluon}{f3,v2}
        \fmf{gluon}{f4,v2}
        \fmf{gluon}{v2,e4}
        \fmf{gluon}{v2,e3}
        \fmf{gluon,tension=3}{vx,v3}
        \fmf{gluon}{v3,g3}
        \fmf{gluon}{v3,g4}
        \fmf{gluon}{e1,v3}
        \fmf{gluon}{e2,v3}
        \fmffreeze
        \fmf{phantom}{v2,n3,n4,m3,f4}
        \fmf{phantom}{v2,o3,o4,m4,e4}
        \fmf{dots,right=0.3,tension=0.5}{m4,m3}
        \fmf{phantom}{v3,n5,n6,m5,g4}
        \fmf{phantom}{v3,o5,o6,m6,e1}
        \fmf{dots,right=0.3,tension=0.5}{m5,m6}
        \fmfv{decor.size=0.05w,decor.shape=circle,decor.filled=full,
          label.angle=120,label.dist=0.1w,label=$V_3$}{vx}
        \fmfv{decor.size=.10w,decor.shape=circle,decor.filled=30}{v2}
        \fmfv{decor.size=0,label.dist=.05w,label.angle=90,label=$\sst 1$}{f3}
        \fmfv{decor.size=0,label.dist=.05w,label.angle=60,label=$\sst 2$}{f4}
        \fmfv{decor.size=0,label.dist=.05w,label.angle=0,label=$\sst i-1$}{e4}
        \fmfv{decor.size=0,label.dist=.05w,label.angle=0,label=$\sst i$}{e3}
        \fmfv{decor.size=.10w,decor.shape=circle,decor.filled=30}{v3}
        \fmfv{decor.size=0,label.dist=.05w,label.angle=0,label=$\sst i+1$}{e2}
        \fmfv{decor.size=0,label.dist=.05w,label.angle=0,label=$\sst i+2$}{e1}
        \fmfv{decor.size=0,label.dist=.05w,label.angle=-60,label=$\sst n-2$}{g4}
        \fmfv{decor.size=0,label.dist=.05w,label.angle=-90,label=$\sst n-1$}{g3}
      \end{fmfgraph*}
    \end{center}} 
}
\newcommand{\bgrrb}[4]{
  \parbox{#3mm}{\begin{center}#4
      \begin{fmfgraph*}(#1,#2)
        \fmfcurved
        \fmfleft{l1}
        \fmfright{d7,d3,d4,e1,e2,d1,r1,d2,e3,e4,d5,d6,d8}
        \fmftop{h3,h4,f1,f2,h1,t1,h2,f3,f4,h5,h6}
        \fmfbottom{i3,i4,g1,g2,i1,b1,i2,g3,g4,i5,i6}
        \fmf{gluon}{vx,l1}
        \fmf{phantom}{vx,r1}
	\fmffreeze
        \fmf{gluon,tension=3}{v2,vx}
        \fmf{gluon}{f1,v2}
        \fmf{gluon}{f2,v2}
        \fmf{gluon}{v2,f3}
        \fmf{gluon}{v2,f4}
        \fmf{gluon,tension=3}{v1,vx}
        \fmf{gluon}{v1,e1}
        \fmf{gluon}{v1,e2}
        \fmf{gluon}{e3,v1}
        \fmf{gluon}{e4,v1}
        \fmf{gluon,tension=3}{vx,v3}
        \fmf{gluon}{v3,g1}
        \fmf{gluon}{v3,g2}
        \fmf{gluon}{g3,v3}
        \fmf{gluon}{g4,v3}
        \fmffreeze
        \fmf{phantom}{v2,n3,n4,m3,f2}
        \fmf{phantom}{v2,o3,o4,m4,f3}
        \fmf{dots,right=0.3,tension=0.5}{m4,m3}
        \fmf{phantom}{v1,n1,n2,m1,e2}
        \fmf{phantom}{v1,o1,o2,m2,e3}
        \fmf{dots,right=0.3,tension=0.5}{m1,m2}
        \fmf{phantom}{v3,n5,n6,m5,g2}
        \fmf{phantom}{v3,o5,o6,m6,g3}
        \fmf{dots,right=0.3,tension=0.5}{m5,m6}
        \fmfv{decor.size=.05w,decor.shape=circle,decor.filled=full,
          label.angle=120,label.dist=0.1w,label=$V_4$}{vx}
        \fmfv{decor.size=.10w,decor.shape=circle,decor.filled=30}{v2}
        \fmfv{decor.size=0,label.dist=.05w,label.angle=120,label=$\sst 1$}{f1}
        \fmfv{decor.size=0,label.dist=.05w,label.angle=90,label=$\sst 2$}{f2}
        \fmfv{decor.size=0,label.dist=.05w,label.angle=90,label=$\sst i-1$}{f3}
        \fmfv{decor.size=0,label.dist=.05w,label.angle=60,label=$\sst i$}{f4}
        \fmfv{decor.size=.10w,decor.shape=circle,decor.filled=30}{v1}
        \fmfv{decor.size=0,label.dist=.05w,label.angle=60,label=$\sst i+1$}{e4}
        \fmfv{decor.size=0,label.dist=.05w,label.angle=0,label=$\sst i+2$}{e3}
        \fmfv{decor.size=0,label.dist=.05w,label.angle=0,label=$\sst j-1$}{e2}
        \fmfv{decor.size=0,label.dist=.05w,label.angle=-60,label=$\sst j$}{e1}
        \fmfv{decor.size=.10w,decor.shape=circle,decor.filled=30}{v3}
        \fmfv{decor.size=0,label.dist=.05w,label.angle=-60,label=$\sst j+1$}{g4}
        \fmfv{decor.size=0,label.dist=.05w,label.angle=-90,label=$\sst j+2$}{g3}
        \fmfv{decor.size=0,label.dist=.05w,label.angle=-90,label=$\sst n-2$}{g2}
        \fmfv{decor.size=0,label.dist=.05w,label.angle=-120,label=$\sst n-1$}{g1}
      \end{fmfgraph*}
    \end{center}} 
}
\newcommand{\clusterexample}{
  \begin{picture}(150,175)(5,-175)
    \smallfeynmf
    \put(10,-55){\begin{fmfgraph*}(40,50)
      \fmftop{t1,t2,t3,t4}
      \fmfbottom{b1}
      \fmf{dots,tension=2}{b1,v1}
      \fmf{fermion}{t1,v1}
      \fmf{fermion}{v1,t4}
      \fmffreeze
      \fmf{gluon}{v1,t2}
      \fmf{gluon}{v1,t3}
      \fmfv{decor.shape=circle,decor.size=.5w,decor.filled=0,
        label.dist=0w,label=$\sf ?$}{v1}
    \end{fmfgraph*}}
    \put(60,-35){\begin{fmfgraph*}(40,45)
      \fmfleft{l1}
      \fmftop{t1}
      \fmfbottom{b1}
      \fmf{phantom,tension=.2}{b1,v2,v1,t1}
      \fmf{phantom_arrow}{v1,b1}
      \fmffreeze
      \fmf{plain,left=.4,label=\parbox{2cm}{\sf
        \scriptsize cluster once\\find $t,,z,,\phi$}}{l1,v2}
    \end{fmfgraph*}}
    \put(60,-90){\begin{fmfgraph*}(40,50)
      \fmftop{t1,t2,t3,t4}
      \fmfbottom{b1}
      \fmf{dots,tension=2}{b1,v1}
      \fmf{fermion}{t1,v1}
      \fmf{fermion,tension=2}{v1,v2,t4}
      \fmffreeze
      \fmf{gluon}{v1,t2}
      \fmf{gluon}{t3,v2}
      \fmfv{label.dist=.05w,label.angle=-60,label=$\sf\sst t$}{v2}
      \fmfv{decor.shape=circle,decor.size=.37w,decor.filled=0,
        label.dist=0w,label=$\sst ?$}{v1}
    \end{fmfgraph*}}
    \put(57,-135){\begin{fmfgraph*}(45,40)
      \fmftop{l1}
      \fmfright{t1}
      \fmfleft{b1}
      \fmf{phantom,tension=.2}{b1,v2,v1,t1}
      \fmf{phantom_arrow}{v1,b1}
      \fmffreeze
      \fmf{plain,left=.4,label=\parbox{2cm}{\sf
        \scriptsize cluster twice\\find $t',,z',,\phi'$}}{l1,v2}
    \end{fmfgraph*}}
    \put(10,-120){\begin{fmfgraph*}(40,50)
      \fmftop{t1,t2,t3,t4}
      \fmfbottom{b1}
      \fmf{dots,tension=2}{b1,v1}
      \fmf{fermion,tension=1}{t1,v3}
      \fmf{fermion,tension=2}{v3,v1}
      \fmf{fermion,tension=2}{v2,t4}
      \fmf{fermion,tension=1}{v1,v2}
      \fmffreeze
      \fmf{gluon}{v3,t2}
      \fmf{gluon}{t3,v2}
      \fmfv{label.dist=.05w,label.angle=-60,label=$\sf\sst t$}{v2}
      \fmfv{label.dist=.05w,label.angle=-120,label=$\sf\sst t'$}{v3}
      \fmfv{decor.shape=circle,decor.size=.3w,decor.filled=0}{v1}
    \end{fmfgraph*}}
  \end{picture}
}
\newcommand{\examplesplit}{
  \begin{picture}(175,90)(2,-90)
    \smallfeynmf
    \put(75,-55){\begin{fmfgraph*}(40,50)
      \fmftop{t1,t2,t3,t4}
      \fmfright{r1,r2,r3,r4}
      \fmfbottom{b1}
      \fmf{dots,tension=2}{b1,v1}
      \fmf{fermion}{t1,v1}
      \fmf{fermion}{v1,t4}
      \fmffreeze
      \fmf{gluon}{v1,t2}
      \fmf{gluon}{v1,t3}
      \fmf{gluon,foreground=red,label.dist=0.15w,
        label.side=right}{v1,r3}
      \fmfv{decor.shape=circle,decor.size=.5w,decor.filled=0}{v1}
    \end{fmfgraph*}}
    \put(52,-40){\begin{fmfgraph*}(20,10)
      \fmfright{r1}
      \fmfleft{l1}
      \fmf{phantom}{l1,v1,m1,v2,r1}
      \fmffreeze
      \fmf{plain}{v1,v2}
      \fmf{phantom_arrow}{m1,r1}
      \fmf{phantom_arrow}{m1,l1}
    \end{fmfgraph*}}
    \put(10,-55){\begin{fmfgraph*}(40,50)
      \fmftop{t1,t2,t3,t4}
      \fmfright{r1,r2,r3,r4}
      \fmfbottom{b1}
      \fmf{dots,tension=2}{b1,v1}
      \fmf{fermion,tension=1}{t1,v3}
      \fmf{fermion,tension=2}{v3,v1}
      \fmf{phantom,tension=1}{v6,t4}
      \fmf{phantom,tension=2}{v1,v6}
      \fmffreeze
      \fmf{fermion,tension=1.33}{v2,t4}
      \fmf{fermion,tension=1.33}{v5,v2}
      \fmf{fermion,tension=1}{v1,v5}
      \fmffreeze
      \fmf{gluon}{v3,t2}
      \fmf{gluon}{t3,v2}
      \fmf{gluon,foreground=red,label.dist=0.15w,
        label.side=right}{v5,r3}
      \fmfv{label.dist=.075w,label.angle=0,label=$\sst t$}{v2}
      \fmfv{label.dist=.05w,label.angle=-120,label=$\sst t'$}{v3}
      \fmfv{decor.shape=circle,decor.size=.3w,decor.filled=0}{v1}
    \end{fmfgraph*}}
  \end{picture}
}

\hypersetup{
  pdfauthor={Stefan Hoeche},
  pdftitle={Introduction to parton-shower event generators}
}
\preprint{SLAC-PUB 16160}
\author{Stefan H{\"o}che$^1$}
\title{Introduction to parton-shower event generators}
\institute{$^1$ SLAC National Accelerator Laboratory, 
  Menlo Park, CA 94025, USA}
\usepackage{feynmf}
\begin{document}
\begin{fmffile}{tasi_fg}
\maketitle
\begin{abstract}
  This lecture discusses the physics implemented by 
  Monte Carlo event generators for hadron colliders.
  It details the construction of parton showers
  and the matching of parton showers to fixed-order 
  calculations at higher orders in perturbative QCD.
  It also discusses approaches to merge calculations for
  a varying number of jets, the interface to the 
  underlying event and hadronization.
\end{abstract}

\section{Introduction}
Hadron colliders are discovery machines. The fact that proton beams can be
accelerated to higher kinetic energy than electron beams favors hadron colliders
over lepton colliders when it comes to setting the record collision energy.
But it comes at a cost: Because of the composite nature of the beam particles,
the event structure at hadron colliders is significantly more complex than at
lepton colliders, and the description of full final states necessitates 
involved multi-particle calculations. The high-dimensional phase space
leaves Monte-Carlo integration as the only viable option. Over the past three decades, 
this led to the introduction and development of multi-purpose Monte-Carlo event generators
for hadron collider physics~\cite{Webber:1986mc,Buckley:2011ms}. This lecture series discusses
some basic aspects in the construction of these computer programs.

\subsection{Stating the problem}
The search for new theories of nature which explain the existence of dark matter
and dark energy is the focus of interest of high-energy particle physics
today. One possible discovery mode is the creation of dark matter candidates at 
ground-based collider experiments like the Large Hadron Collider (LHC).

However viable a given new physics scenario might be, all potential hadron collider
signatures have in common that they will be hidden by overwhelming Standard Model
backgrounds. The large phase space at the LHC, for example, typically leads to the
creation of $\mc{O}(100)-\mc{O}(1000)$ particles. Their momenta can span several orders 
of magnitude, and they may be subject to intricate kinematical restrictions 
imposed by the detector geometry. The most pressing problem preventing the accurate 
prediction of such final states is the non-abelian nature of Quantum Chromodynamics (QCD),
which leads to color confinement at long distances. For the complex final states 
in question, a first-principles approach to this phenomenon is currently out of reach. 
The two main problems which arise are the description of hadron formation
and the evolution of QCD final states from short to long distances. Both can,
however, be tackled to a good approximation by Monte-Carlo event generators.

In addition to QCD effects, electroweak interactions will complicate the event structure.
Most notably, the emission of soft photons in Bremsstrahlung processes may occur whenever
charged particles are produced in the final state. The computation of such processes
will not be discussed in these lectures. The interested reader is referred to the many 
excellent reviews in the literature~\cite{Buckley:2011ms}.

\subsection{Factorization of the cross section}
The production of a high invariant-mass final state, or a reaction with large invariant 
momentum transfer, can be described using the factorization Ansatz~\cite{Bodwin:1984hc,
  Collins:1985ue,Collins:1988ig}. The inclusive cross section for the production of the final
state $X$ (for example a Drell-Yan lepton pair, or a Higgs boson) in the collision of
hadron $h_1$ and $h_2$, is then given by the convolution
\begin{equation}\label{eq:factorization}
  \sigma_{h_1h_2\to X}=\sum_{a,b\in\{q,g\}}\int\dt x_a\int\dt x_b\,
  f^{h_1}_{a}(x_a,\mu_F^2)\,f^{h_2}_{b}(x_b,\mu_F^2)\;
  \int\dt\Phi_{ab\to X}\;\frac{\dt\hat{\sigma}_{ab}(\Phi_{ab\to X},\mu_F^2)}{\dt\Phi_{ab\to X}}\;.
\end{equation}
The functions $f^{h}_{a}(x,\mu_F^2)$ are the parton distribution functions (PDFs) 
in collinear factorization. At leading order in perturbative QCD they represent the probability 
for resolving a parton of flavor $a$ with momentum fraction $x$ in the parent hadron $h$
at the factorization scale $\mu_F$.
$\dt\sigma_{ab}/\dt\Phi$ denotes the differential cross section for the production of
the final state $X$ from the partonic initial state, and $\dt\Phi_{ab\to X}$ is the 
corresponding differential final-state phase-space element.

Equation~\eqref{eq:factorization} determines the total cross section for the
production of the final state $X$, but it does not specify anything beyond. 
This means in particular that any number of particles may emerge alongside $X$
and that $X$ can assume any kinematical configuration.
We therefore call Eq.~\eqref{eq:factorization} the inclusive cross section
for $X$-production. Exclusive cross sections can in principle be obtained by 
restricting the phase space for $X$, or by requiring a certain number of additional
particles, or both. The approach followed in event generators
is different: Starting from Eq.~\eqref{eq:factorization}, an inclusive 
final state is first produced, which consists only of $X$. This configuration is 
augmented by additional particles in a Markov process, where four-momentum 
and probability are conserved in the creation of each new particle.\footnote{
A slight exception to this rule is the transition from the perturbative 
to the non-perturbative regime, which will be discussed in Sec.~\ref{sec:hadronization}.}
Eventually, a high-multiplicity final state emerges which still respects
the inclusivity requirement with respect to the production of the original
final state of interest. In the following we will identify the relevant
Markov processes. One of them is what is called a parton shower.

\subsection{Collinear factorization and parton showers}
\label{sec:ps_intro}
The factorization of scattering amplitudes in the collinear limit~\cite{Dixon:1996wi,Dixon:2013uaa}
allows to derive evolution equations like the DGLAP equations~\cite{
  Gribov:1972ri,Lipatov:1974qm,Dokshitzer:1977sg,Altarelli:1977zs},
which determine the behavior of the PDFs in collinear factorization 
with changing factorization scale:
\begin{equation}\label{eq:dglap}
  \mu_F^2\frac{\dt f_a(x,\mu_F^2)}{\dt \mu_F^2}=\sum_{b\in\{q,g\}}\,
  \int_x^1\frac{\dt z}{z}\frac{\alpha_s}{2\pi}\,\hat{P}_{ba}(z)\,f_b(x/z,\mu_F^2)\;.
\end{equation}
The functions $\hat{P}_{ba}(z)$ are the regularized Altarelli-Parisi splitting
functions, which describe the collinear splitting of parton $b$ into parton $a$.
They are given by
\begin{equation}\label{eq:ap_kernels}
  \begin{split}
    \hat{P}_{qq}(z)=\;&C_F\bigg[\frac{1+z^2}{(1-z)_+}+\frac{3}{2}\,\delta(1-z)\bigg]&&&
    \hat{P}_{qg}(z)=\;&C_F\bigg[\frac{1+(1-z)^2}{z}\bigg]\\
    \hat{P}_{gq}(z)=\;&T_R\bigg[z^2+(1-z)^2\bigg]&&&
    \hat{P}_{gg}(z)=\;&2\,C_A\bigg[\frac{z}{(1-z)_+}+\frac{1-z}{z}+z(1-z)\bigg]\\
    &&&&&+\delta(1-z)\left(\frac{11}{6}C_A-\frac{2}{3}n_f T_R\right)
  \end{split}
\end{equation}
\begin{figure}
  \begin{center}
    $\dst\frac{\dst\dt}{\dst\dt\log(t/\mu^2)}
    \dglapqa{25}{25}{20}{\smallfeynmf}{x}{z}{t}\hspace*{-3mm}
    =\hspace*{6mm}\int_x^1\frac{\dt z}{z}\frac{\dst\alpha_s}{\dst 2\pi}\quad
    \dglapqba{25}{25}{22}{\smallfeynmf}{x}{z}{t}\hspace*{-3mm}
    +\;\int_x^1\frac{\dt z}{z}\frac{\dst\alpha_s}{\dst 2\pi}\quad
    \dglapqbb{25}{25}{22}{\smallfeynmf}{x}{z}{t}$
    \\
    $\dst\frac{\dst\dt}{\dst\dt\log(t/\mu^2)}
    \dglapga{25}{25}{20}{\smallfeynmf}{x}{z}{t}\hspace*{-3mm}=
    \sum\limits_{i=1}^{2\,n_f}\int_x^1\frac{\dt z}{z}\frac{\dst\alpha_s}{\dst 2\pi}\quad
    \dglapgba{25}{25}{22}{\smallfeynmf}{x}{z}{t}\hspace*{-3mm}
    +\;\int_x^1\frac{\dt z}{z}\frac{\dst\alpha_s}{\dst 2\pi}\quad
    \dglapgbb{25}{25}{22}{\smallfeynmf}{x}{z}{t}$
    \caption{Pictorial representation of the DGLAP evolution of PDFs.
      The white blob represents the incoming hadron.
      \label{fig:Q2_evolution_pdf}}
  \end{center}
\end{figure}
Schematically, the DGLAP evolution equation is shown in Fig.~\ref{fig:Q2_evolution_pdf}.
It can be interpreted in a straightforward manner: Any parton $a$, resolved in the parent
hadron at scale $\mu_F^2$, may have been produced by the branching of parton $b$, 
resolved at scale $\mu_F^2+\dt\mu_F^2$. This is precisely the Markov process we were
looking for. The transition from parton $b$ to parton $a$ is naturally accompanied by
the production of an additional parton, which accounts for momentum and flavor conservation.
The additional particle is ignored when the PDF evolution is computed. In a Monte-Carlo event generator,
it is accounted for as an additional final-state particle, and the production process is 
called initial-state radiation.

It is clear that repeated implementation of Eq.~\eqref{eq:dglap} leads to arbitrarily many
parton splittings, and therefore arbitrarily many particles in the final state. The basic idea 
leading to parton shower Monte Carlo event generators is to use Eq.~\eqref{eq:dglap} to convert 
the inclusive prediction for the occurrence of parton $a$ in the beam hadron $h$ into an
exclusive prediction for parton $a$ and a certain number of additional particles, which
are resolved at smaller and smaller momentum transfer. Two problems remain to be solved.
\begin{itemize}
\item The DGLAP equations are derived in the strict collinear limit, 
  i.e.\ any final-state particles are precisely collinear to the beam particle.
  If four momentum were conserved, this assumption would imply a vanishing 
  virtuality of the $t$-channel propagator, which conflicts with the requirement
  that $\mu_F$ be finite.
\item The DGLAP equations are fully inclusive, in the sense that 
  parton momenta are integrated over the entire available phase space. 
  Quantum Chromodynamics instead imposes a resolution scale set by $\Lambda_{\rm QCD}$. 
\end{itemize}
The first problem is solved in Monte-Carlo event generators by momentum mapping
schemes or ``recoil schemes'', which define unambiguously how the kinematics of
the process is affected when initial-state radiation occurs. This can be interpreted
as a method to assign ``spectators'', which may be a single particle or multiple
particles, that absorb the ``recoil'' when a ``splitter'' particle that was formerly 
on mass-shell branches into two on-shell particles. It is obvious that if the 
splitter has zero on-shell mass, this can only be achieved through absorption
of kinetic energy from another part of the reaction, the spectator.

The second problem is solved by truncating the evolution at a scale of order $\Lambda_{\rm QCD}$.
Due to the fact that parton showers implement four-momentum conservation, this implies
a restricted range in the integral over energy fractions in the DGLAP equation, 
Eq.~\eqref{eq:dglap}. In the following, we discuss the implications of these modifications.

\subsection{Basic parton-shower kinematics}
\label{sec:ps_kin_intro}
\begin{figure}
  \begin{center}
    \includegraphics[width=0.8\textwidth]{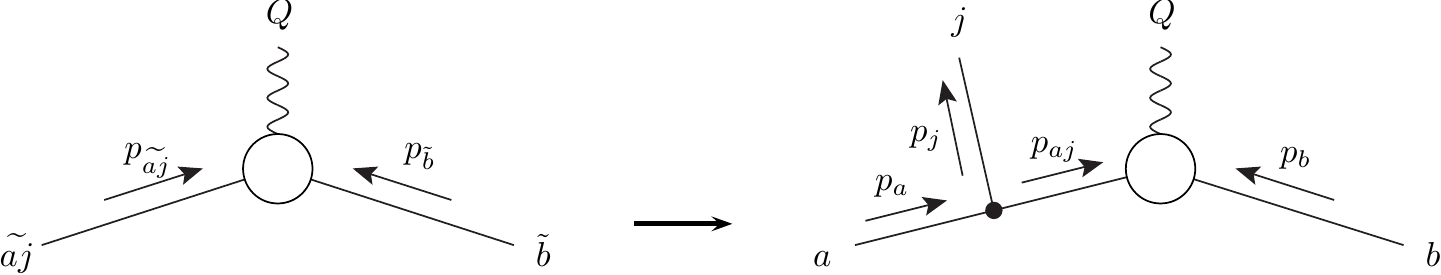}
    \caption{Kinematics in the initial-state parton splitting process $a\to \{aj\}j$.
      The virtuality of parton $\{aj\}$ entering the hard process is given by $t$,
      while its light-cone momentum fraction with respect to the new initial-state
      parton is $z$. The splitting process has an azimuthal symmetry, which 
      may be broken if the $t$-channel parton is a gluon and the hard process
      has a non-trivial Lorentz structure.
      \label{fig:split_kinematics}}
  \end{center}
\end{figure}
Consider the splitting process depicted in Fig.~\ref{fig:split_kinematics}. 
All particles are assumed to have zero on-shell mass. We parametrize their momenta
in terms of the light-cone momenta $p_a$ and $n$, where $n$ is a light-like 
reference vector that satisfies $np_a\neq 0$.
We can then use the Sudakov decomposition~\cite{Sudakov:1954sw}
\begin{align}\label{eq:sud_decomposition}
  p_{aj}^\mu=\;&\tilde{z}\,p_a^\mu+\frac{-2\,p_ap_j+k_\perp^2}{\tilde{z}}\frac{n^\mu}{2np_a}+k_\perp^\mu\;,
  &p_j^\mu=\;&(1-\tilde{z})\,p_a^\mu+\frac{k_\perp^2}{1-\tilde{z}}\frac{n^\mu}{2np_a}-k_\perp^\mu\;.
\end{align}
In parton shower generators, $n$ is identified with the spectator momentum, i.e.\
the momentum of the particle (or set of particles), which recoils against the splitter.
If the spectator has zero mass, like in our example, $2\,np_a$ can be identified with
the invariant mass of the radiating color dipole. We will use this concept extensively
in the future. A more precise definition of color dipoles will be given in 
Sec.~\ref{sec:hard_process}. For now we work under the assumption that the final state 
$X$ does not carry color charge, hence the spectator parton is the opposite-side
beam particle. We can then replace $n$ with $p_b$.

Clearly, any recoil scheme must satisfy the condition that the new initial-state momentum 
after splitting, $p_a$, is parallel to the beam direction. We compute it by rescaling:
\begin{equation}
  p_a^\mu=\frac{p_ap_b}{p_{\wt{a\jm}}p_{\tilde{b}}}\,p_{\wt{a\jm}}^\mu\;.
\end{equation}
A peculiarity of initial-state evolution is that the spectator momentum must also remain
aligned with the beam axis. In the simplest possible scheme we have $p_b=p_{\tilde{b}}$.
Since $p_c$ is not collinear to the beam, the final state $X$ must absorb all transverse
momentum generated in the splitting, but it does not change its invariant mass. This leads
to a Lorentz transformation, which acts on all final-state momenta $p_{\tilde{\im}}$
as~\cite{Catani:1996vz}
\begin{equation}
  p_i^\mu=p_{\tilde{\im}}^\mu
  -\frac{2\,p_{\tilde{\im}}(K+\tilde{K})}{(K+\tilde{K})^2}\,(K+\tilde{K})^\mu
  +\frac{2\,p_{\tilde{\im}}\tilde{K}}{\tilde{K}^2}\,K^\mu\;,
  \quad\text{where}\quad
  K^\mu=p_a^\mu-p_j^\mu+p_b^\mu\;,\quad
  \tilde{K}^\mu=p_{\wt{a\jm}}^\mu+p_b^\mu
\end{equation}
It is the repeated application of this Lorentz transformation which resums large
logarithmic corrections to the transverse momentum of the Higgs boson, for example.
The relation between the transverse momentum, $k_T$, generated in a single splitting
and the corresponding light-cone momentum fraction, $\tilde{z}$, is given by
\begin{equation}\label{eq:is_kt_z_relation}
  k_T^2=-t\,(1-\tilde{z})\;,\qquad\text{where}\qquad t=-2\,p_ap_j\;.
\end{equation}
Both the transverse momentum and the light-cone momentum fraction are Lorentz invariants,
as can be inferred from multiplying Eq.~\eqref{eq:sud_decomposition} by $n_\mu$. 
It follows that the kinematics reconstruction can be achieved in any Lorentz frame.
We will now connect the kinematical variables $t$ and $\tilde{z}$ to the evolution.

\subsection{Exclusive evolution equations and Sudakov factors}
Partons are bound by confinement into color-neutral hadrons at momentum scales of order
$\Lambda_{\rm QCD}$. This implies that both experimentally and theoretically, partons which 
are closer than about 1~GeV in transverse momentum cannot be separately resolved.
This condition introduces a natural cutoff scale for the transverse momentum in
Eq.~\eqref{eq:is_kt_z_relation}, which we call the infrared cutoff scale of the
parton-shower, or the parton-shower cutoff, $t_c$, for short. Since four-momentum 
is conserved in each splitting, the cutoff leads to an upper bound on $\tilde{z}$
\begin{equation}
  \tilde{z}=1-\frac{k_T^2}{|t|}<1-\frac{t_c}{|t_{\rm max}|}\;.
\end{equation}
If we identify $\tilde{z}$ with the energy fraction $z$ in the DGLAP evolution
equations, Eq.~\eqref{eq:dglap}, then the Altarelli-Parisi splitting functions,
Eq.~\eqref{eq:ap_kernels} may be replaced by their unregularized counterparts,
$P_{ba}(z)$, which are obtained by simply dropping the $+$-prescription and
the term proportional to $\delta(1-z)$.

If we made no further modifications, unitarity would be 
violated, as we have effectively removed all singularities in the higher-order 
real-emission contributions to the hard cross section, but also all virtual
corrections. This can be remedied by adding an additional term to the 
DGLAP equations, which reinstates the difference.
\begin{equation}\label{eq:dglap_excl}
  \frac{\dt f_a(x,t)}{\dt\log t}=\sum_{b\in\{q,g\}}\,
  \int_x^{z_{\rm max}}\frac{\dt z}{z}\frac{\alpha_s}{2\pi}\,P_{ba}(z)\,f_b(x/z,t)
  -f_a(x,t)\sum_{b\in\{q,g\}}\,\int_{z_{\rm min}}^{z_{\rm max}}\dt z\,
  \frac{\alpha_s}{2\pi}\,\frac{1}{2}\,P_{ab}(z)\;.
\end{equation}
At the same time we have introduced $t$ as the evolution variable of our 
parton shower. We identify this variable with the factorization scale,
such that the $\mu_F^2$ evolution described by Eq.~\eqref{eq:dglap} turns 
into a $t$-evolution. For now we will leave the precise assignment of $t$
an open question. It should be identified with a variable which is linear 
in the virtuality of the intermediate parton, the only dimensionful variable 
in the splitting process.

Equation~\eqref{eq:dglap_excl} may be rewritten in a more convenient 
fashion using the Sudakov form factor
\begin{equation}\label{eq:sudakov_intro}
  \Delta_a(t,t')=\exp\left\{-\sum_{b\in\{q,g\}}\,
  \int_t^{t'}\frac{\dt\bar{t}}{\bar{t}}\int_{z_{\rm min}}^{z_{\rm max}}\dt z\,
  \frac{\alpha_s}{2\pi}\,\frac{1}{2}\,P_{ab}(z)\right\}\;,
\end{equation}
which represents the unconditional survival probability for a parton 
not to undergo a branching process between the two scales $t'$ and $t$. 
In terms of $\Delta$, Eq.~\eqref{eq:dglap_excl} becomes the master equation 
for our parton shower:
\begin{equation}\label{eq:dglap_sud}
  \frac{\dt}{\dt\log t}\,\log\frac{f_a(x,t)}{\Delta_a(t_c,t)}=
  \sum_{b\in\{q,g\}}\,\int_x^{z_{\rm max}}\frac{\dt z}{z}
  \frac{\alpha_s}{2\pi}\,\hat{P}_{ba}(z)\,\frac{f_b(x/z,t)}{f_a(x,t)}\;.
\end{equation}
This equation is solved in one or the other way by any parton-shower Monte-Carlo.
All event generators have in common that they use Sudakov factors to
account for unresolved splittings and virtual corrections, which are assumed
to precisely cancel the real corrections when integrated over phase space.
The computation of the Sudakov factor is therefore the principal task for
any parton-shower Monte-Carlo event generator. We will discuss the related
algorithms in Sec.~\ref{sec:parton_showers}.

Despite all its intricacies, Eq.~\eqref{eq:dglap_sud} still only leads to 
an approximate description of fully exclusive final states containing our
initial process of interest, $pp\to X$. If detailed experimental measurements 
are to probe the precise distribution of hard QCD radiation, then we need to
improve Eq.~\eqref{eq:dglap_sud} by replacing the Altarelli-Parisi splitting
functions by more precise expressions, at least for the most relevant steps in the evolution.
This will be the subject of Sec.~\ref{sec:matching_merging}. 

The concept of infrared-safe observables and QCD jets plays a crucial role in this context.
Both the initial state and many final states at hadron colliders include hard partons.
Initial- and final-state Bremsstrahlung dresses these partons with further radiation, as we have 
seen above. The new particles are found predominantly in the vicinity of the original ones, 
leading to clusters of radiation called QCD jets. The jet structure is preserved
when hadrons are formed. A cluster of hadronic energy in the experimental measurement
can thus be associated with one or more hard initiating partons in the theoretical calculation. 
For this concept to work an algorithm must be defined that unambiguously relates the two.
Crucially, this algorithm must be infrared and collinear safe: if a single parton 
is replaced by a set of collinear partons sharing its original energy, the jet configuration 
must not change. Likewise, if a parton of vanishing energy is added to the original 
event, the identified jet configuration must not change. More details on jet algorithms
can be found in~\cite{Salam:2009jx}.

\begin{figure}
  \begin{center}
    \includegraphics[width=0.55\textwidth]{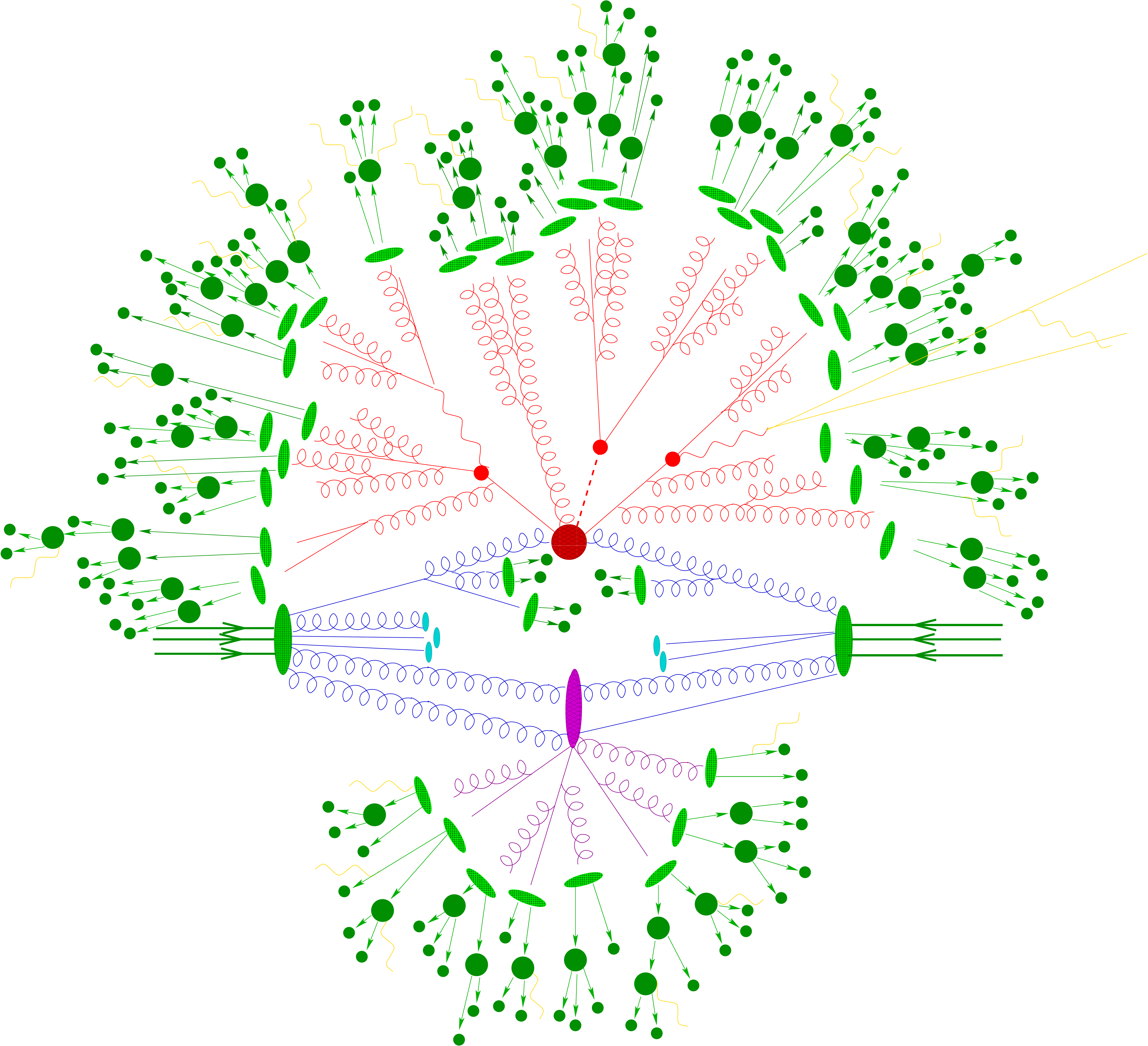}
    \caption{Sketch of a hadron-hadron collision as simulated by a Monte-Carlo
      event generator. The red blob in the center represents the hard collision,
      surrounded by a tree-like structure representing Bremsstrahlung as simulated
      by parton showers. The purple blob indicates a secondary hard scattering event.
      Parton-to-hadron transitions are represented by light green blobs, dark green 
      blobs indicate hadron decays, while yellow lines signal soft photon radiation.
      \label{fig:event}}
  \end{center}
\end{figure}

At hadron colliders, multiple scattering and rescattering effects arise, which
must be simulated by Monte-Carlo event generators in order to reflect the full 
complexity of the event structure. This will be discussed in Sec.~\ref{sec:underlying_event}.
Eventually we need to convert the full partonic final state into a set
of color-neutral hadrons, which is the topic of Sec.~\ref{sec:hadronization}.
The interplay of all these effects makes for the full simulation of hadron-hadron
collisions. This is sketched in Fig.~\ref{fig:event}.

\section{The hard scattering}
\label{sec:hard_process}
Event simulation in parton-shower Monte-Carlo event generators starts with the 
computation of the hard-scattering cross section at some given order in perturbation theory.
Traditionally, this calculation was performed at leading order (LO), but nowadays, 
with next-to-leading-order (NLO) calculations completely automated, it is often 
done at NLO. Computing the hard cross section at NLO requires a dedicated 
matching to the parton shower, which will be discussed in Sec.~\ref{sec:matching_merging}. 
For now we focus on the evaluation of the differential cross sections and the related 
phase-space integrals.

The basis for our calculations is the factorization formula, Eq.~\eqref{eq:factorization}.
We rewrite it here, in order to simplify the discussions in the following sections.
The full initial and final state in a $2\to (n-2)$ reaction can be identified by 
a set of $n$ particles, which is denoted by $\args{a}=\{a_1,\ldots,a_n\}$. 
Their flavors and momenta are similarly specified as $\args{f\,}=\{f_1,\ldots,f_n\}$ 
and $\args{p}=\{p_1,\ldots,p_n\}$. The differential cross section at leading order
is a sum over all flavor configurations, and it depends only on the parton momenta:
\begin{equation}\label{eq:definition_born_xs}
  \dt\sigma^{\rm(LO)}(\args{p})\,=\;
  \sum_{\args{f\,}}\dt\sigma^{\rm(B)}_n(\args{a})\;,
  \qquad\text{where}\qquad
  \dt\sigma^{\rm(B)}_n(\args{a})\,=\;\dt\bar{\Phi}_n(\args{p})\,\mr{B}_n(\args{a})\;.
\end{equation}
Each individual term in the sum consists of the differential phase-space element, 
$\dt{\Phi}_n$, the squared matrix elements, $\abs{\mc{M}_n}^2$, as well as
parton luminosities, $\mc{L}$, flux ($F$) and symmetry factors ($S$) 
\begin{equation}\label{eq:definition_born_xs_parts}
  \begin{split}
  \mr{B}_n(\args{a})\,=&\;\mc{L}(\args{a})\,\mc{B}_n(\args{a})\;,
  &\mc{B}_n(\args{a})\,=&\;
  \frac{1}{F(\args{p})}\,\frac{1}{S(\args{f\,})}\,
  \abs{\mc{M}_n}^2(\args{a})\;,\\
  \dt\bar{\Phi}_n(\args{p})\,=&\;\frac{\dt x_1}{x_1}\frac{\dt x_2}{x_2}\,
  \dt{\Phi}_n(\args{p})\;,
  &\mc{L}(\args{a};\mu^2)\,=&\;x_1f_{f_1}(x_1,\mu^2)\;x_2f_{f_2}(x_2,\mu^2)\;.
  \end{split}
\end{equation}
Two challenges arise in the implementation of these formulae:
\begin{itemize}
\item The squared matrix elements are tedious to compute if more than two 
  particles are involved in the final state. Problems arise both in
  the management of the Lorentz structure and of the color structure, as the number
  of Feynman diagrams grows factorially with the number of external particles.
\item The phase-space integrals are hard to evaluate for processes with 
  high particle multiplicity in the final state. As the number of Feynman diagrams
  grows factorially, so does the number of peaks in the integrand, which are 
  related to particle propagators becoming (close to) on-shell.
\end{itemize}
Many solutions have been proposed to deal with the above problems. We will
review only a few of them here, which are simple and generic and can be
implemented regardless of the process in question.

\subsection{Quantum number management}
For each phase-space point, a sum over unobserved external quantum numbers 
and an averaging over initial states must in principle be performed. This involves
in particular the sum over colors and helicities. At the level of squared amplitudes,
this sum can be performed analytically, using completeness relations and color algebra.
While this leads to structurally simple results and allows analytic insight into
the dynamics of the process, it quickly becomes too cumbersome to be carried out
at large particle multiplicities, even for powerful computer algorithms, and 
it is sometimes more useful to compute the sum in a Monte-Carlo fashion.
Consider an $n$-gluon amplitude: at tree-level, the sum over external states 
involves $\mc{O}(2^n)$ nonvanishing terms of different helicity. The growth
with the number of external states is actually rather mild, and the sum can
therefore be computed explicitly. In contrast, the sum over color involves
$\mc{O}(8^n)$ terms, which is clearly beyond computational capabilities
as soon as $n$ gets close to eight. It is therefore worth thinking about
a method to sample color space efficiently.

A powerful organizing principle for calculations of QCD matrix elements is 
the expansion in the number of colors, $N_C$. It relies on the fact that any
octet state of $SU(3)$ may be represented as a nonet state minus the overcounted
singlet, or $3\otimes\bar{3}=8\oplus1$. More specifically
\footnote{Note that we normalize the $SU(3)$ generators as
  $T^a_{i\bar\jmath}T^b_{\bar\jmath i}\,=\;\delta^{ab}$.
  The color-ordered Feynman rules then include an additional factor $\sqrt{2}$.
  Details on this convention are found in~\cite{Dixon:1996wi}.}
\begin{equation}
  \begin{split}
  T^a_{i\bar\jmath}T^a_{k\bar l}\,=&\;
  \delta_{i\bar l}\delta_{k\bar\jmath}-
    \frac{1}{N_C}\,\delta_{i\bar\jmath}\delta_{k\bar l}\quad\leftrightarrow
  \text{\coldecga{25}{15}{25}{\smallfeynmf}}-\frac{1}{N_C}
  \text{\coldecgb{25}{15}{25}{\smallfeynmf}}\;.
  \end{split}
\end{equation}
This equation relates the octet in color space to two triplet/antitriplet terms.
The second term in the expression, which can be interpreted as the overcounted
$U(1)$ gluon, is suppressed by one over the number of colors. In many approximate 
hard cross section calculations, this term is (partially) dropped in order to ease 
the computation. This can lead to substantial simplifications, and in some cases is needed to
make the calculation possible. Crucially, neglecting this term is also the
basis for any standard parton-shower algorithm (although improvements exist, 
see for example~\cite{Platzer:2012np}). We will return to this subject 
in Sec.~\ref{sec:large_nc_ps}.

Consider the computation of an all-gluon amplitude at tree-level.
We assume fixed color assignments of the external gluons, i.e.\ for each external
gluon the color index in the adjoint representation of $SU(3)$ is fixed. 
We call this amplitude $\mc{M}_n(\args{a})$.
It can be factorized into phase-space independent coefficients which are functions
of the color structure only, and phase-space dependent partial amplitudes, 
also called color-ordered amplitudes~\cite{Mangano:1987xk}:
\begin{equation}\label{eq:fund_decomposition} 
  \mc{M}_n(\args{a})\,=\;\sum_{\args{\sigma} \in S_{n-1}}
    {\rm Tr}\sbr{\,T^{a_1} T^{a_{\sigma_2}}\ldots T^{a_{\sigma_n}}\,}\,
    A_n(1,\sigma_2,\ldots,\sigma_n)\;.
\end{equation}
The sum is over all $(n-1)!$ permutations of the indices $(2,\ldots,n)$.
Each trace corresponds to a particular color structure. $A_n(1,\sigma_2,\ldots,\sigma_n)$
are the partial amplitudes, which depend on the four-momenta $\args{p}$ of the gluons
permuted according to $\args{\sigma}$. The color-ordered amplitudes are much easier 
to calculate than the full amplitude, as they contain only planar Feynman diagrams. 

The decomposition in Eq.~\eqref{eq:fund_decomposition} is not unique. 
A method better suited to Monte-Carlo treatment is the color-flow decomposition.
As the name suggests, it corresponds to identifying the color flow in terms of 
fundamental $3$ and $\bar{3}$ indices, which then also define the color state 
of external gluons in $\mc{M}_n$~\cite{Kanaki:2000ey,Maltoni:2002mq}.
The main advantage is that color factors are products of delta functions:
\begin{equation}\label{eq:colorflow_decomposition}
  \mc{M}_n(\args{a})\,=\;\sum_{\sigma \in S_{n-1}}
    \delta_{i_1}^{\bar\jmath_{\sigma_2}} 
    \delta_{i_{\sigma_2}}^{\bar \jmath_{\sigma_3}} \cdots 
    \delta_{i_{\sigma_{n}}}^{\bar \jmath_1}\,A_n(1,\sigma_2,\ldots,\sigma_n)\;.
\end{equation}
Equation~\eqref{eq:colorflow_decomposition} is straightforwardly implemented in
a computer program, since no costly matrix multiplications of complex valued matrices 
has to be performed, but only integer comparisons. Similar decompositions exist 
for all tree-level parton amplitudes including any number of quark pairs, 
gluons and color-singlet objects.

In order to compute the color-summed matrix element squared, we need to square
Eq.~\eqref{eq:fund_decomposition} and sum over adjoint color indices assigned
to the external gluons. Alternatively, we can square Eq.~\eqref{eq:colorflow_decomposition}
and sum over $3\otimes\bar{3}$ indices assigned to the external gluons. However,
Eq.~\eqref{eq:colorflow_decomposition} squared already is a theoretically meaningful
result (although not a measurable one). Instead of explicitly summing color flows
we may sample them using Monte-Carlo methods. In this case the computational
effort per phase space point is much reduced, because a single color configuration
leads to fewer partial amplitudes on average than the sum over all, 
which by definition includes all $(n-1)!$ permutations\footnote{
  A third decomposition of the all gluon amplitude exists, which makes
  the Kleiss-Kuijf relations manifest and requires the evaluation of only
  $(n-2)!$ partial amplitudes~\cite{DelDuca:1999ha,DelDuca:1999rs}.
  However, this does not invalidate our argument, as the factorial
  growth in the number of partial amplitudes is still present.}~\cite{Maltoni:2002mq}.
The additional degrees of freedom introduced through sampling are usually easier 
to deal with than the large numerical effort of computing the summed squared 
matrix element at every point in phase space.

\subsection{Automatic matrix element generation}
Many techniques have been introduced for the automatic computation of tree-level
matrix elements. We will review only one of them, which is particularly suited to
the implementation in a computer algorithm, due to its generality and simplicity.
External wavefunctions in this method are computed in the Weyl-van der Waerden 
formalism~\cite{Weyl:1931qm,Waerden:1974qm}, and full amplitudes are obtained 
by means of the Berends-Giele recursive relations. A detailed discussion 
of this and other techniques can be found in reviews of amplitude
calculations~\cite{Dixon:1996wi,Dittmaier:1998nn}.

Left- and right-handed Weyl spinors are defined by dotted and undotted spinor indices, 
such that $\psi_a$ is a covariant (right-handed) and  $\psi^{\dot a}$ is a contravariant 
(left-handed) spinor. Complex conjugation amounts to dotting and undotting indices, 
according to
\begin{align}\label{eq:weyl_conjugation}
  \psi_{\dot a}\,=&\;\rbr{\psi_a}^*\;,
  &\psi^a\,=\;\rbr{\psi^{\dot a}}^*\;.
\end{align}
Spinor indices are lowered and raised using the spinor metric, given in terms 
of the $\varepsilon$ tensor as
\begin{equation}\label{eq:definition_epsilon}
  \epsilon^{ab}=\epsilon^{\dot a\dot b}\,=\;
  \epsilon_{ab}=\epsilon_{\dot a\dot b}\,=\;
  \rbr{\begin{array}{cc}0&1\\-1&0\end{array}}\;.
\end{equation}
Extensions of the Pauli matrices are defined in terms of the $2\times 2$ 
unit matrix $\sigma^0={\rm I}$ and the Pauli matrices $\vec{\sigma}$ as
$\sigma^{\mu\,\dot{a}b}=\rbr{\sigma^0,\vec{\sigma}}$ and
$\sigma^\mu_{\,a\dot{b}}=\rbr{\sigma^0,-\vec{\sigma}}$.
Using this definition, an arbitrary real-valued four vector $k^\mu$ 
can be written as a bispinor:
\begin{equation}\label{eq:decomposition_vector}
  k_{\dot a b}\,=\;\sigma^\mu_{\dot a b}k_\mu\,=\;
  \rbr{\begin{array}{cc}k^+&k_\perp\\k_\perp^*&k^-\end{array}}\;,
  \quad{\rm where}\quad\quad
  \begin{array}{l}k^\pm\,=\;k^0\pm k^3\\k_\perp\,=\;k^1+ik^2\end{array}\;.
\end{equation}
For massless vectors, $\vec{k}_\perp^2=k^+k^-$, and a spinor 
$\xi(k)$ can be determined such that $k_{\dot a b}=\xi_{\dot a}\xi_b$:
\begin{equation}\label{eq:decomposition_mvs}
  k_{\dot a b}\,=\;\xi_{\dot a}(k)\,\xi_b(k)\;,
  \quad{\rm where}\quad\quad
  \xi_a(k)\,=\;\rbr{\begin{array}{c}\sqrt{k^+}\\\sqrt{k^-}\,e^{i\phi_k}\end{array}}\;,
  \quad\phi_k=\arg k_\perp\;.
\end{equation}
Note that this definition is by no means unique, as an arbitrary phase can be 
added without changing the physics. Equation~\eqref{eq:decomposition_vector} 
is also ambiguous, because the $x$-, $y$- and $z$-direction along which $k_\perp$ 
and $k^\pm$ are defined can be changed through a rotation of the Pauli matrices.
The final result for the squared matrix element must not depend on these choices.
This fact can be used in automated computer programs to automatically test 
the consistency of the calculation.

In order to decompose massive vectors in terms of bispinors they must first be 
reduced to massless components. Using an auxiliary vector, $a^\mu$, we obtain
\begin{equation}\label{eq:decomposition_massive_vector_pre}
  k^\mu\,=\;b^\mu-\kappa a^\mu\;,
  \quad{\rm where}\quad\quad
  \kappa\,=\;\frac{k^2}{2\,ak}\;.
\end{equation}
and therefore
\begin{equation}\label{eq:decomposition_massive_vector}
  k^\mu\,=\;\frac{1}{2}\,\sigma^\mu_{\dot a b}\;b^{\dot a}b^b
    -\frac{\kappa}{2}\,\sigma^\mu_{\dot a b}\;a^{\dot a}a^b\;.
\end{equation}
We introduce the standard shorthand notation which denotes the 
spinor $\xi_a(k_i)$ as $|k_i\rangle$ or $|k_i^+\rangle$ and the 
corresponding spinor $\xi^{\dot a}(k_i)$ as $|k_i]$ or $|k_i^-\rangle$. 
The inner product in spinor space is then given by
\begin{align}\label{eq:def_spinor_product}
  \abr{\xi\eta}\,=&\;\abr{\xi^+|\eta^+}=\,\xi_a\eta^a\;,
  &\sbr{\xi\eta}\,=&\;\abr{\xi^-|\eta^-}=\,\xi_{\dot a}\eta^{\dot a}\;,
\end{align}
Due to the spinor metric $\epsilon$, the inner product is antisymmetric 
in its arguments and the Schouten identity holds. 
Equation~\eqref{eq:weyl_conjugation} implies $\sbr{\xi\eta}=\abr{\xi\eta}^*$.
The invariant mass of a pair of massless particles described by the four vectors
$k_i$ and $k_j$ is obtained in terms of spinor products as
$2\,k_ik_j=\abr{i^+|\sigma^\mu|i^+}\abr{j^+|\sigma_\mu|j^+}/2=\abr{ij}\sbr{ji}$.
Hence, up to a phase, spinor products are square roots of Lorentz invariants.

To compute full matrix elements we need explicit Dirac spinors and polarization vectors
for external states. They can easily be derived in the Weyl-van der Waerden formalism.
Dirac spinors are represented in terms of Weyl spinors as
\begin{equation}\label{eq:def_dirac_spinor_wvdw}
  \Psi\,=\;\rbr{\begin{array}{c}\phi^{\dot a}\\\psi_a\end{array}}\;.
\end{equation}
The corresponding Dirac matrices read
\begin{align}\label{eq:def_dirac_matrices_wvdw}
  \gamma^\mu\,=&\;\rbr{\begin{array}{cc}0&\sigma^{\mu\,\dot a b}\\
    \sigma^\mu_{a\dot b}&0\end{array}}\;,
  &\gamma^5\,=&\;i\gamma^0\gamma^1\gamma^2\gamma^3\,=\;
  \rbr{\begin{array}{cc}-\sigma^0&0\\0&\sigma^0\end{array}}\;.
\end{align}
Covariant and contravariant spinors $\psi_a$ and $\phi^{\dot a}$ can be 
singled out using the projectors $P_\pm=(1\pm\gamma^5)/2$.
A complete set of Eigenspinors of the Dirac equation can be computed 
in terms of the variables $\bar p={\rm sgn}\rbr{p_0}\,\abs{\vec p\,}$
and $\hat p=(\,\bar p,\vec p\;)$, as~\cite{Hagiwara:1985yu}
\begin{equation}\label{eq:eigenspinors_wvdw}
  \begin{split}
    u_+(p,m)&=\frac{1}{\sqrt{2\,\bar p}}\left(\begin{array}{r}
      \sqrt{p_0-\bar p}\;\chi_+(\hat p)\\
      \sqrt{p_0+\bar p}\;\chi_+(\hat p)\end{array}\right)\;,
    &v_-(p,m)&=\frac{1}{\sqrt{2\,\bar p}}\left(\begin{array}{r}
      -\sqrt{p_0-\bar p}\;\chi_+(\hat p)\\
      \sqrt{p_0+\bar p}\;\chi_+(\hat p)\end{array}\right)\;,\\
    u_-(p,m)&=\frac{1}{\sqrt{2\,\bar p}}\left(\begin{array}{r}
      \sqrt{p_0+\bar p}\;\chi_-(\hat p)\\
      \sqrt{p_0-\bar p}\;\chi_-(\hat p)\end{array}\right)\;,
    &v_+(p,m)&=\frac{1}{\sqrt{2\,\bar p}}\left(\begin{array}{r}
      \sqrt{p_0+\bar p}\;\chi_-(\hat p)\\
      -\sqrt{p_0-\bar p}\;\chi_-(\hat p)\end{array}\right)\;.
  \end{split}
\end{equation}
The Weyl spinors $\chi_\pm(p)$ are given by $\chi_+(p)=\xi_a(p)$ and $\chi_-(p)=\xi^{\dot{a}}(p)$.
The above definition has the apparent advantage, that massless Dirac spinors have 
two nonzero components only. This fact, together with the definition of the Dirac
matrices, Eq.~\eqref{eq:def_dirac_matrices_wvdw}, greatly simplifies computations 
in massless theories.

Polarization vectors for massless external vector bosons can be constructed as
\begin{align}\label{eq:definition_polarisations_ml}
  \varepsilon_\pm^\mu\rbr{p,k}\,=\;\pm\frac{\langle k^\mp|\gamma^\mu
    |p^\mp\rangle}{\sqrt{2}\,\langle k^\mp|p^\pm\rangle\;}\;.
\end{align}
In this context, $k$ is an arbitrary light-like auxiliary vector, which satisfies 
$pk\neq 0$. This definition leads to the polarization sum of a light-like 
axial gauge~\cite{Dixon:1996wi,Dittmaier:1998nn}.
For massive vector bosons the wave function must satisfy Proca's equation,
and we obtain one additional polarization:
\begin{align}\label{eq:definition_polarisations_ms}
  \varepsilon_\pm^\mu\rbr{p,k}&=\pm\frac{\langle k^\mp|\gamma^\mu
    |b^\mp\rangle}{\sqrt{2}\,\langle k^\mp|b^\pm\rangle\;}\;,
  &\varepsilon_0^\mu\rbr{p,k}&=\frac{1}{m}
    \rbr{\,\langle b^-|\gamma^\mu|b^-\rangle-
      \kappa\langle k^-|\gamma^\mu|k^-\rangle\;}\;,
\end{align}
where $b=p-\kappa k$ and $\kappa=p^2/2pk$. Again, $k$ is an arbitrary light-like 
gauge vector.

Equations~\eqref{eq:eigenspinors_wvdw},~\eqref{eq:definition_polarisations_ml}
and~\eqref{eq:definition_polarisations_ms} are sufficient to construct all relevant
eigenstates of the external particles in the Standard Model and a wide range of
theories beyond it. We will now explain how the full matrix element is efficiently
computed using this information.

\begin{figure}
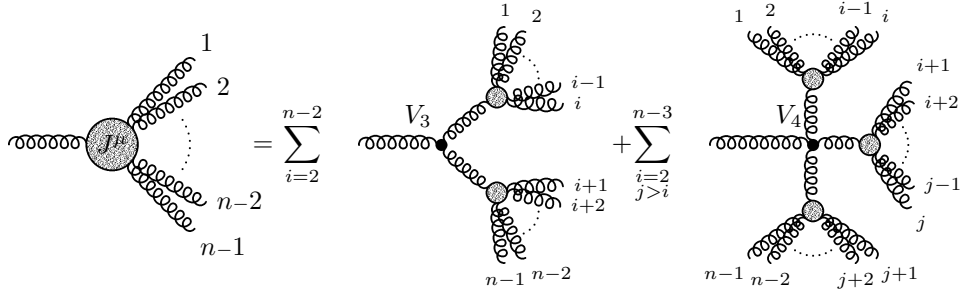

  \begin{center}
    \hspace*{-0.5cm}\bgrrr{55}{45}{55}{\smallfeynmf}\hspace*{-1cm}$=$
    $\dst\sum_{i=2}^{n-2}$\hspace*{-1cm}\bgrra{55}{65}{55}{\smallfeynmf}\hspace*{-1cm}
    $+\dst{\sum_{\substack{i=2\\j>i}}^{n-3}}$\hspace*{-1cm}\bgrrb{55}{65}{55}{\smallfeynmf}
    \hspace*{-0.5cm}
    \caption{Pictorial representation of the Berends-Giele 
      recursion relations, Eq.~\eqref{eq:BG}.
      \label{fig:BG}}
  \end{center}
\end{figure}
Berends and Giele introduced an efficient algorithm~\cite{Berends:1987me}
which generates the color-ordered $n-1$-point gluon off-shell current, $J^\mu$,
defined as the sum of all color-ordered all-gluon Feynman diagrams with $n-1$ 
external legs and a single off-shell leg with polarization $\mu$. 
The recursion relation defining this current reads
\begin{equation}\label{eq:BG}
  \begin{split}
 J^\mu(1,2,\ldots,n-1)\,=&\;\frac{-i}{P_{1,n-1}^2}\left\{\sum_{k=1}^{n-2}
   V_3^{\mu\nu\rho}\left(P_{1,k},P_{k+1,n-1}\right)
   J_\nu(1,\ldots,k)J_\rho(k+1,\ldots,n-1)\right.\\
   &\left.+\sum_{j=1}^{n-3}\sum_{k=j+1}^{n-2}V_4^{\mu\nu\rho\sigma}
   J_\nu(1,\ldots,j)J_\rho(j+1,\ldots,k)J_\sigma(k+1,\ldots,n-1)\right\}\;,
  \end{split}
\end{equation} 
where the momentum sum $P_{i,j}$ is defined as $P_{i,j}\,=\;\sum_{k=i}^{j-1}p_k$,
and where $V_3^{\mu\nu\rho}\left(P_{1,k},P_{k+1,n-1}\right)$ and $V_4^{\mu\nu\rho\sigma}$ 
are the color-ordered three and four-gluon vertices~\cite{Dixon:1996wi}:
\begin{equation}\label{eq:definition_co_gluon_vertices}
  \begin{split}
    V_3^{\mu\nu\rho}(P,Q)\,=&\;i\,\frac{g_s}{\sqrt{2}}\,\rbr{\vphantom{\sum}\,
      g^{\nu\rho}(P-Q)^\mu+2g^{\rho\mu}Q^\nu-2g^{\mu\nu}P^\rho\,}\;,\\
    V_4^{\mu\nu\rho\sigma}\,=&\;i\,\frac{g_s^2}{2}\,\rbr{\vphantom{\sum}\,
      2g^{\mu\rho}g^{\nu\sigma}-g^{\mu\nu}g^{\rho\sigma}-g^{\mu\sigma}g^{\nu\rho}\,}\;.\\
  \end{split}
\end{equation}
The algorithm is schematically depicted in Fig.~\ref{fig:BG}.
The full $n$-gluon amplitude is obtained by amputating the off-shell propagator and
contracting the remaining quantity with the polarization of gluon $n$:
\begin{equation}\label{eq:definition_amplitude}
  A_n(1,\ldots,n)\,=\;\varepsilon_{n}^\mu\,\frac{P_{1,n-1}^2}{i}\,J_\mu(1,\ldots,n-1)\;.
\end{equation}
Similar recursions exists for the off-shell quark currents~\cite{Berends:1987me},
and they can be defined for any gauge theory. Berends-Giele type recursion relations
can straightforwardly be implemented into computer programs and are therefore 
particularly suited for numerical calculations.
They are not limited to color-ordered amplitudes, but can be extended to include
color information, which makes the full result equivalent to the Dyson-Schwinger
approach used in~\cite{Kanaki:2000ey,Cafarella:2007pc}.

The power of recursion relations lies in the fact that for each individual 
phase-space point, each off-shell current in the calculation must be computed 
exactly once. It can be reused as a numerical value whenever the computation 
of Eq.~\eqref{eq:BG} necessitates it. This is obviously true not only for a 
single helicity or color configuration. Currents with a certain assignment
of external-particle quantum numbers can be reused no matter what the final 
amplitude is that needs to be computed. Therefore, the Berends-Giele recursion
is a maximally efficient common subexpression elimination for brute force 
tree-level calculations.

\subsection{Efficient phase space integration}
\label{sec:ps_integration}
Generic methods to deal with the problem of high-dimensional phase-space integrals
were proposed long ago~\cite{Byckling:1969sx}. The crucial observation is that the 
integral factorizes into components, which are associated with $2\to 2$ 
scattering processes, $1\to2$ decay processes, and $2\to1$ annihilation
processes. The knowledge of propagators and vertices in Feynman diagrams
then permits the construction of a Monte-Carlo integrator which precisely
maps onto the peak structure of a certain diagram squared. However, each diagram 
can lead to a different integrator. The multi-channel method is used
to combine those and find the optimal balance, such that the full matrix element
squared is integrated with maximum efficiency~\cite{Kleiss:1994qy}.

The differential final-state phase space element for $2\to (n-2)$ scattering is
\begin{equation}\label{eq:npart_phasespace}
  \dt\Phi_n\rbr{\args{p}}=
    \sbr{\,\prod\limits_{i=3}^n\frac{\dt^4 p_i}{\rbr{2\pi}^3}\,
    \delta\rbr{p_i^2-m_i^2}\Theta\rbr{E_i}\,}\,
    \rbr{2\pi}^4\delta^{(4)}\rbr{p_1+p_2-\sum_{i=3}^n p_i}\;,
\end{equation}
where $m_i$ are the on-shell masses of outgoing particles.
Equation~\eqref{eq:npart_phasespace} factorizes as~\cite{James:1968gu}
\begin{equation}\label{eq:split_ps}
  \dt\Phi_n\rbr{p_1,p_2;p_3,\ldots,p_n}=
    \dt\Phi_{n-m+1}\rbr{p_1,p_2;p_3,\ldots,p_{n-m},P}\,\frac{\dt P^2}{2\pi}\,
    \dt\Phi_m\rbr{P;p_{n-m+1},\ldots,p_n}\;,
\end{equation}
where $P$ denotes a virtual intermediate particle. Even though this particle 
has no direct physical interpretation, it may be associated with an $s$-channel
propagator formed by the set of external states $\{p_{n-m+1},\ldots,p_n\}$. 
If a corresponding propagator exists in any Feynman diagram, the peak structure 
of this diagram squared can efficiently be mapped out by distributing
Monte-Carlo points according to the shape of the propagator squared. This technique 
is also very efficient for the full matrix element, where the diagram containing 
the propagator interferes with other diagrams.

Equation~\eqref{eq:split_ps} allows one to decompose the complete phase space
into only three elementary building blocks that are given by
\begin{equation}\label{eq:ps_building_blocks}
  \begin{split}
    \dt\Phi_2(p_a,p_b;p_i,p_j)&=
    \frac{\lambda\rbr{s_{ab},s_i,s_j}}{
      16\pi^2\,2\,s_{ab}}\;\dt\cos\theta_i\,\dt\phi_i\;,\\
    \dt\Phi_2(p_{ij};p_i,p_j)&=
    \frac{\lambda\rbr{s_{ij},s_i,s_j}}{
      16\pi^2\,2\,s_{ij}}\;\dt\cos\theta_i\,\dt\phi_i\;,\\
    \dt\Phi_1(p_a,p_b;p_i)&=\rbr{2\pi}^4\,\dt^4 p_i\;
      \delta^{(4)}\rbr{p_a+p_b-p_i}\;.
  \end{split}
\end{equation}
We have introduced the K{\"a}llen function
\begin{equation}\label{eq:kallen}
  \lambda\rbr{s_a,s_b,s_c}=\sqrt{\rbr{s_a-s_b-s_c}^2-4s_bs_c}\;.
\end{equation}
Equation~\eqref{eq:ps_building_blocks} may interpreted as elementary 
$t$- and $s$-channel ``vertices'', while the integral $\dt P^2/2\pi$ 
in Eq.~\eqref{eq:split_ps} corresponds to a ``propagator''. 
This makes the correspondence to tree-level matrix elements manifest.
Note that $\dt\Phi_2(p_a,p_b;p_i,p_j)$ and $\dt\Phi_2(p_{ij};p_i,p_j)$
are formally identical, since they represent a solid angle integration.
However, in practice one chooses different sampling strategies~\cite{Byckling:1969sx}
in order to reflect the peak structure of the integrand.
The $s$-channel annihilation vertex $\dt\Phi_1(p_a,p_b;p_i)$ is needed 
only for bookkeeping. It corresponds to overall momentum conservation 
and the associated overall weight factor $(2\pi)^4$.

Let us investigate the situation where multiple diagrams contribute 
to a given process, like for example $gg\to gg$ scattering. In this case
we have three different production channels, and therefore three
different integrators, called integration channels. They are 
$\dt\Phi_2(p_1,p_2;p_3,p_4)$, $\dt\Phi_2(p_1,p_2;p_4,p_3)$ and
$\dt\Phi_2(p_{12};p_3,p_4)$. The task is to find the optimal balance 
between them. The azimuthal angle integration can be carried out trivially.
We can then map the situation onto a single-dimensional integral of a function
$f(x)$ with unknown peak structure, and three ``guesstimates'' with known
integrals, which we call $g_1(x)\ldots g_3(x)$. We assume that $f(x)$ 
is a linear combination of the $g_i(x)$.  Therefore the primitive of
$f(x)$, $F(x)$, is also a linear combination of the (known) primitives
$G_i(x)=\int\dt x g_i(x)$:
\begin{equation}\label{eq:assumption_mc}
  f(x)\approx g(x)=\sum_i \alpha_i g_i(x)\;.
\end{equation}
The set of numbers $\alpha_i$, which must be normalized as $\sum_i\alpha_i=1$,
is called the a-priori weights of the multi-channel integrator. The task is 
to adjust these weights automatically, such that the variance of the Monte-Carlo 
integral is minimized. This procedure is a variant of importance sampling. 
For it to work it is vital that the Monte-Carlo integral is independent of 
the integration variable while its variance is not:
\begin{equation}
  \begin{split}
  I[f]=&\abr{f(x)}_x=\int\dt x\,f(x)
  =\int\dt G(x)\,\frac{f(x)}{g(x)}=\abr{w(x)}_{G(x)}=I_g[f]\;,
  \quad\text{where}\quad
  w(x)=\frac{f(x)}{g(x)}\\
  V[f]=&\abr{f^2(x)}_x=\int\dt x\,f^2(x)
  \neq \int\dt G(x)\,\rbr{\frac{f(x)}{g(x)}}^2=\abr{w^2(x)}_{G(x)}=V_g[f]\\
  \end{split}
\end{equation}
The extremum of the variance $V_g[f]$ is obtained when $V_{g,i}[f]=V_g[f]$ 
for all i, where
\begin{equation}
  V_{g,i}[f]=-\frac{\partial}{\partial\alpha_i}\,V_g[f]=
  \abr{\frac{g_i(x)}{g(x)}\,w^2(x)}_{G(x)}\;.
\end{equation}
This means that all integration channels should contribute equally to the variance.
By setting $\alpha_i\to\alpha_i\sqrt{V_{g,i}[f]}$ after a certain number of points 
in the integration, we obtain the best possible approximation of this situation.
This example can be extended trivially to the case where $x$ is a multi-dimensional
random variable.

The multi-channel integrator described above can be further refined by using
adaptive stratified sampling techniques like Vegas~\cite{Lepage:1980dq}. 
The factorization of each integration channel into basic building blocks
allows for the independent optimization of the grid for each propagator
and each vertex. Challenging situations like non-factorizable integrands can be
imagined and have been investigated in great detail~\cite{Ohl:1998jn,Jadach:1999sf}.
However, in practice the combination of factorization, multi-channel integration 
and adaptive stratified sampling performs reasonably well in most cases.

\subsection{Next-to-leading order calculations}
\label{sec:nlo_calculations}
With the advent of general procedures for the treatment of infrared 
singularities in QCD~\cite{Frixione:1995ms,Catani:1996vz,Catani:2002hc},
existing tree-level matrix element generators became tools to organize
ever-more complicated NLO calculations~\cite{Binoth:2010xt,Alioli:2013nda}.
Crucially, their combination with modern Monte-Carlo event generators enables
an automatic matching to the approximate higher-order corrections implemented
by parton showers, and it allows one to generate particle-level events at high
theoretical accuracy~\cite{Frixione:2002ik,Nason:2004rx,Frixione:2007vw}.
A full review of modern techniques for NLO QCD calculations is beyond the scope
of these lectures. In the following we will focus only on the key components 
needed at the interface between NLO calculations and parton shower simulations.

Cross sections calculated at NLO accuracy consist of four parts: The Born contribution,
the virtual and the real corrections, and the collinear mass factorization counterterms.
A genuine obstacle in the calculation is the
appearance of ultraviolet and infrared divergences. Ultraviolet terms are dealt with
in a rather straightforward manner: Loop amplitudes are regularized in dimensional 
regularization, and the theory is renormalized by adding counterterms.
Infrared divergences are more complicated to handle, as cancellations between the
virtual and the real corrections, which are guaranteed by the Bloch-Nordsieck~\cite{Bloch:1937pw}
and Kinoshita-Lee-Nauenberg~\cite{Kinoshita:1962ur,Lee:1964is} theorems, occur only 
after integration over the final-state phase space.

We start by discussing the real-emission contribution.
In full analogy to Eqs.~\eqref{eq:definition_born_xs} and~\eqref{eq:definition_born_xs_parts}
we write the differential cross section as a sum, depending on parton configurations 
$\{a_1,\ldots,a_{n+1}\}$. The Born-level matrix elements $\mc{B}_n$ are replaced 
by the real-emission matrix elements $\mc{B}_{n+1}$, and the Born-level phase space 
$\dt\Phi_n$ is replaced by the real-emission phase-space $\dt\Phi_{n+1}$. 
We introduce a notation for mapping from real-emission parton configurations 
to Born-level configurations:
\begin{align}\label{eq:parton_map_rtob}
  \bmap{ij}{k}{\args{a}}\,=&\;\left\{\begin{array}{c}
    \args{f\,}\setminus\{f_i,f_j\}\cup\{f_{\wt{\im\jm}}\}\\
    \args{p\,}\to\args{\tilde{p}}
    \end{array}\right.\;.
\end{align}
The map $\bmap{ij}{k}{\args{a}}$ combines partons $a_i$ and $a_j$ into a 
common ``mother'' parton $a_{\wt{\im\jm}}$, in the presence of the 
spectator $a_k$ by defining a new flavor $f_{\wt{\im\jm}}$ and by 
redefining the particle momenta. This is the exact inverse to
the splitting process discussed in Sec.~\ref{sec:ps_intro}.

When two partons become collinear, the real-emission matrix element squared factorizes as 
\begin{equation}\label{eq:coll_factorization}
  |\mc{M}_{n+1}|^2(\args{a})\;\overset{ij\;\text{collinear}}{\longrightarrow}\;
  \frac{8\pi\alpha_s\,\mu^{2\varepsilon}}{2\,p_ip_j}\;
  \mc{M}_{n}(\bmap{ij}{\centerdot}{\args{a}})\otimes
  \tilde{P}_{\wt{\im\jm}\;i}(z,\varepsilon)\otimes
  \mc{M}_{n}^*(\bmap{ij}{\centerdot}{\args{a}})\;,
\end{equation}
where the $\otimes$ indicates spin correlations between the Born matrix elements
and the spin-dependent Altarelli-Parisi splitting functions, $\tilde{P}_{ij}(z)$.
In the strict collinear limit, the map does not depend on the spectator parton
(cf.\ Sec.~\ref{sec:ps_intro}), which is denoted by the open index marked as $\centerdot$.

If a single gluon becomes soft, the real-emission matrix element squared behaves as
\begin{equation}\label{eq:soft_factorization}
  |\mc{M}_{n+1}|^2(\args{a})\;\overset{j\;\text{soft}}{\longrightarrow}\;
  -8\pi\alpha_s\,\mu^{2\varepsilon}\;\sum_{k>i}
  \mc{M}_{n}(\bmap{ij}{k}{\args{a}})\otimes
  {\bf T}_i{\bf T}_k\,\frac{p_ip_k}{(p_ip_j)(p_jp_k)}\otimes
  \mc{M}_{n}^*(\bmap{kj}{i}{\args{a}})\;,
\end{equation}
where ${\bf T}_i$ and ${\bf T}_k$ are the color charge operators of the 
external partons~\cite{Bassetto:1984ik}.

The collinear and soft singularities can be treated individually, after the 
final-state phase space has been separated into sectors where only one divergent 
term contributes~\cite{Frixione:1995ms,Frixione:1997np}. Alternatively, the collinear 
and soft factorization can be rewritten as a dipole factorization, using splitting
functions which capture the singularity structure in both limits, after partial 
fractioning the soft eikonals~\cite{Catani:1996vz,Catani:2002hc}. This scheme
is called the Catani-Seymour (CS) dipole subtraction method. It allows one to
fully regularize the real-emission contribution by adding a set of local counterterms,
$\mr{S}^{ij,k}_n(\args{a})$, which are called the dipole subtraction terms. 
They are defined as
\begin{equation}\label{eq:def_cs_real_subterms}
  \mc{S}^{ij,k}_{n+1}(\args{a})=-\frac{1}{F(\args{p})}\frac{1}{S(\args{f})}
  \frac{8\pi\alpha_s\,\mu^{2\varepsilon}}{2\,p_ip_j}\;\mc{M}_{n}(\bmap{ij}{k}{\args{a}})\otimes
  \frac{{\bf T}_{ij}{\bf T}_k}{{\bf T}_{ij}^2}\,\mr{V}_{ij,k}(a_i,a_j,a_k)\otimes
  \mc{M}_{n}^*(\bmap{kj}{i}{\args{a}})\;,
\end{equation}
Using these terms one can compute arbitrary infrared- and collinear-safe observables,
in particular jet observables. This will become important in Sec.~\ref{sec:matching_merging}.
We will denote such an observable by $O$. Calculating the expectation value of this observable,
$\abr{O}$, is equivalent to an experimental measurement. At NLO QCD we obtain
\begin{equation}\label{eq:cs_subtraction}
  \begin{split}
    \abr{O}^{\rm(NLO)}\,=&\;
    \sum_{\args{f\,}}\int\dt\bar{\Phi}_n(\args{p})\,
    \Big(\mr{B}_n(\args{a})+\tilde{\mr{V}}_n(\args{a})+
    \sum_{\{\wt{\im\jm},\tilde{k}\}}\mr{I}_n^{\,\wt{\im\jm},\tilde{k}}(\args{a})\Big)\,O(\args{p})\\
    &\quad+\sum_{\args{f\,}}\int\dt\bar{\Phi}_{n+1}(\args{p})\,
    \Big(\mr{B}_{n+1}(\args{a})\,O(\args{p})-
    \sum_{\{ij,k\}}\mr{S}_{n+1}^{\,ij,k}(\args{a})\,O(\bmap{ij}{k}{\args{p}})\,\Big),
  \end{split}
\end{equation}
where, in analogy to Eq.~\eqref{eq:definition_born_xs_parts}, $\mr{S}^{ij,k}=\mc{L}\,\mc{S}^{ij,k}$.
The integrated subtraction terms $\mr{I}_n^{\,\wt{\im\jm},\tilde{k}}$ are determined
by the analytically integrated insertion operators $\mr{V}_{ij,k}(a_i,a_j,a_k)$, 
multiplied by Born matrix elements, similar to Eq.~\eqref{eq:def_cs_real_subterms}~\cite{Catani:1996vz,Catani:2002hc}.
$\tilde{\mr{V}}(\args{a})$ represents the virtual corrections after ultraviolet renormalization,
which also include the collinear mass-factorization counterterms. 
Note that the cancellation of $\mr{I}_n^{\,\wt{\im\jm},\tilde{k}}$ and 
$\sum_{\{ij,k\}}\mr{S}_{n+1}^{\,ij,k}(\args{a})$, integrated over the one-parton emission
subspace, must be guaranteed locally in the phase-space of the Born process. This is ensured
by the observable dependence $O(\bmap{ij}{k}{\args{p}})$ in the last term.

The integrated subtraction terms contain poles in the dimensional regularization parameter
$\varepsilon$, which cancel the poles in the virtual corrections, such that the first and
second sum in Eq.~\eqref{eq:cs_subtraction} are separately infrared finite. This is crucial
as the phase-space integrals to be evaluated have a different number of dimensions.
Equation~\eqref{eq:cs_subtraction} therefore permits computation of any process at NLO in an 
automated fashion using the integration techniques of Sec.~\ref{sec:ps_integration}. 
The computation of the real-emission differential cross sections and the corresponding 
dipole subtraction terms can been fully automated, in the same manner as any tree-level 
matrix element calculation~\cite{Gleisberg:2007md,Czakon:2009ss,Frederix:2008hu,Frederix:2010cj}. 
The same is true for the integrated subtraction terms. The only missing ingredients to a 
full NLO calculation are the virtual corrections $\tilde{\mr{V}}(\args{a})$. 
They are typically provided to the Monte-Carlo event generator by specialized 
programs~\cite{Binoth:2010xt,Alioli:2013nda}, and we will not detail their computation here.

\section{The parton shower}
\label{sec:parton_showers}
Parton showers approximate higher-order real-emission corrections to the hard scattering
by simulating the branching of a single external parton into two partons. They locally conserve 
flavor and four momentum, and they respect unitarity, which simply means that a parton 
may either split into two partons, or it may not. These few very basic requirements are 
in principle enough to define a parton shower. 
Many choices can however be made in its precise implementation, and the quality of parton-shower 
predictions often depends significantly on these choices. A prime example is the selection of an
evolution variable representing angular ordering, which itself is a consequence of color coherence.
In parton showers using Altarelli-Parisi splitting functions, this choice (or an explicit
angular-veto requirement) is needed in order to recover the correct soft anomalous dimension
in the evolution. However, angular ordering is not the only way in which color coherence
can be ensured. This section will first introduce the basics of parton shower algorithms, 
including a Monte-Carlo technique known as the veto algorithm, while the choices for evolution
variables and evolution kernels as well as their implications are discussed later.

We start with the next-to-leading order dipole subtraction terms, Eq.~\eqref{eq:def_cs_real_subterms}.
They can be classified according to their Born flavor and momentum configuration, plus
an additional flavor and momentum generated by the $1\to2$ branching process that will
be interpreted as a basic parton shower step. This situation is sketched in Fig.~\ref{fig:split_kinematics}.
We first introduce a notation for mapping from Born parton configurations to real-emission 
configurations, which is the inverse of Eq.~\eqref{eq:parton_map_btor}:
\begin{align}\label{eq:parton_map_btor}
  \rmap{\im\jm}{k}{f_i,\Phi_{+1}^{ij,k}\,;\args{a}}\,=&\;
  \left\{\begin{array}{c}
    \args{f\,}\setminus\{f_{\wt{\im\jm}}\}\cup\{f_i,f_j\}\\
    \args{\tilde{p}}\to\args{p\,}
    \end{array}\right.\;.
\end{align}
Note that while $\bmap{ij}{k}{\args{a}}$ is unambiguously defined by the real-emission
parton configuration $\args{a}$, its inverse, $\rmap{\im\jm}{k}{\args{a}}$ depends on 
additional radiative variables $\Phi_{+1}^{ij,k}$ and a newly-selected flavor.
It is the task of the parton-shower algorithm to select these four variables
using Monte-Carlo methods.

Crucially, the $n+1$-particle phase space factorizes (cf. Sec.~\ref{sec:ps_integration}),
such that the computation of the next-to-leading order dipole subtraction terms can be 
reorganized as
\begin{equation}
  \sum_{\args{f}}\dt\bar{\Phi}_{n+1}\,\mc{S}^{ij,k}_{n+1}(\args{a})\;\to\;
  \sum_{\args{f}}\dt\bar{\Phi}_n\,\bigg[
    \sum_{\{\wt{\im\jm},\tilde{k}\}\in\args{f}}\sum_{f_i}\,\dt\Phi_{+1}^{ij,k}\,
    \mc{S}^{ij,k}_{n+1}(\rmap{\im\jm}{k}{f_i,\Phi_{+1}^{ij,k};\args{a}})
    \bigg]\;.
\end{equation}
Note that the sum over parton configurations is for $n+1$-particle configurations 
on the left-hand side and $n$-particle configurations on the right-hand side.
The one-particle emission phase-space can be parametrized in terms of three variables,
which we will call the evolution variable, $t$, the splitting variable, $z$, and
an azimuthal angle, $\phi$ (cf.\ Sec.~\ref{sec:ps_intro}):
\begin{equation}
  \dt\Phi_{+1}^{ij,k}=\frac{1}{16\pi^2}\,\dt t\,\dt z\,\frac{\dt\phi}{2\pi}\,J(t,z,\phi)\;.
\end{equation}
Among these variables, only the evolution variable is dimensionful. $J(t,z,\phi)$
denotes the Jacobian factor associated with the variable transformation. 
Next we factor out the Born differential cross section:
\begin{equation}\label{eq:dipole_real_approx}
  \dt\sigma_n^{\rm(B)}(\args{a})\,\sum_{\{\wt{\im\jm},\tilde{k}\}\in\args{f}}\sum_{f_i}\,
  \dt\Phi_{+1}^{ij,k}\,\frac{
    \mr{S}^{ij,k}_{n+1}(\rmap{\im\jm}{k}{f_i,\Phi_{+1}^{ij,k};\args{a}})}{
    \mr{B}_n(\args{a})} \; .
\end{equation}
Due to the factorization properties of the real-emission matrix element, 
Eq.~\eqref{eq:dipole_real_approx} is an approximation of real-emission corrections
to the leading-order process with configuration $\args{a}$. It is local in both
phase space and flavor space. We call each term in the sum a dipole, for reasons
which will become apparent later.

\subsection{Branching probabilities}
At this point it is useful to associate the abstract expression, 
Eq.~\eqref{eq:dipole_real_approx} with a concrete implementation. 
Taking the collinear limit, and averaging over helicities, we obtain
\begin{equation}\label{eq:factorisation_ps}
  \dt\sigma_n^{\rm(B)}(\args{a})\,\sum_{\{\wt{\im\jm},\tilde{k}\}\in\args{f}}\sum_{f_i}\,
  \frac{S(\args{f\,})}{S(\rmap{\im\jm}{k}{f_i;\args{f\,}})}\,\;
  \dt t\,\dt z\,\frac{1}{2\,p_ip_j}\,\frac{\alpha_s}{2\pi}\,P_{\wt{\im\jm}i}(z)\;,
\end{equation}
Note that the quantities $S$ in this expression are symmetry factors,
cf.\ Eq.~\eqref{eq:definition_born_xs_parts}.
It is obvious that approximate higher-order corrections in successive 
collinear limits may be computed by simply iterating 
Eq.~\eqref{eq:factorisation_ps}. In this process we would violate unitarity, 
as each integral contributes positively to the total cross section 
for the inclusive process described by $\dt\sigma_n^{\rm(B)}(\args{a})$. 
The solution is to add approximate virtual corrections, which are assumed 
to cancel exactly the real-emission corrections computed above. At the same time, 
a cutoff scale, $t_c$, is introduced, which ensures that no partonic process is 
computed below distances of order $\Lambda_{\rm QCD}$. The virtual corrections
and unresolved emissions below $t_c$ can then be combined into the total
contribution from unresolved emissions and virtual corrections. In the simplest
case of a single radiating quark line, like in $e^+e^-\to q\bar{q}$, we obtain:
\begin{equation}
  \sum_{m=0}^\infty\frac{1}{m!}\Bigg[-\;\int_{t_c}\dt t\int\dt z\,
    \frac{1}{2\,p_ip_j}\,\frac{\alpha_s}{2\pi}\,P_{qq}(z)\Bigg]^m\;
  =\exp\Bigg\{-\;\int_{t_c}\dt t\int\dt z\,
    \frac{1}{2\,p_ip_j}\,\frac{\alpha_s}{2\pi}\,P_{qq}(z)\Bigg\}.
\end{equation}
In full generality, this contribution reads
\begin{equation}\label{eq:sudakov_ijk}
  \Pi_n^{\wt{\im\jm},\tilde{k}}(t,t';\args{a})=\exp
  \Bigg\{\,-\,\frac{1}{16\pi^2}\sum_{f_i\in\{q,g\}}
  \int_t^{t'}\dt\bar{t}\int\dt z\int\frac{\dt\phi}{2\pi}\,\frac{1}{2}\frac{
    \mr{S}^{ij,k}_{n+1}(\rmap{\im\jm}{k}{f_i,\bar{t},z,\phi;\args{a}})}{
    \mr{B}_n(\args{a})}\;\Bigg\}\;.
\end{equation}
In the sum over all possible splittings, we obtain 
the no-branching probability of the parton shower:\footnote{
  In pure final-state parton evolution the no-branching probability 
  $\Pi_n^{\wt{\im\jm},\tilde{k}}$ is equivalent to the Sudakov form factor, 
  because the ratio of parton luminosities, 
  $\mc{L}(\rmap{\im\jm}{k}{f_i,t,z,\phi;\args{a}})/\mc{L}(\args{a})$
  is precisely one.}
\begin{equation}\label{eq:sudakov}
  \Pi_n(t,t';\args{a})=\prod_{\{\wt{\im\jm},\tilde{k}\}}
  \Pi_n^{\wt{\im\jm},\tilde{k}}(t,t';\args{a})\;.
\end{equation}
It represent the probability for no parton branching to occur between
the scales $t'$ and $t$. The probability for any parton to branch at scale
$t$ with evolution starting at $t'$ is then given by Poisson statistics:
\begin{equation}\label{eq:branching_probability}
  \mc{P}_1(t,t')=\frac{\dt\Pi_n(t,t';\args{a})}{\dt\log t}\;.
\end{equation}
Parton showers solve this equation for $t$ using the veto algorithm, which
can be through of as an extension of the hit-or-miss Monte-Carlo method
to Poisson distributions~\cite{Sjostrand:2006za}. In order to do this,
a suitable starting scale for the evolution must be defined, which will
be called the resummation scale in the following. This scale may be identified 
with the factorization scale. After $t$ has been selected in the Monte-Carlo procedure,
a value for the splitting variable and the azimuthal angle is found using standard 
Monte-Carlo techniques.

We can phrase the complete parton shower evolution in term of a generating functional, 
$\mc{F}(t;\args{a})$, such that the expectation value of an observable, $O$, is computed as
\begin{equation}\label{eq:ps_generic}
  \begin{aligned}
  \abr{O}^{\rm(PS)}\,=&\;\sum_{\args{f\,}}\int\dt\bar{\Phi}_n(\args{p})\,
  \bar{\mr{B}}_n(\args{a})\,\mc{F}_n(\mu_Q^2;\args{a},O)\;.
  \end{aligned}
\end{equation}
The generating functional is recursively defined as
\begin{equation}\label{eq:ps_functional}
  \begin{aligned}
  &\mc{F}_n(\mu_Q^2;\args{a},O)\,=\;
    \underbrace{\Pi_n(t_c,\mu_Q^2;\args{a})}_{\text{virtual+unresolved}}\,O(\args{p})\\
  &+\,
    \sum_{\{\wt{\im\jm},\tilde{k}\}}
    \sum_{f_i=q,g}\int\dt\Phi_{+1}^{ij,k}\;\Theta(t(\Phi_{+1}^{ij,k})-t_c)\\
  &\quad\times\,
    \underbrace{\frac{1}{S_{ij}}\,
    \frac{S(\rmap{\im\jm}{k}{f_i;\args{f\,}})}{S(\args{f\,})}\,
    \frac{\mr{D}_{n+1}^{ij,k}(\rmap{\im\jm}{k}{f_i,\Phi_{+1}^{ij,k};\args{a}})}{
      \mr{B}_n(\args{a})}\;
    \Pi_n(t,\mu_Q^2;\args{a})}_{\text{resolved}}\\
    &\quad\times\,\mc{F}_{n+1}(t;\rmap{\im\jm}{k}{f_i,\Phi_{+1}^{ij,k};\args{a}},O)\;.
  \end{aligned}
\end{equation}
The first term is the resummed contribution from virtual and unresolved real-emission
corrections, while the second term comes from a single real-emission and the resummed
virtual and unresolved corrections between the hard scale and the scale of the emission.
$S_{ij}$ is a symmetry factor in the shower approximation. We have replaced
the dipole subtraction terms by a new function, $\mr{D}_{n+1}^{ij,k}$, which accounts 
for the fact that the parton-shower evolution kernels typically do not implement the 
spin and color correlations that are present in Eq.~\eqref{eq:def_cs_real_subterms}.
Further emissions may occur after the first, which is implemented by the generating
functional $\mc{F}_{n+1}$ on the last line. Expanding this formula up to first emission
only, $\mc{F}_{n+1}$ would turn into $O_{n+1}$, which is used in some literature on
matching~\cite{Nason:2004rx,Frixione:2007vw}.

\subsection{The veto algorithm}
\label{sec:veto_algo}
Equation~\eqref{eq:branching_probability} is difficult to solve with Monte-Carlo methods
if the integral of the splitting functions is not known analytically. In practice, 
this is most often the case. One reason is that the evolution kernels may not be
simple Altarelli-Parisi splitting functions but more complicated expressions.
Another reason is that the integral may be hard to compute for a given functional form 
of the evolution variable $t$ and the phase-space boundaries imposed by local 
four-momentum conservation. It would be simpler to find an overestimate of the integrand
and perform a hit-or-miss Monte-Carlo integration.  However, this is hampered by the fact
that we intend to evaluate an integral in the exponent. The solution to this problem
lies in using the Sudakov veto algorithm.

To simplify the notation, let $f(t)$ be the splitting kernel of the parton shower, 
integrated over the splitting variable $z$. We also assume that only one splitting function 
exists, i.e.\ that there is no flavor change of the splitter during the evolution.
The differential probability for generating a single branching at scale $t$ 
when starting from an upper evolution scale $t'$ is then given by
\begin{equation}\label{eq:va_prob}
  \mc{P}_1(t,t')\,=\;f(t)\,\exp\cbr{-\int_t^{t'}\dt\bar{t}\,f(\bar{t})}
  =\;\frac{\dt}{\dt t}\,\exp\cbr{-\int_t^{t'}\dt\bar{t}\,f(\bar{t})}\;.
\end{equation}
A new scale $t$ can in principle be determined as
\begin{align}
  t\,=&\;F^{-1}\sbr{\,F(t')+\log R\,}
  &&\text{where}
  &F(t)\,=\,\int^t\dt t\,f(t)\;,
\end{align}
and where $R$ is a random number between zero and one.
The key point of the veto algorithm is, that even if the integral $F(t)$ is unknown, one
can still generate events according to $\mc{P}$ using an overestimate $g(t)\ge f(t)$
with a known integral $G(t)$. First, a value $t$ is generated as $t=G^{-1}\sbr{\,G(t')+\log R\,}$.
Second, the value is accepted with probability $f(t)/g(t)$. 
A splitting at $t$ with $n$ intermediate rejections is then produced with differential 
probability
\begin{equation}\label{eq:va_mcprob}
  \begin{split}
  \mc{P}_1^{(n)}(t,t')\,=&\;\frac{f(t)}{g(t)}\,g(t)\,\exp\cbr{-\int_t^{t_1}\dt\bar{t}\,g(\bar{t})}\\
  &\times\prod_{i=1}^n\sbr{\,\int_{t_{i-1}}^{t_{i+1}}\dt t_i\rbr{1-\frac{f(t_i)}{g(t_i)}}g(t_i)\,
      \exp\cbr{-\int_{t_i}^{t_{i+1}}\dt\bar{t}\,g(\bar{t})}}\;,
  \end{split}
\end{equation}
where $t_{n+1}=t'$ and $t_c=t$. The nested $t_i$-integrals in Eq.~\eqref{eq:va_mcprob} can be 
symmetrized, which leads to a symmetry factor $1/n!$. The exponentials can be combined into
a single term where the $\bar{t}$-integral runs from $t$ to $t'$. Summing over all possibilities
for the number of intermediate rejections, $n$, then leads to the exponentiation of a factor 
$g(t)-f(t)$, such that Eq.~\eqref{eq:va_prob} is reproduced~\cite{Sjostrand:2006za}.

\subsection{Coherent branching}
\label{sec:color_coherence}
Up to this point we have not chosen the precise form of the splitting kernels
in Eq.~\eqref{eq:sudakov_ijk}. Assume that we use the Altarelli-Parisi splitting
functions, $P_{\wt{\im\jm}i}(z)$. $P_{qq}(z)$ is soft-enhanced when $z\to1$.
However, it does not differentiate between a situation where the soft gluon is
radiated in the direction of the initial quark or in the direction of the spectator.
When considering all radiating partons in the process, a naive integration over the
full phase space available to soft gluon radiation would therefore lead to double
counting of logarithmically enhanced soft (but not collinear) contributions.
This can be circumvented using either an appropriate evolution variable or a variant
of the splitting kernel which includes a regulator that damps the soft singularity
in the anti-collinear region of the emission phase space.

Let us investigate this picture in more detail in $e^+e^-$-annihilation to hadrons.
The differential cross section for $e^+e^-\to q\bar{q}g$ is given by the QCD ``antenna'' 
radiation pattern
 \begin{equation}
  \dt\sigma_3=\dt\sigma_2\,\frac{\dt w}{w}\frac{\dt\Omega}{2\pi}\,C_F\,W_{q\bar{q}}^g\;,
  \qquad\text{where}\qquad
  W_{q\bar{q}}=\frac{1-\cos\theta_{q\bar{q}}}{(1-\cos\theta_{qg})(1-\cos\theta_{\bar{q}g})}\;.
\end{equation}
We can split the antenna $W_{q\bar{q}}$ into two parts, $W_{q\bar{q}}^{(q)}$ and 
$W_{q\bar{q}}^{(\bar{q})}$, which are divergent only if the gluon is collinear to
the quark / antiquark:
\begin{equation}\label{eq:antenna_function}
  W_{q\bar{q}}=W_{q\bar{q}}^{(q)}+W_{q\bar{q}}^{(\bar{q})}\;,
  \qquad\text{where}\qquad
  W_{q\bar{q}}^{(q)}=\frac{1}{2}\bigg(W_{q\bar{q}}
  +\frac{1}{1-\cos\theta_{qg}}
  -\frac{1}{1-\cos\theta_{\bar{q}g}}\bigg)\;.
\end{equation}
Upon azimuthal integration, we obtain~\cite{Ellis:1991qj}
\begin{equation}
  \frac{\dt\phi_{qg}}{2\pi}\,W_{q\bar{q}}^{(q)}=\left\{
  \begin{array}{cc}
    \dst\frac{1}{1-\cos\theta_{qg}} &\text{if}\quad\theta_{qg}<\theta_{q\bar{q}}\\[1em]
    0 &\text{else}
  \end{array}\right.\;.
\end{equation}
This is known as angular ordering: The gluon can only be emitted inside the cone
spanned by the initial directions of the quark/antiquark. If it is emitted outside,
it cannot resolve the individual color charges of the quarks.

For processes with more final-state partons the situation becomes slightly more 
complicated. A convenient method to analyze the situation is to work with
color charge operators ${\bf T}$, cf.~\eqref{eq:soft_factorization}. This amounts to
analyzing the combined color charge, which leads to emission of QCD radiation 
off a parton pair. The color charge operators squared give the Casimir operators,
${\bf T}_i^2=C_F$, if $i$ is a quark and ${\bf T}_i^2=C_A$, if $i$ is a gluon.
For color singlets, ${\bf T}_i^2$ vanishes. Each antenna 
multiplies a corresponding color-charge operator, such that the full contribution 
from the radiating ``color dipole'' formed by partons $i$ and $j$ reads
\begin{equation}
  {\bf W}_{ij}\,=\;-{\bf T}_i\cdot{\bf T}_j\;W_{ij}\;.
\end{equation}
In electron-positron annihilation into quarks, this corresponds exactly to the
situation discussed above. Consider now the radiation from a three-parton final 
state. The radiation pattern is then given by 
\begin{equation}\label{eq:rad_pattern_three_part_fs}
  \begin{split}
    {\bf W}_{ijk}\,=&\;-{\bf T}_i\cdot{\bf T}_j\,W_{ij}
    -{\bf T}_j\cdot{\bf T}_k\,W_{jk}
    -{\bf T}_k\cdot{\bf T}_i\,W_{ik}\\
    =&\;\frac{1}{2}\,\left[\,
      {\bf T}_i^2\left(\,\,W_{ij}+\,W_{ik}-\,W_{jk}\,\right)
      +{\bf T}_j^2\left(\,\,W_{jk}+\,W_{ij}-\,W_{ik}\,\right)
      +{\bf T}_k^2\left(\,\,W_{ik}+\,W_{jk}-\,W_{ij}\,\right)\,\right]\;.
  \end{split}
\end{equation}
If $i$ and $j$ are close to each other they form a combined system $l$,
which carries the net color charge ${\bf T}_i+{\bf T}_j={\bf T}_l$.
For small angles between $i$ and $j$, $W_{ik}\approx W_{jk}\approx W_{lk}$.
Equation~\eqref{eq:rad_pattern_three_part_fs} can then be written as~\cite{Ellis:1991qj}
\begin{equation}\label{eq:rad_pattern_ao}
  {\bf W}_{ijk}\,\approx\;{\bf T}_i^2\, W_{ij}^{(i)}+{\bf T}_j^2\, W_{ij}^{(j)}
  +{\bf T}_k^2\, W_{lk}^{(k)}+{\bf T}_l^2\,\tilde{W}_{lk}^{(l)}\Theta(\theta_{lg}-\theta_{ij})\;.
\end{equation}
This equation has again a simple interpretation. Each parton itself 
radiates proportional to its color charge squared, while additional 
radiation comes from coherent emission off the parton pair $ij$ if the
emission angle $\theta_{lg}$ exceeds the opening angle $\theta_{ij}$ of the pair. 
The partons then radiate proportional to their combined color charge squared, 
${\bf T}_l^2$. The formalism may be extended to higher multiplicity, 
and leads to the coherent-branching formalism. It can be interpreted
as an angular-ordering constraint for the partons emitted in each step 
of a parton shower.

In the parton shower implemented in Herwig, this angular ordering constraint 
is realized through the choice of evolution variable. Alternatively, it may
be implemented by using Eq.~\eqref{eq:antenna_function} instead of the sum 
of two Altarelli-Parisi kernels used in standard parton showers. This choice 
was first advocated in the linked dipole chain model~\cite{Kharraziha:1997dn}.
Partial fractioning the antenna and assigning each term the meaning of a splitting
function in the presence of a spectator parton leads to yet another option for 
implementing effective angular ordering. This is the option we have chosen
to introduce the generic parton-shower model above, because it allows one to 
retain the notion of a splitter parton, which can be associated with the 
collinear direction in the collinear limit.

Schematically, this partial fractioning is performed as~\cite{Catani:1996vz}
\begin{equation}\label{eq:antenna_partfrac}
  \frac{p_ip_k}{(p_ip_j)(p_jp_k)}\to
  \frac{1}{p_ip_j}\frac{p_ip_k}{(p_i+p_k)p_j}+
  \frac{1}{p_kp_j}\frac{p_ip_k}{(p_i+p_k)p_j}\;.
\end{equation}
The terms $1/(p_ip_j)$ and $1/(p_kp_j)$ lead to double-collinear singularities,
while the remaining terms do not contain any two-particle poles. Only the soft
singularity structure is reflected by Eq.~\eqref{eq:antenna_partfrac}. 
The spin-dependent terms of the collinear splitting functions are added explicitly,
leading to the Catani-Seymour dipole splitting functions~\cite{Catani:1996vz}.
In a parton shower, they are mostly used in their spin-averaged form, which reads
\begin{equation}\label{eq:cs_splittings_fs}
  \begin{split}
  \abr{V}_{qg}(\tilde{z},y)=&\;
  C_F\bigg[\frac{2}{1-\tilde{z}(1-y)}-(1+\tilde{z})\bigg]\;,\\
  \abr{V}_{gg}(\tilde{z},y)=&\;2C_A\bigg[\frac{1}{1-\tilde{z}(1-y)}
    +\frac{1}{1-(1-\tilde{z})(1-y)}-2+\tilde{z}(1-\tilde{z})\bigg]\;.
  \end{split}
\end{equation}
Note that $\abr{V}_{gq}(\tilde{z},y)=P_{gq}(\tilde{z})$, as no soft gluon singularity
needs to be taken care of. The variable $y$ is given by $y=(p_ip_j)/(p_ip_j+p_ip_k+p_jp_k)$,
while the light-cone momentum fraction $\tilde{z}$ is defined as 
$\tilde{z}=(p_ip_k)/(p_ip_k+p_jp_k)$, cf.\ Sec.~\ref{sec:ps_kin_intro}.

\subsection{The large-$N_C$ approximation}
\label{sec:large_nc_ps}
Parton showers as Markov-Chain Monte-Carlo algorithms build on the assumption that
the Sudakov factors in Eq.~\eqref{eq:sudakov} are positive numbers, which represent
the probability for a parton not to undergo branching between two scales.
This makes it difficult to accommodate full color coherence, as the color dipoles
discussed above radiate proportional to their color correlators $-{\bf T}_i{\bf T}_k$.
These terms may be negative, which would lead to non-probabilistic Sudakov factors.
This situation can be dealt with in principle~\cite{Nagy:2007ty,Hoeche:2011fd,Platzer:2012np},
and several algorithms have been proposed to accommodate the non-probabilistic terms
in the veto algorithm~\cite{Hoeche:2011fd,Lonnblad:2012hz}. The much more common
solution, however, is to use an approximation similar to the large-$N_C$ approximation.

In the large-$N_C$ limit, color-octet gluons are replaced by a color triplet-antitriplet
pair. $1/N_C$ terms are absent, leading to a simple color topology consisting of a planar
flow. Each branching creating a gluon in the final state leads to a new ``color'', and
each gluon (quark) is connected to two and only two (one and only one) other parton.
QCD radiation in this approximation is always simulated as the radiation from a single
color dipole, rather than a coherent sum from a color multipole. However, the color 
charge for radiation off quarks is still set by $C_F$, and not by $N_C/2$. This accounts
for the leading $1/N_C$ effects, and it matches the result obtained by color conservation
in the collinear limit.

\subsection{Practical implementation}
Up to now we have not specified the precise form of the evolution variable, $t$.
If we choose the propagator virtuality, $\dt t/(2 p_ip_j)$ becomes a logarithmic
integral. At this point, we can perform arbitrary variable transformations without
introducing additional Jacobian factors. In other words, the evolution variables
virtuality, transverse momentum, and polar angle are all formally equivalent, because
\begin{equation}
  \frac{\dt t}{t}=\frac{\dt k_T^2}{k_T^2}=\frac{\dt q_T^2}{q_T^2},
\end{equation}
where $q_T=-t/(1-z)$ for initial-state and $q_T=t/(z(1-z))$ for final-state splittings,
while $k_T^2$ is the relative transverse momentum in the branching process.
We have seen in Sec.~\ref{sec:color_coherence} that angular ordering can effectively
model color coherence, and it might therefore be preferred as an evolution variable,
as long as the evolution kernels are given by the Altarelli-Parisi splitting functions. 

While resumming universal higher-order corrections to the hard process, parton showers
are themselves derived only from the leading real-emission corrections. There are,
however, universal higher-order terms which must be taken into account to make the
parton-shower prediction meaningful: The first is the universal coupling renormalization,
which leads to corrections of the form $\alpha_s/(2\pi)\beta_0\log(k_T^2/\mu_R^2)$, 
where $k_T^2$ is again the relative transverse momentum in the gluon emission~\cite{
  Dokshitzer:1978qu,Amati:1978by,Ellis:1978sf,Libby:1978ig,Mueller:1978xu,Dokshitzer:1978hw}
This term can be absorbed into the running coupling, leading to a particular scale
choice which is different for each branching and depends on the splitting kinematics:
$\alpha_s(k_T^2)$. The other universal term to be incorporated relates the unphysical 
$\overline{\rm MS}$ renormalization scheme to a physical scheme, by including the two-loop
cusp anomalous dimension, $K=(67/18+\zeta_2)\,C_A-10/9\,T_R\,n_f$, into the soft-enhanced
terms of the splitting functions~\cite{Catani:1990rr}. As the term may equally well
be absorbed into the scale of the running coupling, this method is also referred
to as the CMW scale choice.

\begin{table}
  \begin{center}
    \begin{tabular}{ccccc}
      \hline\vphantom{$\int_a^b$}
      & Evolution variable & Splitting variable & Coherence & Reference\\\hline\vphantom{$\int_a^b$}
      Ariadne    & dipole-$k_\perp^2$ & Rapidity         & Antenna   & \cite{Gustafson:1987rq,Kharraziha:1997dn,Lonnblad:1992tz}\\
      Dire       & dipole-$k_\perp^2$ & LC mom fraction  & Dipole    & \cite{Hoche:2015sya} \\
      Herwig     & $E^2\theta^2$      & Energy fraction  & AO        & \cite{Marchesini:1987cf,Corcella:2000bw} \\
      Herwig++   & $(t-m^2)/z(1-z)$   & LC mom fraction  & AO/Dipole & \cite{Gieseke:2003rz,Platzer:2009jq} \\
      Pythia $<$6.4 & $t$             & Energy fraction  & Enforced  & \cite{Sjostrand:1985xi,Bengtsson:1986et} \\
      Pythia $\ge$6.4 & $k_\perp^2$   & LC mom fraction  & Enforced  & \cite{Sjostrand:2004ef} \\
      Sherpa $<$1.2& $t$              & Energy fraction  & Enforced  & \cite{Kuhn:2000dk} \\
      Sherpa $\ge$1.2 & $k_\perp^2$   & LC mom fraction  & Dipole    & \cite{Schumann:2007mg} \\
      Vincia     & variable           & variable         & Antenna   & \cite{Giele:2007di,Giele:2011cb} \\
      \hline
    \end{tabular}
    \caption{Choice of evolution/splitting variable and evolution kernels
      in common parton-shower programs.
      \label{tab:ps_programs}}
  \end{center}
\end{table}
The choices made in presently-available parton shower programs are listed in 
Tab.~\ref{tab:ps_programs}. The quality of a certain choice of splitting parameters 
can often be judged only after comparing the results of the simulation to experimental
measurements.

\section{Matching and Merging}
\label{sec:matching_merging}
Parton showers implement approximate higher-order real-emission corrections to arbitrary
hard processes using the universal soft and collinear factorization properties of the
hard cross sections. They estimate virtual corrections through unitarity conditions.
In order to improve the description of observables, it is often necessary to go beyond
these approximations. One possibility in so doing is to replace the parton shower
approximation at given orders in the strong coupling expansion by exact perturbative
QCD results. This can be done in two different ways.
\begin{itemize}
\item {\em Matching}\\
  The parton-shower expression at fixed order is computed and subtracted from the 
  higher-order calculation to remove double counting. The subtracted result is
  processed by the parton shower.
\item {\em Merging}\\
  A separate tree-level calculation is performed for each parton multiplicity of interest.
  Soft and collinear divergences of the hard matrix elements are regulated by resolution cuts.
  The parton shower is combined with all these calculations, and double-counting is removed
  by appropriate vetoes on shower branchings.
\end{itemize}
The concept of infrared-safe observables and QCD jets plays a crucial role for merging.
Both the initial state and many final states at hadron colliders include hard partons.
Initial- and final-state Bremsstrahlung dress these partons with further radiation. 
The new particles are found predominantly in the vicinity of the original ones, leading 
to clusters of radiation called QCD jets. The hadronization mechanism preserves the jet
structure, such that it can be observed experimentally. Theoretically, a cluster of hadronic
energy can thus be identified with one or more hard initiating partons. For this concept
to work an algorithm must be defined which unambiguously relates the two scenarios.
Crucially, this algorithm must be infrared and collinear safe: if a single parton 
is replaced with a set of collinear partons sharing its original energy, the jet
configuration must not change. Likewise, if a soft parton is added to the original
jet configuration, the jet configuration according to the jet algorithm must not change.
Reviews of jet algorithms are provided elsewhere~\cite{Salam:2009jx}. In the following
we will make use of jets only as theoretical tools for the merging of parton showers
with multiple higher-order tree-level calculations.

\subsection{Matching}
\label{sec:matching}
Using the subtraction formalism introduced in Sec.~\ref{sec:nlo_calculations}, 
an arbitrary infrared and collinear safe observable can be computed at NLO QCD as
\begin{equation}\label{eq:reorder_nlo}
  \begin{split}
    \abr{O}^{\rm(NLO)}\,=&\;
    \sum_{\args{f\,}}\int\dt\bar{\Phi}_n(\args{p})\,
    \Big(\mr{B}_n(\args{a})+\tilde{\mr{V}}_n(\args{a})+
    \sum_{\{\wt{\im\jm},\tilde{k}\}}\mr{I}_n^{\,\wt{\im\jm},\tilde{k}}(\args{a})\Big)\,O(\args{p})\\
    &\quad+\sum_{\args{f\,}}\int\dt\bar{\Phi}_{n+1}(\args{p})\,
    \Big(\mr{B}_{n+1}(\args{a})\,O(\args{p})-
    \sum_{\{ij,k\}}\mr{S}_{n+1}^{\,ij,k}(\args{a})\,O(\bmap{ij}{k}{\args{p}})\,\Big).
  \end{split}
\end{equation}
Note that the configurations $\args{f}$, $\args{p}$ and $\args{a}$ on the second line
 each include one more particle than the term on the first line, because they represent
the real-emission momentum and flavor configuration. The infrared and collinear safety
of the observable guarantees that the expectation value computed in this manner
is physically meaningful.

We now subtract the parton-shower approximation of the NLO result, which consists of 
the resolved real-emission corrections, and the unresolved corrections 
(virtual and real-emission contribution below $t_c$). To distinguish the parton-shower
expression from the original subtraction terms, and because the actual form of the dipole
terms used in the parton shower may differ, we call these contributions $\mr{D}(\args{a})$.
We also re-order the real-emission term according to the principles in Sec.~\ref{sec:parton_showers}:
\begin{equation}\label{eq:nlo_addsub}
  \begin{split}
    &\sum_{\args{f\,}}\int\dt\bar{\Phi}_n(\args{p})\,
    \bar{\mr{B}}_n(\args{a})\,O(\args{p})
    +\sum_{\args{f\,}}\int\dt\bar{\Phi}_{n+1}(\args{p})\,\mr{H}_{n+1}(\args{a})\,O(\args{p})\;,
  \end{split}
\end{equation}
The function $\bar{\mr{B}}(\args{a})$ represents the NLO cross section differential in the 
phase space of the Born process, up to a hard correction. We will call this term the NLO-weighted
Born differential cross section. The function $\mr{H}_{n+1}(\args{a})$ represents the difference 
between the real-emission correction and the parton-shower approximation, expanded to first order 
in the strong coupling:
\begin{equation}\label{eq:def_bbar_a}
  \begin{split}
    \bar{\mr{B}}_n(\args{a})\,=&\;\mr{B}_n(\args{a})+\tilde{\mr{V}}_n(\args{a})+
      \sum_{\{\widetilde{\im\jm},\tilde{k}\}}\mr{I}_n^{\wt{\im\jm},\tilde{k}}(\args{a})\\
    &\qquad+\sum_{\{\wt{\im\jm},\tilde{k}\}}
      \sum_{f_i=q,g}\int\dt\Phi_{+1}^{ij,k}\;
      \sbr{\,\mr{D}_{n+1}^{ij,k}(\rmap{\im\jm}{k}{\args{a}})
        -\mr{S}_{n+1}^{ij,k}(\rmap{\im\jm}{k}{\args{a}})\,}\\
    \mr{H}_{n+1}(\args{a})\,=&\;\mr{B}_{n+1}(\args{a})-\sum_{\{ij,k\}}\mr{D}_{n+1}^{ij,k}(\args{a})\;.
  \end{split}
\end{equation}
Equation~\eqref{eq:nlo_addsub} alone cannot be used to compute physical observables, 
because of the non-local nature of the subtraction. Once the parton shower is added, 
this mismatch is canceled to NLO accuracy, and only corrections of higher-order 
in the strong coupling expansion remain. The first working proposal for NLO matching,
MC@NLO, is therefore also called a modified subtraction method~\cite{Frixione:2002ik}.
The full matched result reads
\begin{equation}\label{eq:master_nlomc}
  \begin{split}
  \abr{O}^{\rm(NLOMC)}\,=&\;\sum_{\args{f\,}}\int\dt\bar{\Phi}_n(\args{p})\,
  \bar{\mr{B}}_n(\args{a})\,\mc{F}_n(\mu_Q^2;\args{a},O)\\
  &+\;\sum_{\args{f\,}}\int\dt\bar{\Phi}_{n+1}(\args{p})\,\mr{H}_{n+1}(\args{a})\,
  \mc{F}_{n+1}(\mu_Q^2;\args{a},O)\;.
  \end{split}
\end{equation}
$\mc{F}_{n+1}(\mu_Q^2)$ is defined in the sense of a truncated parton shower, 
which will be discussed in Sec.~\ref{sec:merging}. Expanded up to the first emission, 
we obtain the formula used to describe the POWHEG method~\cite{Nason:2004rx,Frixione:2007vw}
\begin{equation}\label{eq:master_nlomc_oneem}
  \begin{split}
  \abr{O}^{\rm(NLOMC)}\,\to&\;
  \sum_{\args{f\,}}\int\dt\bar{\Phi}_n(\args{p})\,
  \bar{\mr{B}}_n(\args{a})\Bigg[\,
    \underbrace{\Pi_n(t_c,\mu_Q^2;\args{a})}_{\text{virtual+unresolved}}\,O(\args{p})\\
  &\qquad+\,
    \sum_{\{\wt{\im\jm},\tilde{k}\}}
    \sum_{f_i=q,g}\int\dt\Phi_{+1}^{ij,k}\;\Theta(t(\Phi_{+1}^{ij,k})-t_c)\,
    O(\rmap{\im\jm}{k}{\Phi_{+1}^{ij,k};\args{p}})\\
  &\qquad\qquad\times\,
    \underbrace{\frac{1}{S_{ij}}\,
    \frac{S(\rmap{\im\jm}{k}{f_i;\args{f\,}})}{S(\args{f\,})}\,
    \frac{\mr{D}_{n+1}^{ij,k}(\rmap{\im\jm}{k}{f_i,\Phi_{+1}^{ij,k};\args{a}})}{
      \mr{B}_n(\args{a})}\;
    \Pi_n(t,\mu_Q^2;\args{a})}_{\text{resolved, singular}}\,\Bigg]\\
  &+\;\sum_{\args{f\,}}\int\dt\bar{\Phi}_{n+1}(\args{p})\,
    \underbrace{\Bigg[\,\mr{B}_{n+1}(\args{a})
      -\sum_{ij,k}\mr{D}_{n+1}^{ij,k}(\args{a})\,\Bigg]}_{\text{resolved, non-singular}}\,
    O(\args{p})\;.
  \end{split}
\end{equation}
Equation~\eqref{eq:master_nlomc} can be used to describe the matching of NLO QCD
calculations to parton showers in both the MC@NLO and POWHEG methods~\cite{
  Frixione:2002ik,Nason:2004rx,Frixione:2007vw,Nason:2012pr}. Monte-Carlo events 
are generated in the following way.
\begin{itemize}
\item  A seed event is produced according to either the first or the second
  line of Eq.~\eqref{eq:nlo_addsub}.
\item If the second line is chosen, the event has real-emission kinematics and is kept as-is.
  This generates the ``resolved, non-singular'' term of Eq.~\eqref{eq:master_nlomc}.
  Such events are called hard remainder events, or $\mb{H}$-events.
\item If the first line is chosen, the event has Born-like kinematics and is processed
  by the parton shower. Such events are called standard events, or $\mb{S}$-events.
  An emission might or might not occur, as indicated by the
  ``resolved, singular'' and ``unresolved'' terms of Eq.~\eqref{eq:master_nlomc}.
\end{itemize}
For this method to work in processes with non-trivial color structure, it is vital that
the dipole terms used in the shower, $\mr{D}_{n+1}^{ij,k}$, have the exact same soft
and collinear limits as the real-emission matrix elements. Otherwise the hard remainder
contribution will be divergent, as will the NLO weighted Born differential cross section.
Typical parton showers, however, do not correctly account for the soft singularities in
processes with non-trivial color structure at Born level (cf.\ Sec.~\ref{sec:large_nc_ps}).
The problem can be solved by adding a soft-suppression factor to both the second 
line in Eq.~\eqref{eq:def_bbar_a} and the second line in Eq.~\eqref{eq:nlo_addsub}~\cite{Frixione:2002ik}.
Alternatively, one may correct the parton-shower approximation with the exact dipole terms as defined
by Eq.~\eqref{eq:def_cs_real_subterms}, which necessitates the computation of 
non-probabilistic Sudakov factors~\cite{Hoeche:2011fd,Lonnblad:2012hz}.

Note that Eq.~\eqref{eq:master_nlomc} describes only a single parton-shower step.
In principle one can therefore change the parton-shower generator after the matching
has been performed. This idea is used in the POWHEG method~\cite{Nason:2004rx,Frixione:2007vw},
which can be thought of as a matching to a matrix-element corrected parton shower.
In this case the parton-shower dipoles are defined as:
\begin{equation}\label{eq:def_powheg}
  \mr{D}_{n+1}^{ij,k}\to\rho_{ij,k}(\args{a})\,\mr{B}_{n+1}(\args{a})
  \qquad\text{where}\qquad
  \rho_{ij,k}(\args{a})=\frac{\mc{D}_{n+1}^{ij,k}(\args{a})}{\sum_{mn,l}\mc{D}_{n+1}^{mn,l}(\args{a})}\;.
\end{equation}
This means the full radiative corrections are exponentiated into a Sudakov form factor.
After the first emission (or no-emission), the parton shower is used to implement 
further splittings.

One can construct a mixed scheme, where $\mr{D}_{n+1}^{ij,k}$ is defined as
\begin{equation}\label{eq:def_powheg_zhsplit}
  \mr{D}_{n+1}^{ij,k}\to\rho_{ij,k}(\args{a})\,
  \sbr{\,\mr{B}_{n+1}(\args{a})-\mr{B}_{n+1}^{(r)}(\args{a})\,}\;,
\end{equation}
with $\mr{R}^{(r)}$ an arbitrary infrared-finite part of the real-emission 
cross section and $\rho_{ij,k}$ given by Eq.~\eqref{eq:def_powheg}.
This method can be used in particular to deal with the problem of 
zeros in the Born process~\cite{Alioli:2008gx}.

\begin{figure}
  \begin{center}
    \includegraphics[width=0.5\textwidth]{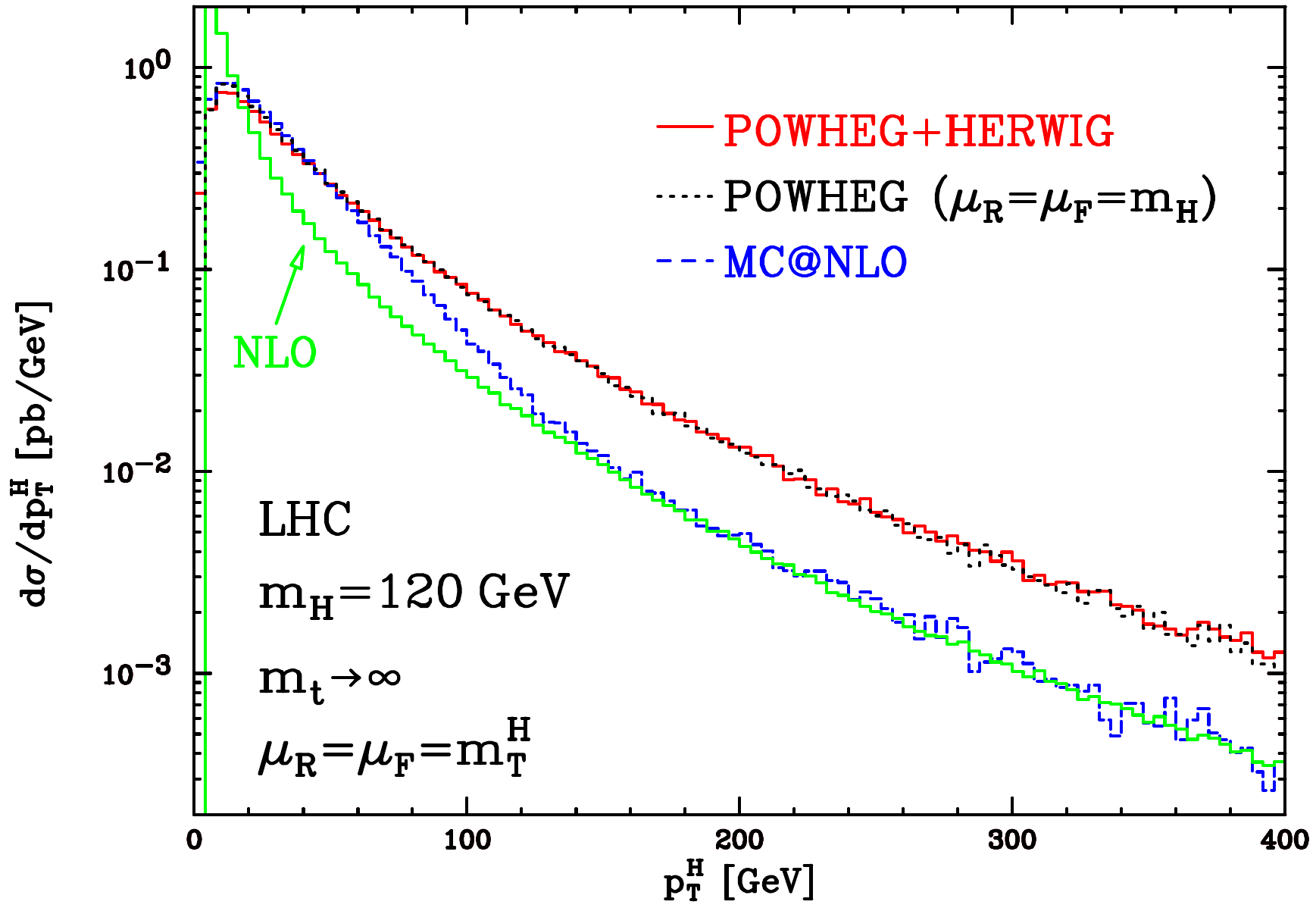}
    \caption{Comparison of predictions from MC@NLO and POWHEG for the transverse
      momentum of the Higgs boson in inclusive Higgs-boson production.
      Figure taken from~\cite{Alioli:2008tz}.
      \label{fig:mcnlo_powheg_comparison}}
  \end{center}
\end{figure}
Figure~\ref{fig:mcnlo_powheg_comparison} shows a comparison between predictions
from MC@NLO~\cite{Frixione:2002ik} and POWHEG~\cite{Nason:2004rx,Frixione:2007vw}
for the transverse momentum spectrum of the Higgs boson in inclusive Higgs-boson
production. The MC@NLO result shows a feature around the resummation scale, 
$\mu_Q=120~GeV$, while the POWHEG result does not. This can be attributed to the
large NLO $K$-factor in this process, which multiplies the NLO-weighted Born
cross section, Eq.~\eqref{eq:def_bbar_a}. In the MC@NLO method, this contribution 
generates resolved radiation only up to a transverse momentum of $\mu_Q$, 
cf.\ Eq.~\eqref{eq:master_nlomc}. In contrast, in the original POWHEG method 
shown here it generates radiation up to the hadronic center of mass energy,
as $\mu_Q\to E_{\rm cms}$. The difference between the full matrix-element 
corrections used in POWHEG (cf.\ Eq.~\eqref{eq:def_powheg}) and the DGLAP 
splitting kernels used in MC@NLO has a much smaller effect.

This example demonstrates that NLO matching techniques do not account for all possible
higher-order effects, as the matching condition only guarantees the preservation of
NLO and parton-shower accuracy. If higher-order corrections to the process are still
large, as in the case of Higgs-boson production, improvements to NLO matching
must be found.

\subsection{Merging}
\label{sec:merging}
\begin{figure}
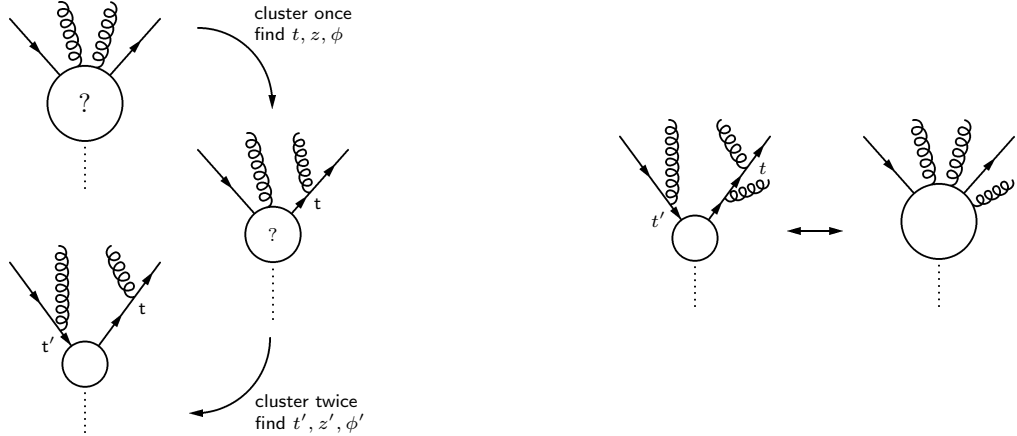

  \begin{center}
    \clusterexample\hskip 2.5cm
    \raisebox{40pt}{\examplesplit}
    \caption{Left: Sketch of a clustering sequence identifying the 
      $e^+e^-\to q\bar{q}$ inclusive reaction in a $e^+e^-\to q\bar{q}gg$ final state.
      Right: Sketch of a truncated parton-shower emission in the existing
      history and its correspondence to the higher-order tree-level matrix element.
      \label{fig:ps_clustering}}
  \end{center}
\end{figure}
Merging algorithms combine tree-level calculations for multi-jet configurations
with parton showers. In this context, the leading-order calculation must be 
interpreted in terms of a parton-shower branching history, in order to identify
the Sudakov factors that need to be included in order to maintain (approximate)
unitarity of the inclusive cross section. It is therefore vital that the parton
shower is invertible, in other words, that any particular $(n+k)$-particle 
final state can be mapped onto the $n$-particle final state of the inclusive
process by successive recombination of two partons into one. This is sketched 
in Fig.~\ref{fig:ps_clustering} (left). At each step of the clustering four-momentum 
is conserved locally, according to the exact inverted parton-shower kinematics.
Standard parton showers have multiple coverage of the full phase space through
different splitting processes, and therefore several possible clustered configurations
exist for one and the same final state. To obtain a definite configuration that
can be used as a starting condition for the parton shower, the cluster algorithm
must select one of the options with the correct probability.

Once a parton-shower ``history'' corresponding to the hard-scattering configuration
has been identified, it is dressed with further radiation by the parton shower.
Radiation can occur at any point in the configuration. This is exemplified in 
Fig.~\ref{fig:ps_clustering} (right), where the emission of a gluon from an 
intermediate quark propagator is sketched. The situation is called ``truncated
showering'', because the evolution is truncated at scale $t$, where radiation of
the additional gluon already present in the hard scattering calculation must be
re-implemented. At the same time, we need to test whether the newly-emitted parton
is hard enough to constitute a jet on its own. In this case the configuration
should be simulated using a hard-scattering calculation with one more jet,
as indicated in Fig.~\ref{fig:ps_clustering} (right). To avoid double counting,
the event with hard truncated showering must then be vetoed, which reduces the cross
section and effectively generates a Sudakov factor for the evolution between 
the hard scale and the scale of jet resolution. This scale, called $Q_{\rm cut}$, 
is a technical parameter of the merging algorithm, and the dependence of the 
final result on it must cancel to the logarithmic accuracy of the parton shower.

To formalize the above ideas we first introduce $Q_{\rm cut}$ as a criterion 
for the real-emission configuration to be of $l$-jet type. For final states 
with $(n+l+1)$ particles, we identify an $(n+l)$-particle final state by clustering 
according to the kinematics and probability defined by the parton shower. 
We define the $l$-jet inclusive and exclusive expectation values
\begin{equation}\label{eq:omean_jet}
  \begin{split}
  \abr{O}_{l}^{\rm incl}\,=&\;\sum_{k=0}^\infty\abr{O_{l+k}\,\Theta(Q_{l+k}-Q^{\,l}_{\rm cut})}\;,\\
  \abr{O}_{l}^{\rm excl}\,=&\;
  \sum_{k=0}^\infty\abr{O_{l+k}\,\Theta(Q_{l+k}-Q^{\,l}_{\rm cut})\,\Theta(Q^{\,l}_{\rm cut}-Q_{l+k+1})}\;.
  \end{split}
\end{equation}
They include contributions from all final states with at least $l$ partons, 
which must form $l$ (but not $l+1$ in the case of $\abr{O}_l^{\rm excl}$) 
parton-level jets according to the definition of the jet resolution criterion $Q$.
Note that the identification of jet configurations is used only to separate
the phase space for hard QCD radiation into regions described by different hard
matrix elements. Eventually, a fully inclusive simulation will be obtained 
by summing all jet configurations as
\begin{equation}\label{eq:omean_total}
  \abr{O}\,=\;\sum_{l=0}^{N_{\rm max}-1}\abr{O}_{l}^{\rm excl}
  +\abr{O}_{N_{\rm max}}^{\rm incl}\;.
\end{equation}
$N_{\rm max}$ is the highest possible jet multiplicity that can practically
be computed using the matrix-element generator. It depends on the order of 
the hard-scattering calculation and on the process, and typically ranges 
between 4 and 6 for LO calculations and 2 and 3 for NLO calculations.

If the parton shower were to evolve only the final state, and not any
of the intermediate states, we would obtain, after the first emission,
\begin{equation}\label{eq:mlm_merging}
  \begin{split}
    &\abr{O}_{l}^{\rm excl}\,\to\;\sum_{\args{f\,}}\int\dt\bar{\Phi}_{n+l}\,
    \mr{B}_{n+l}\,\Theta(Q_{n+l}-Q^{\,l}_{\rm cut})\,
    \Pi_{m}(t_c,\mu_Q^2;>\!Q^{\,l+1}_{\rm cut})\,
    \mc{F}_{n+l}(\mu_Q^2;O,<\!Q^{\,l+1}_{\rm cut})\;.
  \end{split}
\end{equation}
For ease of notation we dropped the arguments $\args{a}$, $\args{f}$, and $\args{p}$. 
From here on we also drop the explicit notation of the kinematics mappings,
$\bmap{ij}{k}{\args{a}}$ and $\rmap{\im\jm}{k}{\args{a}}$. 
In addition we define Sudakov factors for the vetoed parton shower,
using Eq.~\eqref{eq:sudakov} and
\begin{equation}\label{eq:sudakov_ijk_vetoed}
  \begin{split}
  \Pi_m^{\wt{\im\jm},\tilde{k}}(t,t';>\!Q^{\,l+1}_{\rm cut})=\exp
  \Bigg\{\,-\,&\frac{1}{16\pi^2}\sum_{f_i\in\{q,g\}}
  \int_t^{t'}\dt\bar{t}\int\dt z\int\frac{\dt\phi}{2\pi}\,\\
  &\times\frac{1}{2}\frac{
    \mr{D}^{ij,k}_{m+1}}{
    \mr{B}_m}\;\Theta(Q_{m+1}-Q^{\,l+1}_{\rm cut})\;\Bigg\}\;.
  \end{split}
\end{equation}
We also introduce the vetoed generating functional
\begin{equation}\label{eq:ps_functional_veto}
  \begin{split}
    &\mc{F}_m(\mu_Q^2;O,<\!Q^{\,l+1}_{\rm cut})\,=\;
    \Pi_m(t_c,\mu_Q^2;<\!Q^{\,l+1}_{\rm cut})\,O_m
    +\sum_{\{\wt{\im\jm},\tilde{k}\}}
    \sum_{f_i=q,g}\int\dt\Phi_{+1}^{ij,k}\;\Theta(t(\Phi_{+1}^{ij,k})-t_c)\\
  &\qquad\qquad\times\,
    \frac{1}{S_{ij}}\,\frac{S_{m+1}}{S_m}\,
    \frac{\mr{D}_{m+1}^{ij,k}}{\mr{B}_m}\;
    \Pi_m(t,\mu_Q^2;<\!Q^{\,l+1}_{\rm cut})\,\Theta(Q^{\,l+1}_{\rm cut}-Q_{m+1})\,
    \mc{F}_{m+1}(t;O,<\!Q^{\,l+1}_{\rm cut})\;.
  \end{split}
\end{equation}
Unitarity requires that $\mc{F}_m(\mu_Q^2;1,<\!Q^{\,l+1}_{\rm cut})=1$. 
Therefore, the naive merging algorithm, Eq.~\eqref{eq:mlm_merging}, reduces
the $l$-jet cross section computed by hard matrix elements by the factor
$\Pi_{n+l}(t_c,\mu_Q^2;>\!Q^{\,l+1}_{\rm cut})$.
In other words, it turns the matrix-element result, which is an inclusive
result for $l$-jet production, into an exclusive result by resumming the
leading corrections from virtual and unresolved real-emission contributions.
Qualitatively this observation holds also for the correct merging procedure.

Equation~\eqref{eq:mlm_merging} does not take truncated showers into account 
and therefore violates the parton-shower evolution equations. To correct the problem
without further complicating the notation, we extend our definition of the parton shower 
evolution kernels:
\begin{equation}\label{eq:all_kernels}
  \begin{split}
    \tilde{\mr{D}}_{n+l+1}\,=&\;
    \sum_{\{\wt{\im\jm},\tilde{k}\},f_i}\frac{S_{n+l+1}}{S_{ij}S_{n+l}}\,
    \mr{D}_{n+l+1}^{ij,k}\,\Theta(t_{n+k}-t)\\
    &\qquad+\,\mr{B}_{n+l}\sum_{m=n}^{n+l-1}
    \sum_{\{\wt{\im\jm},\tilde{k}\},f_i}
    \frac{S_{m+1}}{S_{ij}S_{m}}\frac{\mr{D}_{m+1}^{ij,k}}{\mr{B}_m}\,
    \Theta(t_m-t)\,\Theta(t-t_{m+1})\;,
  \end{split}
\end{equation}
Equation~\eqref{eq:all_kernels} is called the compound subtraction term.
The $\Theta$-functions guarantee that the evolution kernel for the $m$-particle
state is active only in the regions where the newly emitted particle has an
evolution parameter, $t$, ranging between the ones in the previous and in the
subsequent splitting. Only the evolution of the full final state proceeds
unrestricted. We also define a compound Sudakov factor and the related generating
functional of the parton shower:
\begin{equation}\label{eq:sudakov_ijk_compound}
  \tilde{\Pi}_{n+l}(t,t')=
  \exp\Bigg\{\,-\,\frac{1}{16\pi^2}\int_t^{t'}
  \dt\bar{t}\int\dt z\int\frac{\dt\phi}{2\pi}\,
  \frac{\tilde{\mr{D}}_{n+l+1}}{\mr{B}_{n+l}}\;\Bigg\}\;,
\end{equation}
and
\begin{equation}\label{eq:ps_functional_compound}
  \begin{split}
    \tilde{\mc{F}}_{n+l}(\mu_Q^2;O)\,=&\;\tilde{\Pi}_{n+l}(t_c,\mu_Q^2)\,O_{n+l}
    +\,\int_{t_c}^{\mu_Q^2}\dt\Phi_{+1}\;
      \frac{\tilde{\mr{D}}_{n+l+1}}{\mr{B}_{n+l}}\;
      \tilde{\Pi}_{n+l}(t,\mu_Q^2)\,\tilde{\mc{F}}_{n+l+1}(t;O)\;.
  \end{split}
\end{equation}
Correspondingly, $\tilde{\mc{F}}_{n+l}(\mu_Q^2;O,<\!Q^{\,l+1}_{\rm cut})$
is defined with an additional cut on the real-emission term, in complete
analogy to Eq.~\eqref{eq:sudakov_ijk_vetoed}.
The $l$-jet contribution to $O$ finally is
\begin{equation}\label{eq:meps_merging}
  \begin{split}
    \abr{O}_{l}^{\rm excl}\,=&\;\sum_{\args{f\,}}\int\dt\bar{\Phi}_{n+l}\,
    \mr{B}_{n+l}\,\Theta(Q_{n+l}-Q^{\,l}_{\rm cut})\,
    \tilde{\Pi}_{n+l}(t_c,\mu_Q^2;>\!Q^{\,l+1}_{\rm cut})\,
    \tilde{\mc{F}}_{n+l}(\mu_Q^2;O,<\!Q^{\,l+1}_{\rm cut})\;.
  \end{split}
\end{equation}
Expanded up to the first emission, we have
\begin{equation}\label{eq:meps_merging_oneem}
  \begin{split}
    \abr{O}_{l}^{\rm excl}\,\to&\sum_{\args{f\,}}\int\dt\bar{\Phi}_{n+l}\,
    \mr{B}_{n+l}\,\Theta(Q_{n+l}-Q^{\,l}_{\rm cut})\,
    \tilde{\Pi}_{n+l}(t_c,\mu_Q^2;>\!Q^{\,l+1}_{\rm cut})\,
    \Bigg[\,\tilde{\Pi}_{n+l}(t_c,\mu_Q^2;<\!Q^{\,l+1}_{\rm cut})\,O_{n+l}\\
    &\qquad +\,\int_{t_c}^{\mu_Q^2}\dt\Phi_{+1}\;
      \frac{\tilde{\mr{D}}_{n+l+1}}{\mr{B}_{n+l}}\;
      \tilde{\Pi}_{n+l}(t,\mu_Q^2;<\!Q^{\,l+1}_{\rm cut})\,
      \Theta(Q^{\,l+1}_{\rm cut}-Q_{n+l+1})\,O_{n+l+1}\,\Bigg]\;.
  \end{split}
\end{equation}

The logarithmic structure of the parton-shower is maintained by the merging
only if the clustering procedure described above is an exact inversion 
of the shower evolution, and if truncated parton shower evolution, 
in particular the jet veto on the truncated shower emissions, is implemented.
Note that if the hardness measure $Q$  is identical to the evolution variable, 
truncated parton showers cannot produce an emission, but only lead to vetoes 
on events. A method to not implement truncated shower emissions while still
maintaining formal accuracy and keeping the evolution variable and jet 
resolution parameter distinct also exists~\cite{Lonnblad:2001iq}.

Note that Eq.~\eqref{eq:omean_total} violates the unitarity of the parton-shower
simulation, unless the splitting kernels used in the shower are given precisely
by the real-emission matrix elements.\footnote{This case is irrelevant, because
  we would not need to perform any merging at all.} The reason for the unitarity
violation is that we replaced resolved real-emission corrections by full tree-level
matrix elements, while not accounting for the corresponding change in the unresolved
and virtual corrections. This situation can be remedied by adding a correction
to Eq.~\eqref{eq:meps_merging}, which accounts for the difference in the unresolved
region. Using the unitarity condition, the relevant terms can be computed with
the same tree-level matrix elements which are also employed in the 
merging~\cite{Lonnblad:2012ng,Platzer:2012bs}.

\begin{table}
  \begin{center}
    \begin{tabular}{cccccc}\hline
      Method & Shower Generator & Unitary & Accuracy & References \\\hline
      MLM & Herwig/Pythia & No & unknown & \cite{Mangano:2001xp,Alwall:2007fs}\\
      CKKW & Apacic & No & LO$\otimes$PS & \cite{Catani:2001cc,Krauss:2002up}\\
      CKKW-L & Ariadne/Pythia & No & LO$\otimes$PS & \cite{Lonnblad:2001iq,Lonnblad:2011xx}\\
      METS & Sherpa & No & LO$\otimes$PS & \cite{Hoeche:2009rj}\\
      CKKW' & Herwig++ & No & LO$\otimes$PS & \cite{Hamilton:2009ne}\\
      UMEPS & Pythia & Yes & LO$\otimes$PS & \cite{Lonnblad:2012ng,Platzer:2012bs}\\
      \hline\end{tabular}
    \caption{Practically implemented LO merging schemes.
      \label{tab:merging_schemes}}
  \end{center}
\end{table}
The available merging schemes implemented in standard Monte-Carlo event generators
are classified in Tab.~\ref{tab:merging_schemes}. They differ in their use of
parton showers, Sudakov factors (analytic/numerical) and jet resolution criterion.
A thorough comparison of the available methods showed that their results are 
well comparable within the expected theoretical uncertainty~\cite{Alwall:2007fs}.

\subsection{Merging matched simulations}
The merging methods described above can be extended to accommodate NLO calculations
for the $l$-jet cross sections. In this process the matching must be modified,
because the radiation probability of the full shower simulation has changed, and
the $\mc{O}(\alpha_s)$ expansion of the truncated parton shower approximation
must be subtracted from the NLO result before it can be processed by the shower.
This is done formally by defining the modified NLO-weighted Born cross section
and the modified hard remainder:
\begin{equation}\label{eq:def_bbar_match}
  \begin{split}
    \tilde{\mr{B}}_{n+l}\;=&\,\mr{B}_{n+l}+\tilde{\mr{V}}_{n+l}+
      \sum_{\{\wt{\im\jm},\tilde{k}\}}\mr{I}_{n+l}^{\wt{\im\jm},\tilde{k}}\,
      +\sum_{\{\wt{\im\jm},\tilde{k}\},f_i}\int\dt\Phi_{+1}^{ij,k}\;
      \sbr{\,\mr{D}_{n+l+1}^{ij,k}\Theta(t_{n+l}-t)-\mr{S}_{n+l+1}^{ij,k}\,}\\
      &\qquad+\,\mr{B}_{n+l}\sum_{m=n}^{n+l-1}
      \sum_{\{\wt{\im\jm},\tilde{k}\},f_i}\int\dt\Phi_{+1}^{ij,k}\;
      \frac{\mr{D}_{m+1}^{ij,k}}{\mr{B}_m}\,
      \Theta(t_m-t)\,\Theta(t-t_{m+1})\;,\\
    \tilde{\mr{H}}_{n+l+1}\;=&\,\mr{B}_{n+l+1}
    -\sum_{ij,k}\mr{D}_{n+l+1}^{ij,k}\Theta(t_{n+l}-t)
    -\mr{B}_{n+l}\sum_{m=n}^{n+l-1}
    \sum_{ij,k}\frac{\mr{D}_{m+1}^{ij,k}}{\mr{B}_m}\,
    \Theta(t_m-t)\,\Theta(t-t_{m+1})\;.
  \end{split}
\end{equation}
The $l$-jet contribution to the observable $O$ from an NLO-calculation
is then given by
\begin{equation}\label{eq:mepsnlo_merging}
  \begin{split}
    \abr{O}_{l}^{\rm excl}\,=&\;\sum_{\args{f\,}}\int\dt\bar{\Phi}_{n+l}\,
    \tilde{\mr{B}}_{n+l}\,\Theta(Q_{n+l}-Q^{\,l}_{\rm cut})\,
    \tilde{\Pi}_{n+l}(t_c,\mu_Q^2;>\!Q^{\,l+1}_{\rm cut})\,
    \tilde{\mc{F}}_{n+l}(\mu_Q^2;O,<\!Q^{\,l+1}_{\rm cut})\\
  &\hspace*{-12mm}+\sum_{\args{f}}
    \int\dt\bar{\Phi}_{n+l+1}\,\tilde{\mr{H}}_{n+l+1}\,
    \Theta(Q_{n+l}-Q^{\,l}_{\rm cut})\,
    \tilde{\Pi}_{n+l}(t_c,\mu_Q^2;>\!Q^{\,l+1}_{\rm cut})\,
    \tilde{\mc{F}}_{n+l+1}(\mu_Q^2;O,<\!Q^{\,l+1}_{\rm cut})\;.
  \end{split}
\end{equation}
Expanded up to the first emission, this becomes
\begin{equation}\label{eq:mepsnlo_merging_oneem}
  \begin{split}
    \abr{O}_{l}^{\rm excl}\,\to&\;\sum_{\args{f\,}}\int\dt\bar{\Phi}_{n+l}\,
    \tilde{\mr{B}}_{n+l}\,\Theta(Q_{n+l}-Q^{\,l}_{\rm cut})\,
    \tilde{\Pi}_{n+l}(t_c,\mu_Q^2;>\!Q^{\,l+1}_{\rm cut})\,
    \Bigg[\,\tilde{\Pi}_{n+l}(t_c,\mu_Q^2;<\!Q^{\,l+1}_{\rm cut})\,O_{n+l}\\
  &\quad+\,\int_{t_c}^{\mu_Q^2}\dt\Phi_{+1}\;
      \frac{\tilde{\mr{D}}_{n+l+1}}{\mr{B}_{n+l}}\;
      \tilde{\Pi}_{n+l}(t,\mu_Q^2;<\!Q^{\,l+1}_{\rm cut})\,
      \Theta(Q^{\,l+1}_{\rm cut}-Q_{n+l+1})\,O_{n+l+1}\,\Bigg]\\
  &\hspace*{-12mm}+\sum_{\args{f}}
    \int\dt\bar{\Phi}_{n+l+1}\,\tilde{\mr{H}}_{n+l+1}\,
    \Theta(Q_{n+l}-Q^{\,l}_{\rm cut})\,
    \tilde{\Pi}_{n+l}(t_c,\mu_Q^2;>\!Q^{\,l+1}_{\rm cut})\,
    \Theta(Q^{\,l+1}_{\rm cut}-Q_{n+l+1})\;O_{n+l+1}\;.
  \end{split}
\end{equation}
Like in the case of matching, the complete local cancellation of infrared singularities 
can only be guaranteed if the parton shower for the first step in the non-truncated
evolution (below $t_{n+l}$) contains full color and spin effects. If this is not the case,
then a suppression factor must be implemented, damping the integrand in the single-soft
region of gluon emission. However, a standard parton shower may be used for all emissions
(or non-emissions) in the truncated shower, because soft and collinear singularities
are regulated by the finite cutoff scales ($t_{m+1}$ in Eqs.~\eqref{eq:def_bbar_match}).

If the definition of the jet measure, $Q$, and the evolution parameter of the 
parton shower, $t$, coincide, we can use this assumption to rewrite
Eq.~\eqref{eq:mepsnlo_merging_oneem} as
\begin{equation}
  \begin{split}
    \abr{O}_{l}^{\rm excl}\,\to&\;
    \int\dt\bar{\Phi}_{n+l}\,\tilde{\mr{B}}_{n+l}\,\Theta(t_{n+l}-t_{\rm cut})
    \Bigg(\prod\limits_{i=n}^{n+l-1}\,\Pi_{\,i}^{\rm(PS)}(t_{i+1},t_i)\Bigg)\\
    &\hspace*{-12mm}\times\;
    \Bigg[\,\Pi_{n+l}(t_c,t_{n+l})\;O_{n+l}+
    \int_{t_c}^{\mu_Q^2}\dt\Phi_{+1}\;\frac{\tilde{\mr{D}}_{n+l+1}}{\mr{B}_{n+l}}\,
    \Pi_{n+l}(t_{n+l+1},t_{n+l})\,\Theta(t_{\rm cut}-t_{n+l+1})\,
    \;O_{n+l+1}\,\Bigg]\\
    &+\,\int\dt\bar{\Phi}_{n+l+1}\,\Theta(t_{n+l}-t_{\rm cut})\,
    \mr{H}_{n+l+1}\Bigg(\prod\limits_{i=n}^{n+l}\,\Pi_{\,i}^{\rm(PS)}(t_{i+1},t_i)\Bigg)\;
    \Theta(t_{\rm cut}-t_{n+l+1})\;O_{n+l+1}\;.
  \end{split}
\end{equation}
We have indicated the potentially different nature of the no-branching 
probabilities for truncated showers using the superscript $(PS)$.
We can now combine the subtraction terms in $\tilde{\mr{B}}_{n+l}$ 
and the no-branching probabilities on the first line. This makes the
$\mc{O}(\alpha_s)$ corrections to the leading-order merging formula,
Eq.~\eqref{eq:meps_merging_oneem}, explicit
\begin{equation}\label{eq:nlo_term_tisq}
  \begin{split}
    \abr{O}_{l}^{\rm excl}\,\to&\;
    \int\dt\bar{\Phi}_{n+l}\,\Theta(t_{n+l}-t_{\rm cut})\;
    \bar{\mr{B}}_{n+l}\;
    \Bigg[\prod\limits_{i=n}^{n+l-1}\,\Pi_{\,i}^{\rm(PS)}(t_{i+1},t_i)
    \Bigg(1+\frac{\mr{B}_{n+l}}{\bar{\mr{B}}_{n+l}}
    \int\limits_{t_{i+1}}^{t_i}\dt\Phi_{+1}\,
    \frac{\mr{D}_{i+1}^{\rm(PS)}}{\mr{B}_i}\Bigg)\Bigg]\\
    &\hspace*{-12mm}\times\;\Bigg[\,\Pi_{n+l}(t_c,t_{n+l})\;O_{n+l}
      +\int_{t_c}^{\mu_Q^2}\dt\Phi_{+1}\;\frac{\mr{D}_{n+l+1}}{\mr{B}_{n+l}}\,
    \Pi_{n+l}(t_{n+l+1},t_{n+l})\,\Theta(t_{\rm cut}-t_{n+l+1})\,
    \;O_{n+l+1}\,\Bigg]\\
    &+\,\int\dt\bar{\Phi}_{n+l+1}\,\Theta(t_{n+l}-t_{\rm cut})\,
    \mr{H}_{n+l+1}\,
    \Bigg(\prod\limits_{i=n}^{n+l}\,\Pi_{\,i}^{\rm(PS)}(t_{i+1},t_i)\Bigg)\;
    \Theta(t_{\rm cut}-t_{n+l+1})\;O_{n+l+1}\;.\hspace*{-20mm}
  \end{split}
\end{equation}
We have added arbitrary higher-order terms, which allow one to include the sum
over truncated shower subtractions in the product of no-branching probabilities.
Each term in the product can then be interpreted as the $\mc{O}(\alpha_s)$-subtracted, 
truncated, vetoed parton-shower no-branching probabilities for a particular final state.
In practice, these expressions can be generated by running a truncated vetoed shower
and skipping the first veto, depending on the ratio $\mr{B}_{n+l}/\bar{\mr{B}}_{n+l}$.
The remainder of the expression corresponds to an ordinary MC@NLO simulation, 
consisting of standard and hard remainder events, including the jet veto. 
This scheme is particularly easy to implement in practice, because no emissions 
need to be generated in the truncated shower.

\begin{table}
  \begin{center}
    \begin{tabular}{ccccc}\hline
      Method & Shower Generator & Unitary & Accuracy & References \\\hline
      NL$^3$ & Ariadne/Pythia & No & NLO$\otimes$PS & \cite{Lavesson:2008ah}\\
      MEPS@NLO & Sherpa & No & NLO$\otimes$PS & \cite{Hoeche:2009rj}\\
      FxFx & Herwig(++)/Pythia & No & unknown & \cite{Frederix:2012ps}\\
      UNLOPS & Pythia & Yes & NLO$\otimes$PS & \cite{Lavesson:2008ah,Lonnblad:2012ix}\\
      \hline\end{tabular}
    \caption{Practically implemented NLO merging schemes.
      \label{tab:nlo_merging_schemes}}
  \end{center}
\end{table}
There are several implementations of next-to-leading order merging methods at present.
They differ in their theoretical accuracy and in the underlying 
parton-shower generator. They are listed in Tab.~\ref{tab:nlo_merging_schemes}.
An alternative scheme, based on known analytic resummed results combined with
the parton shower also exists~\cite{Alioli:2012fc}. It allows to reach higher
logarithmic accuracy for certain observables than attainable in the other methods.

\begin{figure}
  \begin{center}
    \includegraphics[width=0.6\textwidth]{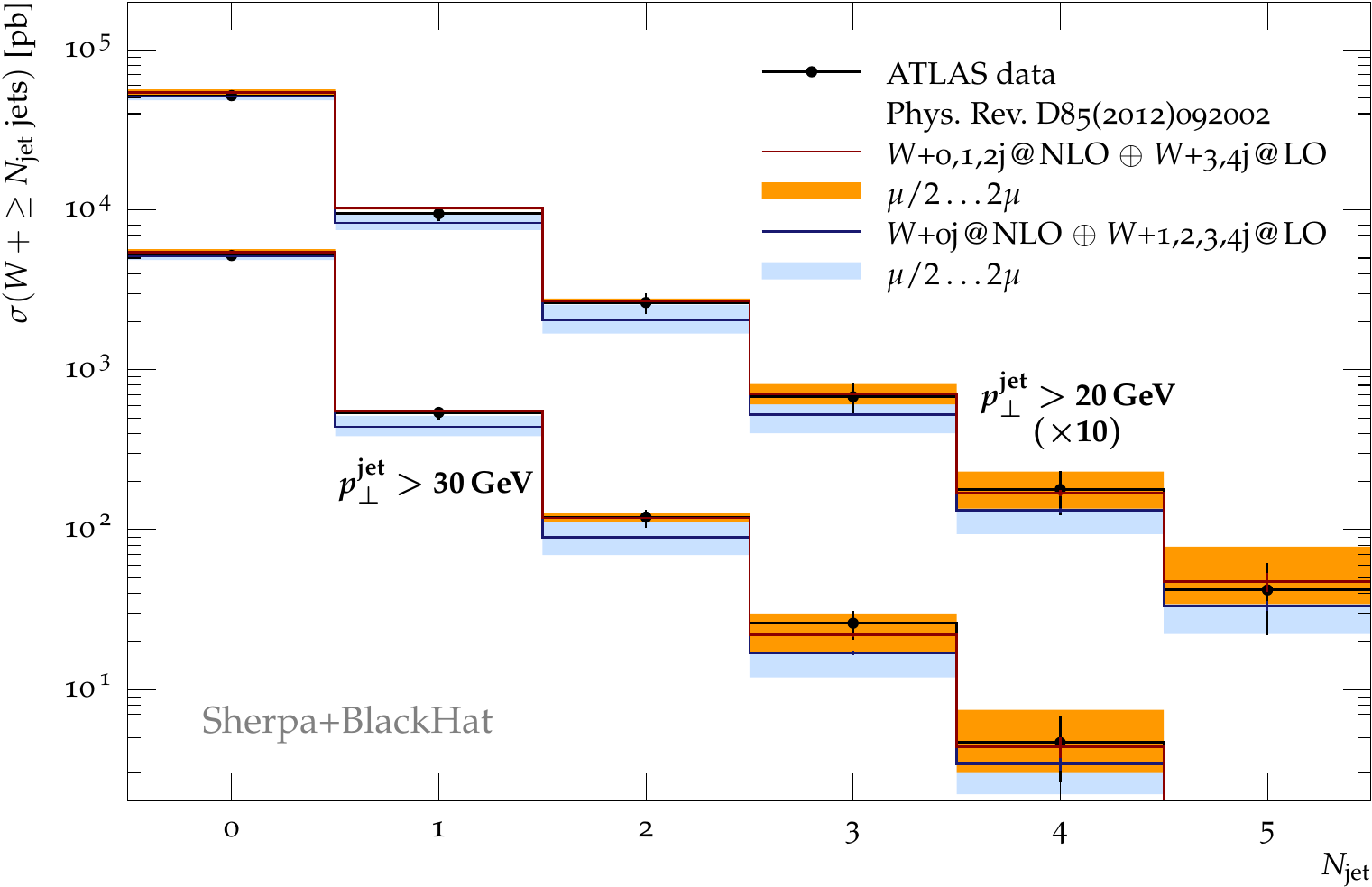}
    \caption{Comparison between LO and NLO merged results for $W$+jets production
      at the Large Hadron Collider. Figure taken from~\cite{Hoeche:2012yf}.
      \label{fig:nlo_merging}}
  \end{center}
\end{figure}
Figure~\ref{fig:nlo_merging} shows a comparison between LO and NLO merged predictions 
for the jet multiplicity distribution in $W$-boson production at the Large Hadron Collider.
The NLO prediction is given by the red histogram. Up to $W$+2 jet final states are
computed at NLO, and up to $W$+4 jet final states are computed at LO. The theoretical 
uncertainty is given by the orange band. The LO prediction is given by the blue histogram,
with the associated blue uncertainty band. Up to $W$+4 jet matrix elements are included
in this calculation.

\section{Underlying events}
\label{sec:underlying_event}
\begin{figure}
  \begin{center}
    \includegraphics[width=0.5\textwidth]{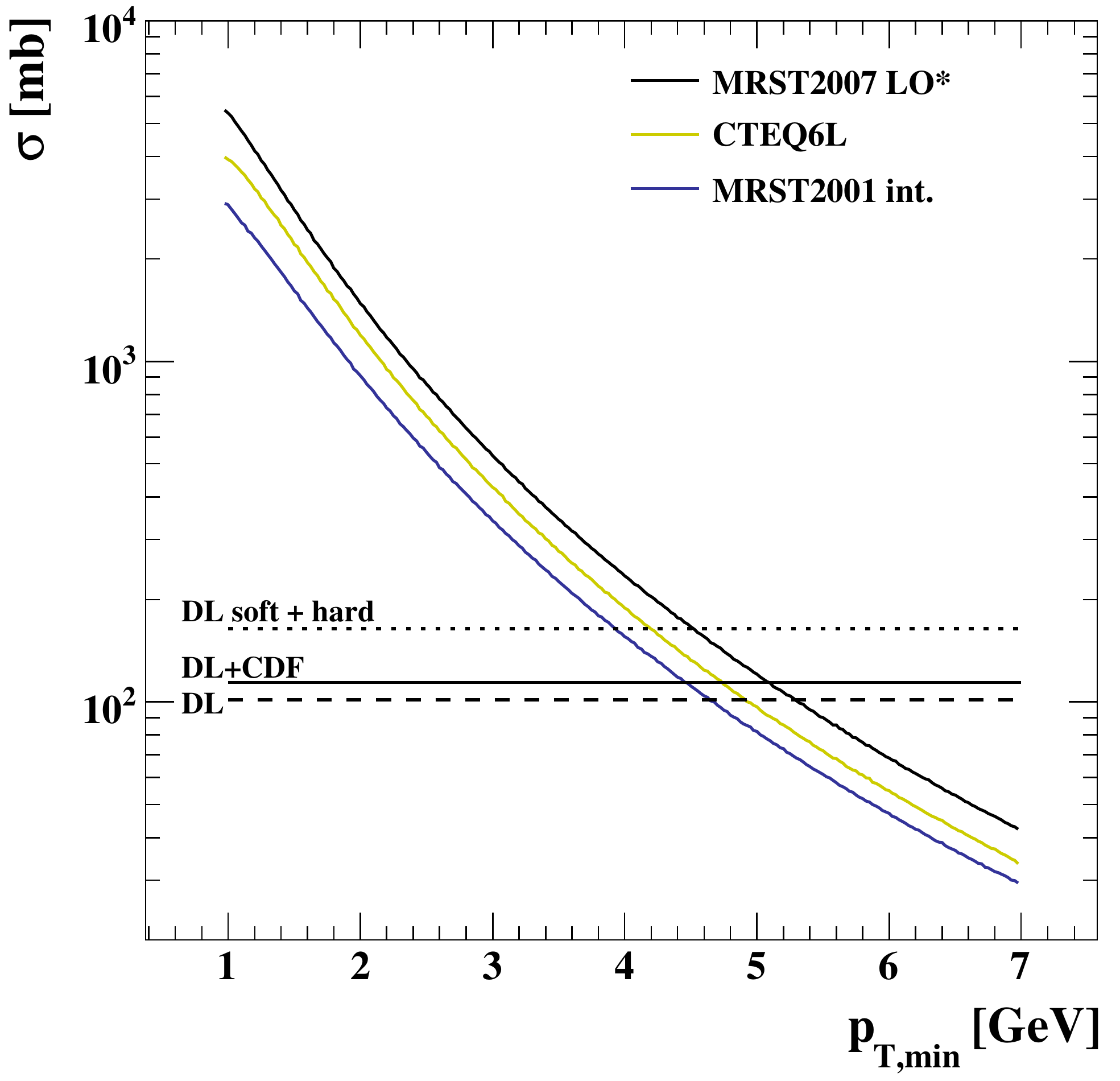}
    \caption{The inclusive jet cross section at LO for different PDF sets,
      compared to various extrapolations of non-perturbative fits to the total 
      proton-proton cross section at 14~TeV. The dashed line gives the prediction
      based on a parametrization by Donnachie and Landshoff~\cite{Donnachie:1992ny},
      the solid line stems from the same fit, but constrained by CDF data~\cite{Abe:1993xy}.
      The dotted line is predicted by the most recent fit~\cite{Donnachie:2004pi}, 
      which includes contributions from both hard and soft Pomerons.
      Figure taken from~\cite{Bahr:2008wk}.
      \label{fig:sigma_nd}}
  \end{center}
\end{figure}
Up to this point we have relied on the strict factorization of the cross section for the
production of a final state of large invariant mass or large invariant momentum transfer,
Eq.~\eqref{eq:factorization}. We have therefore neglected any effects of rescattering
and the exchange of multiple partons between the initial-state protons. Such effects may,
however, play a role in experiments. This can be anticipated by observing that the perturbative 
QCD cross section according to Eq.~\eqref{eq:factorization} is dominated by the exchange of 
$t$-channel gluons, which leads to a $\dt p_T^2/p_T^4$ behavior of the partonic cross section,
where $p_T$ is the transverse momentum of the final-state parton. This behavior, which is
shown in Fig.~\ref{fig:sigma_nd} leads to violations of the Froissart bound at high energies
if the cutoff scale, $p_{T,\rm min}$ is small enough.

The total cross section at hadron colliders consists of different components, which can be 
labeled according to the behavior of the beam particles after the scattering. If both beam
particles survive the collision intact, the collision is called elastic. If one of them
is excited, and the other stays intact, the collision is called single diffractive.
If both are excited, with a large rapidity gap of no activity in between, the collision 
is called double diffractive. Finally, if both beam particles disintegrate and no rapidity
gap is observed, the collision is called non-diffractive. The total hadronic cross section
is then determined as the sum of all these contributions. 

In this section we will focus only on the improved description of the non-diffractive part of the 
total cross section using multiple-parton scattering models. In very rough terms, this means that
the non-diffractive cross section is saturated by more than a single partonic scattering, and that
the number of partonic interactions is determined by a Poisson distribution. This model, which
was originally proposed in~\cite{Sjostrand:1987su} has been very successful in the description
of many experimental measurements at hadron colliders.

The average multiplicity, according to the simple, impact-parameter independent model is given by
\begin{equation}\label{eq:average_n_scatterings}
  \abr{n}\,=\frac{\sigma_{\rm QCD}(p_{\perp \rm min}^2,s)}{\sigma_{\rm ND}(s)}\;.
\end{equation}
Assuming Poisson statistics, we can generate events in a Monte-Carlo simulation by defining
a no-scattering probability that is equivalent to the no-branching probability in
a parton shower:
\begin{equation}\label{eq:diff_probabilitiy_parton_parton}
  \mc{P}_{\rm MPI}(p_T,\mu_{\rm MPI}^2)\,=\;\exp\bigg\{
  -\frac{1}{\sigma_{\rm ND}}\int_{p_T}^{\mu_{\rm MPI}^2}\dt\bar{p}_T^2\,
  \frac{\dt \sigma_{QCD}}{\dt\bar{p}_T^2}\bigg\}\;.
\end{equation}
Events can then be generated using the veto algorithm in Sec.~\ref{sec:veto_algo}.
It is interesting to observe that the functional form of the hard QCD 
cross section is maintained by this formalism, due to the properties of
Poisson distributions. The total probability for any $2\to 2$ QCD scattering 
in this model to occur at hardness scale $p_T$ is given by precisely 
the integrand in the exponent of Eq.~\eqref{eq:diff_probabilitiy_parton_parton}.

\begin{figure}
  \begin{center}
    \includegraphics[width=0.5\textwidth]{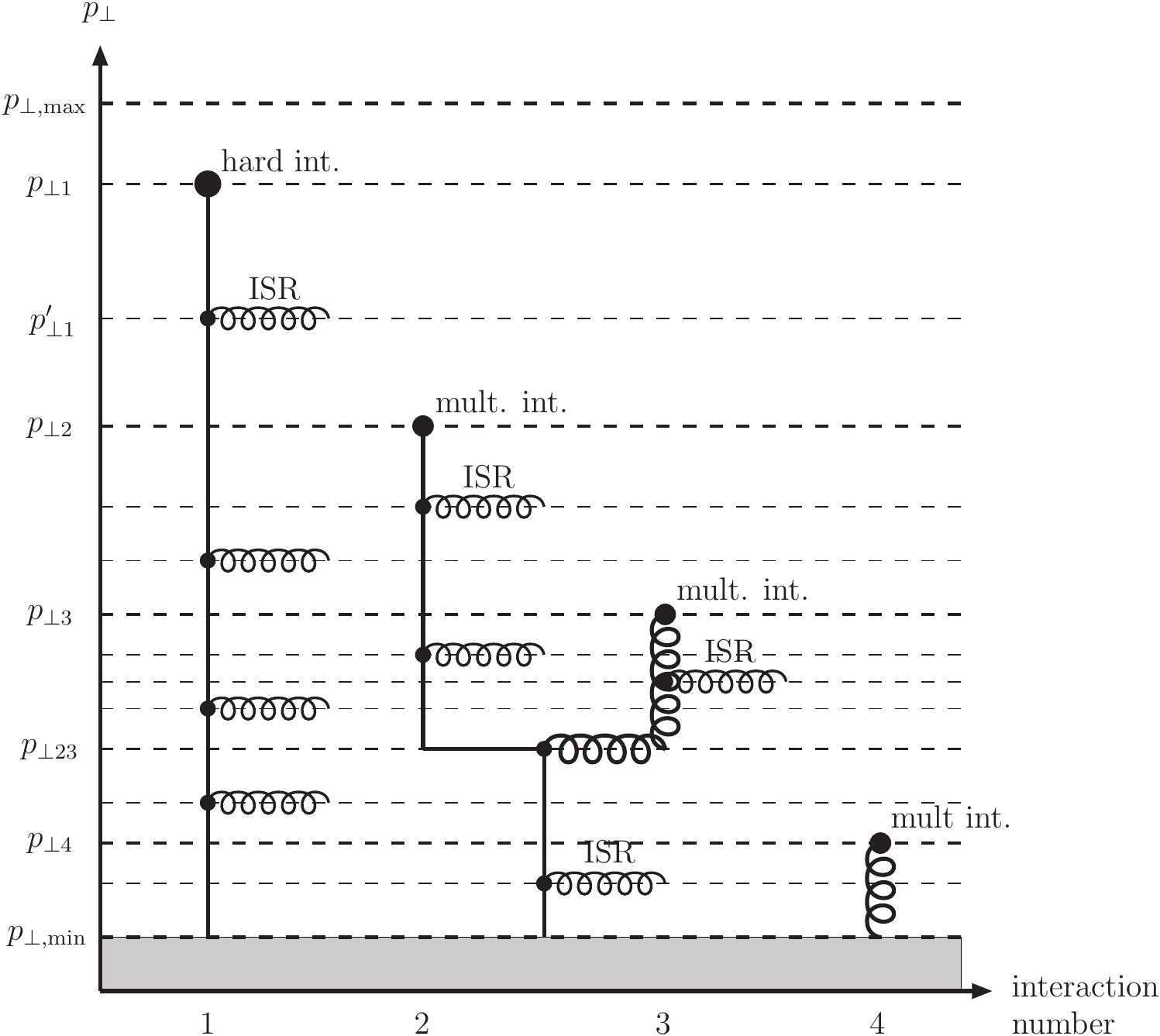}
    \caption{Sketch of interleaved parton shower and multiple scattering evolution.
      The evolution proceeds in the negative $p_\perp$ direction. The hard scattering 
      takes place at $p_{\perp 1}$, followed by parton-shower emissions.
      The first secondary scattering happens at $p_{\perp 2}$. At $p_{T,23}$,
      two parton-shower initiators are generated by splitting of a single initial-state
      parton. Figure taken from~\cite{Sjostrand:2004ef}.
      \label{fig:interleaved_ps_mpi}}
  \end{center}
\end{figure}

The event structure in hadronic interactions may actually be very complex,
leading to situations where a single initial-state parton can split into 
two before both of them enter a hard collision. At the same time an 
independently resolved parton may undergo another collision, while all of
them collectively radiate further gluons. Clearly, this situation is too
complex to be described exactly. But a good fraction of it may be modeled 
in event generators with an interleaved initial-state parton shower and multiple
interaction evolution~\cite{Sjostrand:2004ef}. The combined no-branching
probability for such an evolution is given by
\begin{equation}
  \mc{P}_{\rm MPI+PS}(p_T)\,=\;\mc{P}_{\rm MPI}(p_T)\,\Pi(p_T)\;,
\end{equation}
where $\Pi(p_T)$ is the no-branching probability of the parton shower, 
Eq.~\eqref{eq:sudakov}. Multiple interaction evolution and parton-shower 
evolution must have a common evolution variable for this model to be applicable.
A possible resulting event structure, together with the associated scales
at which the partons are resolved is depicted Fig.~\ref{fig:interleaved_ps_mpi}.

The structure of beam remnants and their connection with the many hard 
and semi-hard scatterings, especially the treatment of baryon number, 
is intricate once multiple parton interactions are included in the 
Monte-Carlo simulation. In simulations without a hard underlying event,
baryon number is normally carried by the diquark remnant of the proton.
This model must be improved when multiple scatterings are 
included~\cite{Sjostrand:2004pf}.

It is also necessary to consider the finite size of hadrons. 
Each proton-proton collision can be characterized by an impact parameter $b$,
which measures the transverse separation of the centers of incoming hadrons 
in position space. If $\rho({\bf x})$ denotes the (suitably Lorentz contracted)
hadronic-matter distribution, the time-integrated overlap between the two 
distributions in the center of mass frame is given by
\begin{equation}
  \tilde{O}(b)\,=\;\int\dt^3{\bf x}\,\dt t\,
  \rho_{1\rm b}\left(x-{\textstyle\frac{1}{2}}\,b,y,z-vt\right)\,
  \rho_{2\rm b}\left(x+{\textstyle\frac{1}{2}}\,b,y,z+vt\right)
\end{equation}
It is natural to assume that there is a linear relationship between the overlap
and the mean number of hard interactions in the event,
$\abr{\tilde{n}(b)}\,=\;k\,\tilde{O}(b)$. However, we also have the requirement
that each event contain at least one hard interaction. For each impact parameter
the number of interactions should be Poisson distributed. This requires that
\begin{equation}\label{eq:average_pre_n_scatterings_ip}
  \abr{\tilde{n}(b)}\,=\frac{k\,\tilde{O}(b)}{1-\exp\{-k\tilde{O}(b)\}}
  =\frac{k\,\tilde{O}(b)}{P_{\rm int}(b)}\;.
\end{equation}
with $P_{\rm int}(b)=1-\exp\{\,-k\tilde{O}(b)\}$ the total interaction
probability. When averaged over all impact parameters, $\abr{n(b)}$ must satisfy
Eq.~\eqref{eq:average_n_scatterings}, requiring that:
\begin{equation}\label{eq:average_n_scatterings_ip}
  \frac{\sigma_{\rm QCD}(p_{T,\rm min},s)}{\sigma_{\rm ND}(s)}\,=\;
  \frac{\displaystyle\int_0^\infty\dt^2 {\bf b}\,k\tilde{O}(b)}
  {\displaystyle\int_0^\infty\dt^2 {\bf b}\,P_{\rm int}(b)}\;.
\end{equation}
This allows one to compute the constant of proportionality, $k$.
As the normalization of $\tilde{O}(b)$ is irrelevant, it is 
convenient to introduce an enhancement factor $f(b)$, gauging 
how the interaction probability for a given impact parameter $b$ 
compares to the suitably defined average~\cite{Sjostrand:1987su}
\begin{equation}\label{definition_enhancement_factor}
  \abr{\tilde{n}(b)}\,=\;f(b)\,\big<k\,\tilde{O}\big>\;.
\end{equation}
This enhancement factor is normalized such that $\left<f\right>\,=\,1$.
The average number of scatterings $\abr{\tilde{n}(b)}$ is
\begin{equation}
  \abr{\tilde{n}(b)}\,=\;f_c f(b)\,
  \frac{\sigma_{\rm QCD}(p_{T,\rm min},s)}{\sigma_{\rm ND}(s)}\;.
\end{equation}
The full no-scattering probability in this model is the given as
\begin{equation}\label{eq:diff_probabilitiy_parton_parton_ip}
  \mc{P}_{\rm MPI}(b,p_T,\mu_{\rm MPI}^2)\,=\;\exp\bigg\{
  -f_cf(b)\,\frac{1}{\sigma_{\rm ND}}\int_{p_T}^{\mu_{\rm MPI}^2}\dt\bar{p}_T^2\,
  \frac{\dt \sigma_{QCD}}{\dt\bar{p}_T^2}\bigg\}\;.
\end{equation}
One can assume different hadronic matter distributions, like exponential,
Gaussian or double Gaussian distributions. More complicated pictures
can also be imagined.

It may also be useful to model saturation effects by requiring a hard 
cross section which has no sharp cutoff at the minimum scale $p_{T,\rm min}$. 
The simplest possible procedure to account for this effect is to regularize
the differential cross section by including a factor
\begin{equation}\label{naive_regularisation_ppxs}
  \frac{p_T^4}{(p_T^2+p_{T\,0}^2)^2}
  \frac{\alpha_s^2(p_T^2+p_{T\,0}^2)}{\alpha_s^2(p_T^2)}\;,
\end{equation}
where $p_{T\,0}$ is the regularization scale.

\begin{figure}
  \begin{center}
    \includegraphics[width=0.5\textwidth]{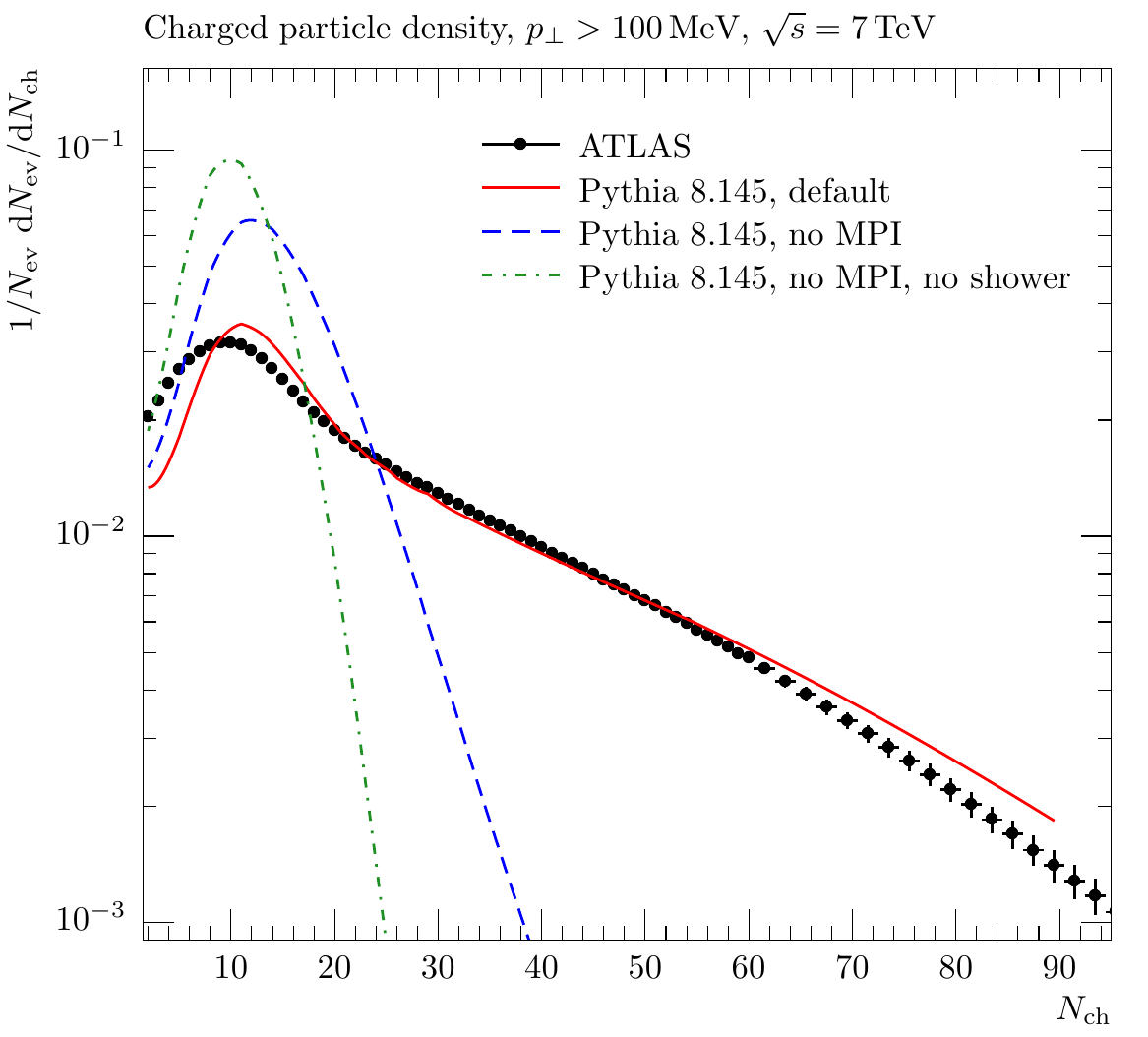}
    \caption{Effects of the simulation of Multiple Parton Scatterings (MPI) on the
      charged particle multiplicity distribution at the LHC as measured by 
      the ATLAS experiment~\cite{Aad:2010ac}. Figure taken from~\cite{Buckley:2011ms}.
      \label{fig:mpi_example}}
  \end{center}
\end{figure}
Figure~\ref{fig:mpi_example} shows the effect of the simulation of multiple 
scattering effects on the prediction for the charged particle multiplicity
spectra at the Large Hadron Collider. It is evident that without a simulation
of multiple interactions, the data cannot be described properly.

Since the original proposal for the simulation of multiple 
interactions~\cite{Sjostrand:1987su}, a variety of other models 
have been implemented in general-purpose event 
generators~\cite{Sjostrand:2004ef,Butterworth:1996zw,Bahr:2008dy}.
A review of the related models and predictions is given in~\cite{Buckley:2011ms}.
Other, more inclusive approaches to hadron-hadron scattering, which
naturally include multiple scattering effects also 
exist~\cite{Flensburg:2011kk,Martin:2012nm}.

\section{Hadronization}
\label{sec:hadronization}
To complete the simulation of realistic event topologies as observable experimentally,
the quarks and gluons from hard scattering simulations, parton showers and multiple 
scattering simulations must be transformed into color-neutral final states. 
In the context of a Monte-Carlo simulation this process is called hadronization or
jet fragmentation. Traditionally, the first hadronization model applicable to Monte-Carlo 
simulation was the Feynman-Field model~\cite{Field:1977fa}. It gives a recipe to 
produce mesons iteratively starting from a single quark. Because the hadronization of
each parton is considered separately in this model, it is also called ``independent
fragmentation''. However, it suffers from frame dependence and collinear unsafety.
The two hadronization models used today are the string and cluster models, which are
based on the ideas pioneered in~\cite{Artru:1974hr}.

\subsection{String model}
\begin{figure}
  \begin{center}
    \includegraphics[width=0.6\textwidth]{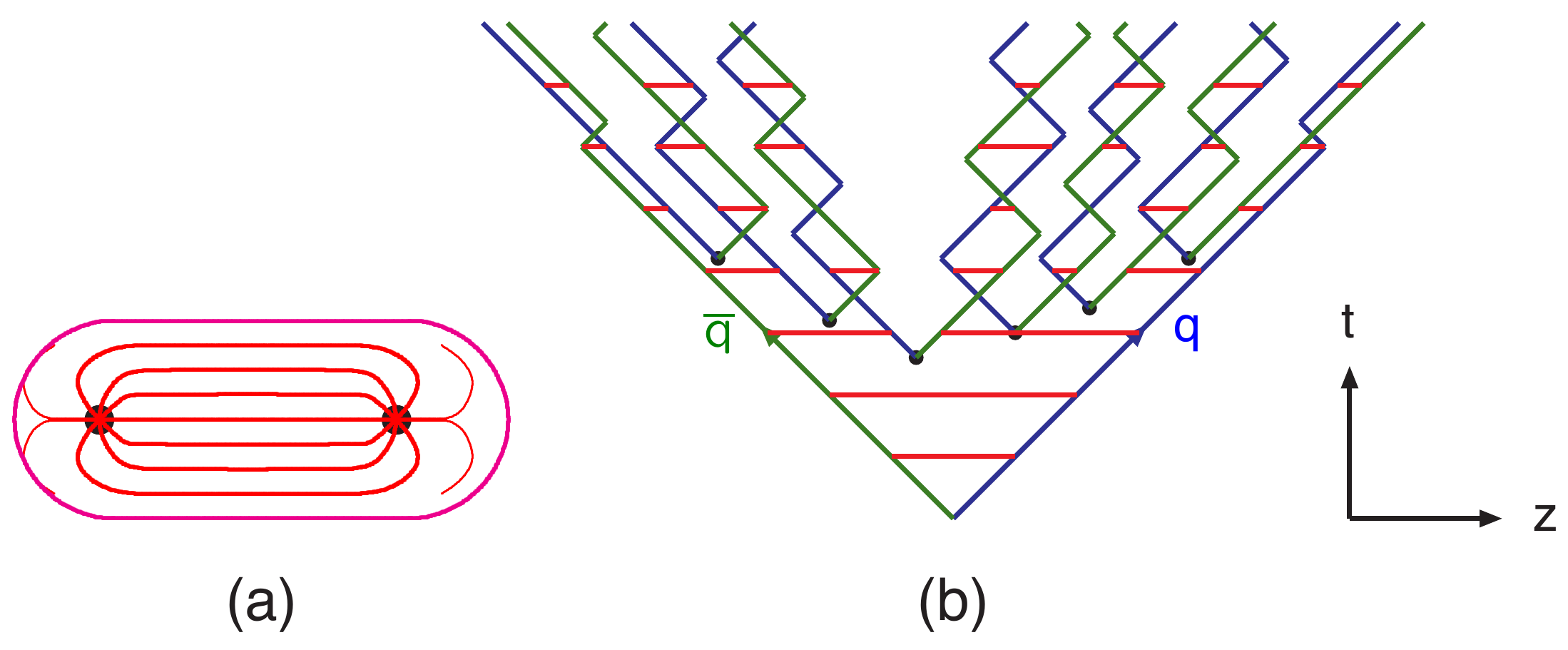}
    \caption{Left: Flux tube spanned between quark and antiquark.
      Right: Motion and breakup of a string system in the longitudinal direction over time.
      Figure taken from~\cite{Buckley:2011ms}.
      \label{fig:lund_model}}
  \end{center}
\end{figure}
The string or Lund model of jet fragmentation~\cite{Andersson:1983ia,Andersson:1998tv}
is based on the observation that the quark-antiquark potential rises linearly 
with the distance between quarks in a meson system. This effect is measurable 
in quarkonium spectra, and it has been computed using lattice QCD in the 
quenched approximation~\cite{Bali:1992ab}. It translates into a physical picture 
where a flux tube is stretched between the two quarks, with constant energy 
per unit length, leading to a potential $V(r)=\kappa r$ where $\kappa\approx 1\,{\rm GeV/fm}$.
A sketch of such a flux tube is shown in Fig.~\ref{fig:lund_model} (left).

A Lorentz covariant and causal description of the energy flow in the flux tube
is obtained by the dynamics of a massless relativistic string with no transverse 
degrees of freedom, which parametrizes the longitudinal axis of the flux tube.
As a quark-antiquark pair produced at high energy moves apart at the speed of light,
the potential energy stored in the string stretched between it can lead to the
creation of a new quark-antiquark pair, such that the system splits into two 
color-neutral strings with a quark/antiquark at either end. If the energy stored
the field between the new quark-antiquark pairs is large enough, further string
breaks may occur until no further partitioning is possible and the quarks enter
into ``yo-yo'' motion about each other. This is shown in Fig.~\ref{fig:lund_model} 
(right). The space-time picture can be mapped
onto a corresponding picture in momentum space, where $\dt p_z/\dt t=\kappa$.
Different string breaks are causally separated. The fragmentation function describing 
the string breakup should therefore exhibit left-right symmetry.

The Lund model proposes the use of the Lund symmetric fragmentation function,
\begin{equation}
  f(z)\varpropto \frac{1}{z}(1-z)^\alpha\exp\rbr{-\frac{b\,m_T^2}{z}}\;.
\end{equation}
where $z$ is the remaining light-cone momentum fraction of the quark (antiquark)
in the $+z$ ($-z$) direction and $a$ and $b$ are free parameters~\cite{Andersson:1983jt}.
A slightly modified form is introduced for heavy quarks~\cite{Bowler:1981sb}.
The transverse motion of the newly-created quarks/antiquarks is parametrized as
a quantum mechanical tunneling effect, with probability proportional to
\begin{equation}\label{eq:lund_tunneling}
  \exp\rbr{-\frac{\pi\,m_T^2}{\kappa}}=
  \exp\rbr{-\frac{\pi\,m^2}{\kappa}}\exp\rbr{-\frac{\pi\,p_T^2}{\kappa}}
\end{equation}
The factorization of mass and transverse momentum dependence then leads to a
flavor-independent transverse-momentum spectra of the hadrons with an average
of $\abr{p_T^2}=2\kappa/\pi$. Equation~\eqref{eq:lund_tunneling} also implies 
a natural heavy-flavor suppression.

In the simplest scheme for baryon production, diquark pairs are produced instead
of quark pairs. A more advanced model is the popcorn approach, where baryons appear
from multiple production of quark pairs.

Gluons are accommodated in the string model as kinks on the flux tube stretched 
between the two initial quarks. As such, the gluon can also be assigned the 
incoherent sum of a color and an anticolor charge, which effectively models
the dynamics of the color field in the large-$N_C$ approximation 
(cf.\ Sec.~\ref{sec:large_nc_ps}). This leads to a genuine prediction of the
Lund string model, called the string effect: Final states containing a quark-antiquark
pair and a gluon should receive enhanced hadron production in the angular regions between
the quark and the gluon and the gluon and the antiquark. This was confirmed in
experiments~\cite{Bartel:1983ii}. Crucially, the string model of jet fragmentation 
is infrared and collinear safe, because a soft or collinear gluon induces 
a vanishingly small kink on the color string~\cite{Sjostrand:1984ic}.

Pythia~\cite{Sjostrand:2006za,Sjostrand:2007gs} is the only Monte-Carlo event generator
which currently implements the string model.

\subsection{Cluster model}
\begin{figure}
  \begin{center}
    \includegraphics[width=0.55\textwidth]{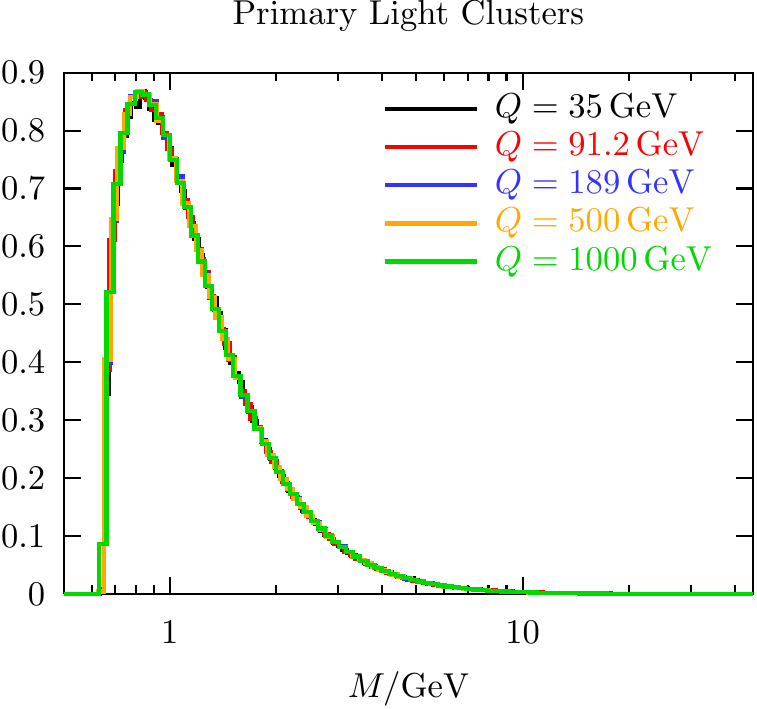}
    \caption{Invariant mass distribution of color singlet clusters in the cluster model.
      Figure taken from~\cite{Buckley:2011ms}.
      \label{fig:cluster_model}}
  \end{center}
\end{figure}
The cluster-hadronization model is based on the so-called preconfinement property
of QCD~\cite{Amati:1979fg}. This means that at each point the parton shower forms
color-singlet combinations of partons, called clusters, which have an asymptotically
universal invariant mass distribution. In this context, universal means that the distribution
only depends on the cutoff scale $t_c$ of the parton shower and on $\Lambda_{\rm QCD}$,
but not on the center-of-mass energy of the collision~\cite{Amati:1979fg,Bassetto:1979vy}.
This is shown in Fig.~\ref{fig:cluster_model} for a variety of center-of-mass energies 
in $e^+e^-$ collisions. 

Preconfinement can be inferred from the topology of parton-shower final states,
where color-adjacent partons, due to the large-$N_C$ approximation, are also adjacent
in phase space, as adjacency implies they likely originated at the same evolution scale.
Therefore clusters of large invariant mass are suppressed.

The first hadronization model based on preconfinement was the Field-Fox-Wolfram
model~\cite{Fox:1979ag,Field:1982dg}. In this model, a non-perturbative splitting of
final-state gluons into quarks is enforced at the end of the parton-shower evolution.
Adjacent color lines then form primary clusters with a mass distribution as shown
in Fig.~\ref{fig:cluster_model}. The mass distribution of these final states is closely
connected to hadron spectra, an effect known as local parton-hadron 
duality~\cite{Azimov:1984np,Azimov:1985by}. The flavor assignment in gluon splitting
is important to obtain the correct heavy flavor suppression. This can be approximated
to a good extent by kinematic effects, which reduce the phase space for heavy flavor
production.

Once primary clusters are formed, the ones with mass below 3-4~GeV are transformed 
into hadrons through a two-body decay according to phase space. Heavier clusters may 
first undergo non-perturbative splitting processes, and decay into two lighter clusters,
or a lighter cluster and a hadron, before the cluster-to-hadron transition is resumed.
This process is repeated until all clusters have been transformed into hadrons.
Very low mass clusters may transition directly into hadrons, in which case another
hadron or cluster must absorb the recoil if the cluster mass is different from the
hadron mass.

Two cluster-hadronization models are currently widely used, which are implemented in 
Herwig++~\cite{Webber:1983if} and Sherpa~\cite{Winter:2003tt}.

\section{Summary}
Parton-shower event generators are indispensable tools for particle physics
phenomenology at hadron colliders. They are used in the planning of new experiments,
detector design and performance studies, and in the extraction of theoretical parameters
from the measurements themselves.

Event simulation in modern generators starts with the computation of hard interactions,
often at higher orders in perturbation theory. QCD Bremsstrahlung is then simulated
using the parton-shower approach, and the resummed higher-order calculation of the 
parton shower is matched to / merged with the higher-order fixed-order calculations
for the hard processes. Multiple scattering effects are simulated by
repeated generation of hard processes according to the hard cross sections for jet
production in perturbative QCD, such that the non-diffractive part of the total cross
section is saturated. Eventually, the perturbatively-computed final state is transformed
into measurable hadrons by means of hadronization models.

Event generators traditionally contain several free parameters, especially in
the simulation of hadronization and multiple scattering effects. The simulation 
of hard QCD radiation, however, is based on perturbative QCD in the parton
shower approximation, which is to a large extent defined by the factorization 
properties of QCD amplitudes and color-coherence effects in soft gluon emission.

\section*{Acknowledgments}
I wish to thank the scientific organizers of TASI 2014, Lance Dixon and Frank Petriello, 
and the local organizers, Tom Degrand and Kalyana Mahanthappa for a great school, and for the
enjoyable time together at UC Boulder. Many thanks to Valentin Hirschi for his help in
organizing and running the Monte-Carlo tutorials and to all the students for 
interesting discussions and a great atmosphere during the lectures and tutorials.
This work was supported by the US Department of Energy under contract DE--AC02--76SF00515.
\end{fmffile}
\bibliographystyle{amsunsrt_modp}
\bibliography{journal}
\end{document}